\documentclass[rmp,aps,twocolumn,eqsecnum,floatfix,showpacs]{revtex4} 
\usepackage{amssymb,latexsym,amsmath,bm}
\usepackage{graphicx,epsfig}

\begin{document}

\title{Rotating Trapped Bose-Einstein Condensates}
\author{Alexander L.~Fetter}
\affiliation{Geballe Laboratory for Advanced Materials and Departments of Physics and Applied Physics, Stanford University, Stanford, CA 94305-4045, USA}
\email{fetter@stanford.edu}
\date{\today}
\begin{abstract}
After reviewing the ideal Bose-Einstein gas in a box and in a harmonic trap, I discuss the effect of interactions on the formation of a Bose-Einstein condensate (BEC), along with the dynamics of small-amplitude perturbations (the Bogoliubov equations).  When the condensate rotates with angular velocity $\Omega$, one or several vortices nucleate, with  many  observable  consequences.  With more rapid rotation, the vortices form a dense triangular array, and the collective behavior of these vortices has additional experimental implications.  For  $\Omega$  near the radial trap frequency $\omega_\perp$, the lowest-Landau-level approximation becomes applicable, providing a simple picture of such rapidly rotating condensates.  Eventually, as $\Omega\to\omega_\perp$, the rotating dilute gas  is expected to undergo a quantum phase transition from a superfluid  to various  highly correlated (nonsuperfluid) states analogous to those familiar from the fractional quantum Hall effect for electrons in a strong perpendicular magnetic field.
\end{abstract}
\maketitle
\tableofcontents
\section{Introduction}
\label{sec:intro}

The remarkable creation of  Bose-Einstein condensates (BECs) in cold dilute alkalai-metal gases \cite{Anderson:1995, Bradley:1995, Davis:1995} has produced a wholly new exciting field that continues to thrive.  Typically, these systems have a macroscopic condensate wave function $\Psi(\bm r, t)$ that characterizes the static and dynamic behavior of the BEC [for a general background, see, for example,~\textcite{Dalfovo:1999,Inguscio:1999,Pethick:2002,Pitaevskii:2003}].  In the usual limit of a dilute gas, $\Psi$ obeys a nonlinear Schr\"odinger equation known as the Gross-Pitaevskii equation~\cite{Gross:1961,Pitaevskii:1961}.  

Subsequent experiments formed rotating condensates, initially with a small number of vortices~\cite{Matthews:1999a,Madison:2000a}, and then with large vortex arrays~\cite{Abo:2001,Haljan:2001b}.  As in the case of rotating superfluid $^4$He~\cite{Donnelly:1991}, the vortices have quantized circulation, arising from the single-valued nature of the condensate wave function $\Psi$.  Nevertheless, the new dilute-gas  BECs differ considerably from superfluid $^4$He because the Gross-Pitaevskii equation here provides a remarkably detailed description of the physics, allowing a careful comparison of theory and experiment.  

This article focuses on the  physics of quantized vortices in ultracold 
dilute trapped quantum gases.  I first review the dilute Bose-Einstein gas, both in a box where the density is uniform and in harmonic traps where the density is nonuniform (Sec.~II).  The presence of (repulsive)  interactions have a  dramatic effect, leading to (1) a collective mode with a linear long-wavelength dispersion relation  and (2) superfluidity of macroscopic flow.  Section III treats the behavior of one vortex (or a small number of vortices) in a rotating condensate.   For more rapid rotations, the vortices form a regular array that has many analogies with rotating $^4$He and with the flux-line lattice in type-II superconductors (Sec.~IV).  As the rotation rate $\Omega$ continues to increase and approaches the radial trapping frequency $\omega_\perp$, the effective radial trapping potential weakens, and condensate expands.  In this limit, the system enters a new two-dimensional  superfluid regime that has close analogies with the lowest-Landau-level description of an electron in a uniform magnetic field (Sec.~V).  Ultimately, in the limit $\Omega/\omega_\perp \gtrsim  0.999$, the rotating Bose gas is predicted to make a quantum phase transition to one of several  highly correlated ground states that are not superfluid (Sec.~VI).  This regime involves a sequence of many-body states that are similar to those developed for the quantum Hall effect (a two-dimensional electron gas in a strong magnetic field).  

Throughout this article, I follow the lead of \textcite{Bloch:2007},  focusing on models and predictions with experimental basis or confirmation.  As a result, I  treat only briefly  two areas that have been studied in some experimental detail  as  nonrotating systems:  (1) dipole condensates that interact with long-range electric or magnetic dipole forces [see, for example, ~\textcite{Griesmaier:2005,Lahaye:2007} and references therein]   (2) spinor condensates [see, for example, \textcite{Stenger:1998,Sadler:2006,Vengalattore:2007} and references therein].  In both cases, however,  rotation and the corresponding vortex structures remain an unexplored experimental capability (see, however, Sec.~VII.B for an introductory discussion).

\section{Physics of Bose-Einstein Condensates in Dilute Trapped Gases}
\label{sec:BEC}

The general topic of Bose-Einstein condensates (BEC) in dilute trapped gases has been the focus of several reviews and books~\cite{Inguscio:1999, Dalfovo:1999,  Pethick:2002, Pitaevskii:2003}.  The current  article will emphasize those aspects most relevant to rotating trapped gases and the associated quantized vortices~\cite{Fetter:2001,Aftalion:2006,Kasamatsu:2007,Parker:2007}.  In particular, I will emphasize the regime of regular vortex arrays, both the ``mean-field Thomas-Fermi" limit when the vortex cores remain small and  the ``lowest-Landau-level'' limit when the cores are comparable to the intervortex spacing~\cite{Ho:2001,Baym:2005}.  

To review the basic facts of Bose-Einstein condensation, let us consider an ideal (noninteracting) uniform three-dimensional gas with $N$ particles in a cubical box (volume $V = L^3$)  with mean density $n = N/V$.  If the gas has a temperature $T$, the mean momentum per particle is  $p_T \approx \sqrt{Mk_B T}$, ignoring numerical factors of order unity, where   $M$ is the atomic mass and $k_B$ is Boltzmann's constant.  The de~Broglie relation then yields the thermal wavelength $\lambda_T \approx h/p_T\approx h/ \sqrt{ Mk_B T}$.  The other relevant length is the interparticle spacing $n^{-1/3}$.  The short-wavelength limit $ \lambda_T\ll n^{-1/3} $ holds when $\hbar \to 0$ or  $T$ is large; it describes the classical regime when  quantum interference and diffraction are  negligible (the analog of ray optics).  As $T$ falls, however, this  inequality eventually fails when $ \lambda_T \approx n^{-1/3}$, or, equivalently, when the ``phase-space density''  defined as $n\lambda_T^3$ becomes of order unity.  This condition   signals the onset of quantum degeneracy;   whenever $n\lambda_T^3 \gtrsim 1$, quantum mechanics plays a central role in the physics.

The preceding description relies on an atomic-physics perspective, but a condensed-matter view yields a similar picture.  In an ideal gas, each atom is effectively confined to a volume $n^{-1}$ by its neighbors, leading to a zero-point energy $\epsilon_{\rm zp}\approx \hbar^2 n^{2/3}/M$.  At high $T$ or low density ($k_BT \gg \epsilon_{\rm zp} $),  the system behaves classically, but when $k_BT \approx \epsilon_{\rm zp}$ (either because of reduced temperature or increased density),  quantum effects again become crucial. These two characterizations  of the onset of quantum degeneracy are essentially equivalent.  

For ideal bosons, the critical temperature $T_c$  for the onset of quantum degeneracy leads to  Bose-Einstein condensation,  whereas for ideal Fermi gases, the similar onset temperature is known as the Fermi temperature $T_F$.  For liquid helium with $n \approx 10^{22}$ cm$^{-3}$, the appropriate transition temperature is of order 1 K for both isotopes ($^4$He is a boson and $^3$He is a  fermion).  A typical bosonic dilute alkalai-metal gas (such as $^7$Li, $^{23}$Na,  and $^{87}$Rb) has  a much lower $T_c$   of order 100-1000 nK, because of the larger atomic mass and  reduced number density ($n\approx 10^{13}$ cm$^{-3}$ is a typical value for dilute cold gases). This article will focus on such  bosonic  systems.  

The presence of Bose-Einstein condensation means that a single quantum state has macroscopic occupation.  This particular quantum state acts like a particle reservoir that can absorb or emit excited particles with negligible change in its own properties.  Thus it is natural to use the grand canonical ensemble in which the system of interest is assumed to be in equilibrium with a reservoir at temperature $T$ and chemical potential $\mu$ that determine the mean total  energy  and  mean total number of particles.

\subsection{Ideal Bose gas}

Consider an ideal Bose gas in an external trap potential $V_{\rm tr}$, with a complete set of quantum-mechanical single-particle energies~$\epsilon_j$, and assume that the system is in equilibrium at temperature $T$ and chemical potential $\mu$.  In the grand canonical ensemble,  the mean occupation of the $j$th state is 
\begin{equation}
\label{eq:bose}
n_j = \frac{1}{\exp\left[\beta\left(\epsilon_j-\mu\right)\right]-1} \equiv f\left(\epsilon_j\right),
\end{equation}
where $\beta = (k_B T)^{-1}$ is proportional to  the inverse temperature.  Here, $f(\epsilon) = \{\exp\left[\beta\left(\epsilon-\mu\right)\right]-1\}^{-1} $ is the usual Bose-Einstein distribution function (it depends explicitly on $T$ and $\mu$), as shown, for example,  in Sec.~5 of \textcite{Fetter:2003}, Chap. V of ~\textcite{Lifshitz:1980a}, Part 1,  or Chap.~7 of~\textcite{Pathria:1996}.  A detailed analysis shows that the mean total  number of particles $N(T,\mu)$ and the mean total energy $E(T,\mu)$ are given by 
\begin{eqnarray}
\label{eq:N}
N(T,\mu) & = & \sum_j f(\epsilon_j),\\
\label{eq:E}
E(T,\mu) & = & \sum_j \epsilon_j f(\epsilon_j).
\end{eqnarray}
In principle, the first relation Eq.~(\ref{eq:N}) can be inverted to give $\mu(T,N)$,
and substitution  into Eq.~(\ref{eq:E}) then  yields the  mean total energy $E(T,N)$ as a function of  temperature and the total number of particles.

It is useful to introduce the ``density of states'' 
\begin{equation}
g(\epsilon) = \sum_j \delta(\epsilon - \epsilon_j),
\end{equation}
so that Eqs.~(\ref{eq:N}) and (\ref{eq:E})  reduce to 
\begin{eqnarray}
\label{eq:Nint}
N(T,\mu) & = & \int d\epsilon\,g(\epsilon)\, f(\epsilon),\\
\label{eq:Eint}
E(T,\mu) & = & \int d\epsilon\, g(\epsilon)\,  \epsilon f(\epsilon).
\end{eqnarray}
Although $g(\epsilon)$ is formally  singular, a smoothed (coarse-grained) version  would be well-defined.

In the classical limit,  the chemical potential   $\mu_{\rm cl} = k_B T \ln(n\lambda_T^3)$ is large and negative since $n\lambda_T^3 \ll 1$.  As the temperature decreases (or the density increases), however, $\mu$ increases toward positive values.   The critical temperature $T_c$ for the onset of BEC is  defined implicitly through the relation  $\mu(T_c,N) =\epsilon_0$, where $\epsilon_0$ is the ground-state energy of the single-particle Hamiltonian with the potential $V_{\rm tr}$.  For $T < T_c$, the chemical potential remains fixed at this value, and the ground state has a macroscopic occupation  $N_0(T)$, whose temperature dependence follows from the conservation of total particles $N = N_0(T) + N'(T)$, with 
\begin{equation}
\label{eq:N'}
N'(T) = \int_{\epsilon_0}^\infty d\epsilon\,\frac{g(\epsilon)}{\exp\left[\beta \left(\epsilon-\epsilon_0\right)\right]-1}
\end{equation}
 defining the total number of particles {\it not} in the condensate.  If $g(\epsilon_0)$ vanishes, then this integral is finite and well defined, and the number of noncondensed particles decreases with decreasing temperature. 
In contrast, if $g(\epsilon_0)$ is finite or singular, then this integral for the number of noncondensed particles diverges,  and the system cannot form a BEC.    It is helpful to examine two specific examples.

\subsubsection{Ideal BEC in a box with periodic boundary conditions}

The textbook example  of BEC is an ideal gas in a three-dimensional cubical box of volume $V = L^3$ with periodic boundary conditions.  The single-particle eigenfunctions are plane waves $\psi_{\bm k} (\bm r) = V^{-1/2} \exp(i\bm k \cdot \bm r)$ with single-particle energy $\epsilon_k = \hbar^2 k^2 /(2M)$, where $\bm k = (2\pi/L) (n_x,n_y,n_z)$ with $n_j$ any integer.  It is not hard to obtain the corresponding density of states
\begin{equation}
\label{eq:dosbox}
g(\epsilon) = \frac{V}{4\pi^2}\,\left( \frac{2M}{\hbar^2}\right)^{3/2}\,\epsilon^{1/2}.
\end{equation}
In a box with periodic boundary conditions, the lowest single-particle energy is $\epsilon_0 = 0$, which corresponds to $\bm k = \bm 0$ (a uniform state with density $|\psi_{\bm 0}|^2 = V^{-1}$).  A standard analysis with the conventional definition $\lambda_T^2 = 2\pi\hbar^2/(Mk_B T)$  yields the condition 
\begin{equation}
\label{eq:BEC}
n\lambda_{T_c}^3 = \zeta(\textstyle{\frac{3}{2}}) \approx 2.612,
\end{equation}
which is the critical phase-space density for the onset of BEC.  Alternatively, the critical temperature $T_c$ is given by 
\begin{equation}
\label{eq:Tc}
k_B T_c =\frac{2\pi}{[\zeta(\textstyle{\frac{3}{2}})]^{2/3}}\,\frac{\hbar^2 n^{2/3}}{M}\approx 3.31 \,\frac{\hbar^2 n^{2/3}}{M},
\end{equation}
as anticipated from the previous qualitative discussion.  Below $T_c$, the (uniform) condensate with zero momentum has a macroscopic occupation number, which is the signature of BEC.  In particular, it is easy to verify that 
\begin{equation}
\label{eq:N'Tunif}
\frac{N'(T)}{N} = \left(\frac{T}{T_c}\right)^{3/2}
\end{equation}
for $T<T_c$, so that the macroscopic occupation of the  uniform ground state has the temperature dependence \begin{equation}
\label{eq:N0unif}
\frac{N_0(T)}{N} = 1 -  \left(\frac{T}{T_c}\right)^{3/2}.
\end{equation}

It is worth generalizing these results to a uniform  ideal Bose gas  in a $d$-dimensional hypercubical box;   the corresponding density of states $g(\epsilon)$ is proportional to $\epsilon^{d/2-1}$.  This alteration means that  the integral in Eq.~(\ref{eq:N'}) diverges  for $d\le 2$.  As a result,  the limit $\mu \to \epsilon_0$ requires a more careful analysis, leading to the conclusion that such a uniform gas in  two or one dimension cannot form a BEC; for a detailed treatment of the two-dimensional uniform Bose gas, see Chap.~2 of~\textcite{Pethick:2002}.

\subsubsection{Ideal BEC in a harmonic trap}

In three dimensions, an anisotropic harmonic trap potential 
\begin{equation}
\label{eq:trap}
V_{\rm tr}(\bm r) = \textstyle{\frac{1}{2}}M\left(\omega_x^2 x^2 +\omega_y^2 y^2 +\omega_z^2 z^2\right)
\end{equation}
 yields  single-particle energies that are labeled by a triplet of non-negative integers $n_x,n_y,n_z$
\begin{equation}
\label{eq:harm}
\epsilon_{n_x,n_y,n_z} = \hbar\left(n_x\omega_x + n_y\omega_y + n_z\omega_z\right) + \epsilon_0,
\end{equation}
where the second term is the zero-point energy $\epsilon_0 = \frac{1}{2} \hbar\left(\omega_x + \omega_y + \omega_z\right)$. The ground-state wave function is a product of three one-dimensional Gaussians with mean widths $d_j = \sqrt{\hbar/(M \omega_j)}$   (here $j = x$, $y$, or $z$).    If the sums in the density of states are approximated by integrals (which holds for large $\epsilon$), the corresponding density of states is 
\begin{equation}
\label{eq:dosharm}
g(\epsilon) =\frac{\epsilon^2}{2\hbar^3 \omega_0^3},
\end{equation}
where $\omega_0^3 = \omega_x\omega_y\omega_z$ defines a geometric-mean trap frequency.

The onset of BEC in a harmonic trap occurs at $\mu = \epsilon_0$, and an elementary analysis yields the transition temperature
\begin{equation}
\label{eq:Tcharm}
k_BT_c = \left[\zeta(3)\right]^{-1/3} \hbar\omega_0 N^{1/3} \approx 0.94 \hbar \omega_0 N^{1/3},
\end{equation}
where $\zeta(3) \approx 1.212$.  
Below $T_c$, a macroscopic number of particles occupies the anisotropic Gaussian ground state of this harmonic trap, which has a macroscopic occupation with  temperature dependence 
\begin{equation}
\label{eq:N0harm}
\frac{N_0(T)}{N} = 1 -  \left(\frac{T}{T_c}\right)^{3}.
\end{equation}
 Images of this temperature-dependent Gaussian condensate density profile provided the first clear evidence for the creation of BEC in $^{87}$Rb~\cite{Anderson:1995}.  Typical traps have $d_0=\sqrt{\hbar/(M\omega_0)}\sim$ a few $\mu$m and $N\sim 10^6$, confirming the previous value $T_c\approx 100$-$1000$ nK (depending on the atomic mass).

For a harmonic trap in $d$ dimensions, the density of states $g(\epsilon)$ is proportional to $\epsilon^{d-1}$, which differs from the $d$ dependence  for a hypercubical box.  As a result, a two-dimensional Bose gas in a trap {\em can}  form a BEC  with a finite transition temperature $T_c$,  in contrast to a uniform two-dimensional gas in a box.  This behavior is especially important in the limit of rapidly rotating Bose-Einstein condensate, when the centrifugal forces flatten the atomic system, producing an effectively two-dimensional trapped gas.   Chapter 2 of \textcite{Pethick:2002} discusses this  situation in some detail, especially the behavior in a one-dimensional trap, where the order of limits $N\to \infty$ and $T\to 0$ becomes crucial.

\subsection{Inclusion of interparticle interactions}

Roughly sixty years ago,~\textcite{Bogoliubov:1947} introduced what is now seen as the essential  physical approximation for  a dilute  Bose gas with repulsive interactions.  To understand his idea, note that the repulsive interparticle interaction can be characterized by the positive $s$-wave scattering length $a$, which is typically a few nm for the dilute alkalai-metal atoms of interest [see Chap.~5 of \textcite{Pethick:2002}].   In the low-density limit ($a\ll n^{-1/3}$), the ``gas parameter'' $na^3$ is small and provides a natural expansion parameter.  Bogoliubov notes that  the ground state of the dilute gas will then be close to that of the ideal gas.  Thus the number of  particles $N_0$  in the condensate will be close to the total number $N$, and the difference can be ignored in first approximation.  The macroscopic occupation of the single-particle mode for the condensate means that the creation and annihilation operators for this particular mode can be approximated as ``classical fields'' similar to the electric fields for a laser mode.   Bogoliubov's seminal physical picture   underlies almost all the subsequent developments in the field and is basic for understanding the behavior of a low-temperature BEC.  

For a uniform Bose gas at rest, the condensate is the zero-momentum state, and the excited states are labeled by wavenumber  $\bm k \neq \bm 0$, as discussed, for example, in Sec.~25 of \textcite{Lifshitz:1980b}, in Sec.~7.2 of \textcite{Pethick:2002}, and Secs.~18 and 35 of \textcite{Fetter:2003}.  In the present case of a trapped condensate, however, the analysis is somewhat more intricate and relies on the macroscopic condensate wave function $\Psi$ (sometimes called the ``order parameter'').  This condensate wave function describes the single-particle mode that has macroscopic occupation, with its squared absolute value $|\Psi(\bm r)|^2 = n(\bm r)$ giving the nonuniform condensate particle density $n(\bm r)$.   For a dilute gas at low temperature with $N_0\approx N$, the  normalization requires \begin{equation}
\label{eq:norm}
 \int dV\,|\Psi(\bm r)|^2 = N_0 \approx N.
\end{equation}

In the limit of a nearly ideal Bose gas at $T = 0$ K, the spatial form of this condensate wave function follows by minimizing the total energy $E$, along with the constraint that the total number of particles $N$ is conserved.  Equivalently, one can simply minimize the grand-canonical thermodynamic potential $ E - \mu N$, where $\mu$ is the chemical potential.  
To understand this approach, I shall generally rely on physical arguments, but many-body quantum field theory provides a  rigorous foundation for the same results~\cite{Fetter:1972,Fetter:1996,Hohenberg:1965}.  In a trap with confining potential $V_{\rm tr}$, the Gross-Pitaevskii~\cite{Gross:1961,Pitaevskii:1961} energy functional for the condensate has the form
\begin{equation}
\label{eq:EGP}
E_{GP}[\Psi] = \int dV \left(\underbrace{\frac{\hbar^2 |\bm \nabla \Psi|^2}{2M}}_{\rm kinetic} +\underbrace{ V_{\rm tr} |\Psi|^2}_{\rm trap}+ \underbrace{\frac{1}{2}g|\Psi|^4}_{\rm interaction}\right)
\end{equation}
containing the kinetic energy, the trap energy [see Eq.~(\ref{eq:trap})], and the interaction energy, respectively. The first two  terms (those quadratic in $\Psi$) are just the one-body energy for an ideal Bose gas in a (usually) harmonic trap.  In contrast, the quartic term (the two-body energy) describes  the effect of interactions, where the interparticle potential has been approximated by a short-range ``pseudopotential'' $V(\bm r-\bm r') \approx g \,\delta^{(3)}(\bm r - \bm r')$, with  $g= 4\pi a\hbar^2/M$  a coupling constant fixed by the $s$-wave scattering length~$a$ [this result follows from standard two-body quantum-mechanical scattering theory, for example, Chap.~5 of \textcite{Pethick:2002} or Secs.~11 and 35 of \textcite{Fetter:2003}].

Comparison of the kinetic and trap energies for a harmonic potential yields the familiar oscillator length 
\begin{equation}
\label{eq:d0}
d_0 = \sqrt{\frac{\hbar}{M\omega_0}}
\end{equation}
 that characterizes the mean size of the noninteracting condensate, with $\omega_0 = (\omega_x\omega_y\omega_z)^{1/3}$.  Similarly, comparison of the kinetic energy and the interaction energy for a uniform condensate yields the ``healing length''
 \begin{equation}
 \label{eq:xi}
\xi = \frac{\hbar}{\sqrt{2Mgn}} = \frac{1}{\sqrt{8\pi a n}},
\end{equation}
which characterizes the length scale over which a localized alteration in the condensate density heals back to its uniform value $n$.   In this context,  recall   the Bogoliubov approximation  of small quantum depletion $N'\ll N$, which  requires  $na^3 \ll 1$. As a result,  Eq.~(\ref{eq:xi}) has the important corollary 
\begin{equation}
\label{eq:dilute}
n^{2/3} \xi^2= \frac{1}{8\pi n^{1/3}a}\gg 1,
\end{equation}
namely  the healing length must exceed the interparticle spacing in any GP description.

Variation of $E_{GP}[\Psi]$ in Eq.~(\ref{eq:EGP}) with respect to $\Psi^*$ at fixed normalization yields the time-independent Gross-Pitaevskii (GP) equation~\cite{Gross:1961,Pitaevskii:1961}
\begin{equation}
\label{eq:GP}
\left[-\frac{\hbar^2\nabla^2}{2M} + V_{\rm tr}(\bm r) + g|\Psi(\bm r)|^2\right]\Psi(\bm r) = \mu \Psi(\bm r),
\end{equation}
Here the chemical potential $\mu$  can either be considered a Lagrange multiplier or a parameter in the zero-temperature grand-canonical thermodynamic potential $E-\mu N$.  In  the present context, this GP equation is essentially a nonlinear Schr\"odinger equation that includes an additional  quadratic selfcoupling $g|\Psi|^2$;  such a  term can  be interpreted as a Hartree potential $V_H(\bm r) = g n(\bm r)$ that represents the interaction with the local nonuniform condensate density. 

Compared to a uniform BEC, a trapped condensate involves an additional characteristic length  $d_0$ (the noninteracting condensate  size).  A simple scaling argument [see Sec.~6.2 of \textcite{Pethick:2002}]  yields a new dimensionless parameter $Na /d_0$ that characterizes the importance of the interaction  in a trapped condensate. As a result, the  three terms in Eq.~(\ref{eq:EGP}) can have very different magnitudes depending on the parameters.  In the usual situation ($N\approx 10^6$, $a \approx$ a few nm, and $d_0 \approx $ a few $\mu$m), this dimensionless parameter  $Na /d_0$ is large, and the resulting regime is known as the ``Thomas-Fermi limit''~\cite{Baym:1996}.\footnote{As seen below, the limit $Na/d_0 \gg 1$ means that the condensate density is effectively  uniform on   relevant length scales,  which explains the name ``Thomas-Fermi.''}   In this case, the repulsive interactions dominate and expand the condensate to a mean radius $R_0$ that greatly exceeds the mean oscillator length $d_0$ (a factor of 10 is typical).  This expansion dramatically reduces   the radial gradient of the density, and the associated kinetic energy thus becomes negligible relative to the trap energy and the interaction energy.  Equation (\ref{eq:EGP}) then reduces to 
\begin{equation}
\label{eq:ETF}
E_{\rm TF}[\Psi]\approx \int dV\left(V_{\rm tr}|\Psi|^2 + {\textstyle\frac{1}{2}} g|\Psi|^4\right), 
\end{equation}
which involves only $|\Psi|^2$ and $|\Psi|^4$.   Minimization with respect to $|\Psi|^2$ at fixed normalization gives the Thomas-Fermi (TF)  approximation
\begin{equation}
\label{eq:nonuniform}
V_{\rm tr}(\bm r)  + g|\Psi(\bm r)|^2 = \mu,
\end{equation}
which also follows by omitting the kinetic energy term in Eq.~(\ref{eq:GP}).
This algebraic equation  can be solved directly for the equilibrium density
\begin{equation}
\label{eq:nTF}
n(\bm r) = \frac{\mu - V_{\rm tr}(\bm r)}{g}\,\theta\left[\mu - V_{\rm tr}(\bm r)\right],
\end{equation}
where $\theta(x) $ is the unit positive step function.  Although this expression
 holds for any reasonable trap potential,  the usual case of a quadratic harmonic trap (\ref{eq:trap}) yields the simple and explicit expression 
\begin{equation}
\label{eq:parabolic}
n(\bm r) = n(0)\left( 1 - \frac{x^2}{R_x^2} - \frac{y^2}{R_y^2} - \frac{z^2}{R_z^2}\right) 
\end{equation}
where the right-hand side is positive and zero otherwise.  Here $n(0) = \mu/g $ is the central density and 
 $R_j^2 = 2\mu/(M\omega_j^2) $ are the squared condensate radii in the three coordinate directions.  This TF density has a  parabolic cross section and fills the interior of an ellipsoid.

Application of the  normalization condition (\ref{eq:norm}) to the TF density (\ref{eq:parabolic})  yields 
$N = 8\pi n(0)R_0^3/15$, where 
$R_0^3 \equiv R_xR_yR_z$ depends on the chemical potential $\mu$. Some algebra shows that 
$n(0) = \mu/g =R_0^2/(8\pi a d_0^4)$, and a combination of these results yields the important dimensionless relation
\begin{equation}
\label{eq:Nad}
\frac{R_0^5}{d_0^5} =15 \, \frac{N a}{d_0}, 
\end{equation}
which is large in the present TF limit.  Correspondingly, the TF chemical potential becomes 
\begin{equation}
\label{eq:muTF}
\mu_{\rm TF} = \frac{1}{2} \,M \omega_0^2 R_0^2 =\frac{1}{2} \,\hbar \omega_0\, \frac{R_0^2}{d_0^2},
\end{equation}
so that $\mu_{\rm TF} \gg \hbar\omega_0$ in the TF limit.  Since $\mu_{\rm TF}$ is proportional to $N^{2/5}$, the thermodynamic relation $\mu = \partial E/\partial N$ yields the TF energy for a trapped condensate 
\begin{equation}
\label{eq:ETFN}
E_{\rm TF} = \textstyle{\frac{5}{7}} \,\mu_{\rm TF} N,
\end{equation}
which also follows by direct integration of Eq.~(\ref{eq:ETF})  for the TF density profile.

It is conventional to use the central density $n(0)$ to define the healing length [Eq.~(\ref{eq:xi})] in a nonuniform trapped condensate, which gives 
\begin{equation}
\label{eq:xia}
\xi^2= \frac{1}{8\pi a n(0)}.\end{equation}
These relations lead to the important conclusion that $\xi^2 =d_0^4/R_0^2$; equivalently,
\begin{equation}
\label{eq:xiTF}
\frac{\xi}{d_0} = \frac{d_0}{R_0}.
\end{equation}
This TF limit  holds when the mean condensate radius $R_0$ is large compared to the mean oscillator length $d_0$. Thus  the TF oscillator length $d_0$  is the geometric mean of $\xi$ and $R_0$, and Eq.~(\ref{eq:xiTF}) then yields a clear separation of TF length scales, with   $\xi \ll d_0 \ll R_0$.

 Equation (\ref{eq:GP}) has the expected time-dependent generalization  (known as the time-dependent GP equation)
\begin{equation}
\label{eq:TDGP}
i\hbar\frac{\partial \Psi(\bm r,t)}{\partial t} = \left[-\frac{\hbar^2\nabla^2}{2M} + V_{\rm tr}(\bm r) + g|\Psi(\bm r,t)|^2\right]\Psi(\bm r,t),
\end{equation}
where $\Psi$ now depends on $t$ as well as on $\bm r$.    Although I shall not discuss the field-theoretical derivation of the time-dependent GP equation, it is worth noting that the condensate wave function  is interpreted as an ensemble average $\Psi(\bm r,t) = \langle\hat \psi(\bm r,t)\rangle$ of the field operator $\hat \psi(\bm r,t)$~\cite{Gross:1961,Pitaevskii:1961,Hohenberg:1965,Fetter:1972,Fetter:1996}.  Comparison of Eqs.~(\ref{eq:GP}) and (\ref{eq:TDGP}) implies that a stationary solution has the time dependence $\exp(-i\mu t/\hbar)$.  This result also  follows directly from the field-theoretical definition because the states on the left-hand side  of the ensemble average have one less particle than those on the right-hand side. 

This nonlinear field equation (\ref{eq:TDGP}) can be recast in an intuitive hydrodynamic form by writing the condensate wave function  
\begin{equation}
\label{eq:phase}
\Psi(\bm r,t) = |\Psi(\bm r,t) |\exp\left[iS\left(\bm r,t\right)\right]
\end{equation}
in terms of its magnitude $|\Psi|$ and phase $S$.  In this picture, the condensate (particle) density is $n(\bm r,t) = |\Psi(\bm r,t)|^2$, whereas the particle-current density becomes 
\begin{equation}
\label{eq:current}
\bm j = \frac{\hbar}{2Mi}\left(\Psi^*\bm \nabla \Psi - \Psi\bm \nabla \Psi^*\right) = |\Psi|^2 \frac{\hbar \bm \nabla S}{M}.
\end{equation}
The usual hydrodynamic relation $\bm j = n\bm v$ identifies the local velocity as 
\begin{equation}
\label{eq:v}
\bm v(\bm r,t) = \frac{\hbar}{M}\bm \nabla S(\bm r,t) = \bm \nabla \Phi(\bm r,t),
\end{equation}
where $\Phi \equiv \hbar S/M$ is the velocity potential, as proposed 
by~\textcite{Feynman:1955} in his discussion of the quantum mechanics of vortices in superfluid He~II.   Note that $\bm \nabla \bm \times \bm v$  vanishes wherever $S$ is not singular, so that the ``superfluid velocity'' $\bm v$ is irrotational, as assumed by~\textcite{Landau:1941} in his phenomenological two-fluid hydrodynamics for superfluid He~II.  For more complete discussions, see, for example, Chap.~III of \textcite{Lifshitz:1980b} and Chap.~XVI of~\textcite{Landau:1987}.

The representation of the velocity $\bm v$ in terms of the gradient of the phase has one major implication.  Consider the ``circulation'' $\kappa$,  defined at a given instant of time in terms of a line integral $\kappa = \oint_{\cal C} d\bm l\cdot \bm v$ around a closed path $\cal C$.   Substitution of Eq.~(\ref{eq:v}) yields
\begin{equation}
\label{eq:circ}
\kappa = \frac{\hbar}{M}\oint_{\cal C} d \bm l\cdot \bm \nabla S = \frac{\hbar}{M} \,\Delta S |_{\cal C},
\end{equation}
where $\Delta S|_{\cal C}$ is the net change in the phase on following $\cal C$.  The single-valuedness of the condensate wave function requires that this quantity must be an integral multiple of $2\pi$, which shows that the circulation in a dilute BEC must be quantized in units of $2\pi \hbar/M$~\cite{Onsager:1949,Feynman:1955}.  For accounts of this remarkable quantum-mechanical  relation, see Sec.~23 of \textcite{London:1954} and Sec.~2.3 of~\textcite{Donnelly:1991}.

Substitute Eq.~(\ref{eq:phase}) into the time-dependent GP equation (\ref{eq:TDGP}).  The imaginary part yields the familiar continuity equation for a compressible fluid
\begin{equation}
\label{eq:cons}
\frac{\partial n}{\partial t} + \bm \nabla\cdot \left(n\bm v\right) = 0.
\end{equation}
Correspondingly, the real part provides a generalized Bernoulli equation for this quantum fluid
\begin{equation}
\label{eq:Bern}
\frac{1}{2} M v^2 + V_{\rm tr} - \frac{\hbar^2 }{2M\sqrt n} \nabla^2 \sqrt n + gn + M\frac{\partial \Phi}{\partial t} = 0,
\end{equation}
where the explicitly quantum-mechanical  term containing $\hbar^2$ is sometimes called a ``quantum pressure.''  The  full structure of this Bernoulli equation (including the quantum pressure) can be understood as that for an irrotational compressible isentropic fluid [see Sec.~2 of \textcite{Landau:1987}] with an enthalpy density obtained from Eq.~(\ref{eq:EGP})~\cite{Fetter:2002}.  Such an isentropic picture is clearly  appropriate for a low-temperature superfluid.

The existence of a Bernoulli equation for the time-dependent GP equation has one basic consequence in the present context of vortices in dilute quantum gases.   All the classical results about vortex dynamics in an irrotational fluid [such as Kelvin's  theorem on the conservation of circulation, proved, for example, in Sec.~8 of~\textcite{Landau:1987}] automatically hold for any vortex configuration in a dilute BEC.   Nevertheless, the presence of a nonuniform trapping potential $V_{\rm tr}(\bm r)$ and the resulting nonuniform density in Eq.~(\ref{eq:nTF}) significantly affect the dynamics of vortices, so that qualitative classical pictures based on uniform incompressible fluids may not always apply.

\subsection{Bogoliubov equations}

The time-dependent GP equation (\ref{eq:TDGP}) describes the dynamics of the condensate, and it is natural to consider the linearized behavior of small perturbations around such solutions $\Psi(\bm r,t)$.  Although a fully quantum-mechanical analysis is feasible~\cite{Bogoliubov:1947,Pitaevskii:1961,Fetter:1972}, I  here rely on a simpler classical approach~\cite{Dalfovo:1999} [see also Sec.~7.2 of \textcite{Pethick:2002}]  that yields the same eigenvalue equations.  Assume a stationary nonuniform condensate with wave function $\Psi(\bm r)\,e^{-i\mu t/\hbar}$.  Since the  nonlinear Hartree term involves $|\Psi|^2$, the  perturbations  must,  for consistency, include  $e^{\mp i\omega t}$ with both signs of the frequency.  Specifically, take 
\begin{equation}
\label{eq:linGP}
\Psi(\bm r,t) = e^{-i\mu t/\hbar}\left[ \Psi(\bm r) + u(\bm r)\,e^{- i\omega t} -v^*(\bm r)\,e^{ i\omega t}\right],
\end{equation}
where the Bogoliubov amplitudes $u(\bm r) $ and $v(\bm r)$  are treated as  small. 

Substitute (\ref{eq:linGP}) into Eq.~(\ref{eq:TDGP}) and collect first-order terms proportional separately to $e^{\mp i\omega t}$.  The result is a pair of    linear eigenvalue equations
\begin{eqnarray}
\label{eq:BOGu}
{\cal L} u_j -g (\Psi)^2 v_j & = & \hbar \omega_j u_j, \\
\label{eq:BOGv}
{\cal L} v_j - g(\Psi^*)^2 u_j & = & -\hbar\omega_j v_j,
\end{eqnarray}
where ${\cal L} = -\hbar^2\nabla^2/(2M)  + V_{\rm tr} -\mu + 2g|\Psi|^2$ is a Hermitian operator.  
These coupled equations are known as the Bogoliubov equations, although \textcite{Pitaevskii:1961} was the first to consider the application to nonuniform systems [sometimes they  are called the Bogoliubov-de Gennes equations because of their close correspondence  to similar equations in the theory of superconductivity, as discussed in Chap.~5 of \textcite{deGennes:1966} and  Sec.~10.1 of \textcite{Tinkham:1996}].  In many ways, the Bogoliubov equations  are analogous to a nonrelativistic version of the Dirac equation, with $u$ and $v$ as the  particle and hole amplitudes, including the ($+,-$) metric seen in the minus sign on the right side of Eq.~(\ref{eq:BOGv}) compared to that of Eq.~(\ref{eq:BOGu}) [see, for example, Chap.~12 of \textcite{Abers:2004}, Sec.~52 of \textcite{Schiff:1968}, and Chap.~20 of \textcite{Shankar:1994}].

For a uniform condensate, the appropriate eigenfunctions are plane waves labeled by $\bm k$.  The Bogoliubov energy eigenvalue is easily determined to be 
\begin{equation}
\label{eq:EBOG}
E_k = \left[\frac{g n\hbar^2 k^2 }{M}+ \left(\frac{\hbar^2 k^2}{2M}\right)^2\right]^{1/2},
\end{equation}
where $n$ is the condensate density and $\mu \approx gn$ for this uniform gas.  For long wavelengths ($k\xi \ll 1$), where $\xi$ is the healing length from Eq.~(\ref{eq:xi}), the Bogoliubov energy  reduces to a phonon  spectrum 
\begin{equation}
\label{eq:phonon}
E_k\approx \hbar s k
\end{equation}
with $s  = \sqrt{gn/M}$  the speed of compressional sound.  For short wavelengths ($k\xi \gg 1$), in contrast, the Bogoliubov spectrum becomes that of a free particle.   The existence of a linear (phonon) spectrum at long wavelengths is crucial for superfluidity because the \textcite{Landau:1941} critical velocity $v_c$ for onset of dissipation is the minimum value of $E_k/(\hbar k)$ with respect to $k$.  In the present case, the Bogoliubov excitation spectrum (\ref{eq:EBOG}) yields $v_c = s$.  Note that  $s\propto \sqrt g$ is real and positive only because of the repulsive interactions ($g > 0$).  Furthermore, $v_c = s$  vanishes for an ideal Bose gas with $g=0$.  Hence  the ideal Bose gas is not strictly superfluid even at zero temperature, although it does indeed have a condensate with $\bm k = \bm 0$.  In this  well-defined sense, the existence of superfluidity in a uniform dilute Bose gas arises from the repulsive interactions.  

Manipulation of the Bogoliubov equations shows that 
$\omega_j\int dV\left(|u_j|^2 - |v_j|^2\right)$
is real.  In most cases, the integral is nonzero, so that $\omega_j$ itself is real.  To understand the basic issue involved in the normalization of the Bogoliubov eigenfunctions, it helps briefly to  recall  the usual particle creation and annihilation operators $a_j^\dagger$ and $a_j$ from quantum field theory.  These particle and hole operators obey the familiar Bose-Einstein commutation relations $[ a_j,a_{j'}^\dagger] = \delta_{j,j'}$.   The solution of the Bogoliubov equations  constitutes a  {\it canonical} transformation to a new set of ``quasiparticle'' operators $\alpha_j^\dagger$ and $\alpha_j$ that are linear combinations of the original particle and hole operators.  The situation is very similar to that for   BCS superconductors, as shown, for example, in Sec.~5-1 of \textcite{deGennes:1966} or Sec.~3.5 of \textcite{Tinkham:1996}.    Almost by definition, any  canonical transformation  preserves the commutation relations, so that the Bogoliubov quasiparticle operators must  obey the  {\it same} Bose-Einstein commutation relations $[ \alpha_j,\alpha_{j'}^\dagger] = \delta_{j,j'}$.  A detailed analysis  shows that  physically relevant Bogoliubov eigenfunctions must satisfy   the  following {\it positive} normalization condition 
\begin{equation}
\label{eq:BOGnorm}
\int dV\left(|u_j|^2 - |v_j|^2\right) = 1.
\end{equation}

Under this canonical transformation,  the Bogoliobov approximate grand-canonical Hamiltonian  assumes the very simple form
\begin{equation}
\label{eq:BOGham}
H-\mu N \approx {\rm const.} + {\sum_j}' \hbar\omega_j \,\alpha_j^\dagger\alpha_j,
\end{equation}
where the sum runs over all states with positive normalization.\footnote{The criterion for inclusion in this sum has been a source of confusion, because  every acceptable solution of the Bogoiubov equation $\{u_j,v_j\}$ with frequency $\omega_j$ and positive normalization has a second ``antiparticle'' solution $\{v_j^*,u_j^*\}$ with frequency $- \omega_j$ and {\it negative}  normalization.  The original study of these relations~\cite{Fetter:1972} incorrectly assumed that the positive-norm solution always had positive energy.  This   result indeed describes the simplest case of a uniform stationary  condensate with plane-wave eigenfunctions.  In contrast, nonuniform condensates, such as   one in uniform motion  (discussed here) or one containing a vortex [see Sec.~III.B.2 and~\textcite{Rokhsar:1997b}]  can have physical states with positive normalization and negative frequency. For a more general discussion, see  \textcite{Fetter:2001}.}
It is clear that the sign of the eigenfrequencies is crucial for the stability of the system.  If all the frequencies $\omega_j$ are positive, then creation of quasiparticles raises the energy, and the system is manifestly stable.  If, however, any of the positive-norm solutions has a negative frequency, then creation of those particular quasiparticles lowers the energy, and the  system is said to display an ``energy instability.'' 

 A simple example of this situation is a uniform condensate  that  moves with velocity $\bm v = \hbar \bm q/M$.  The condensate wave function now has a nonuniform complex  structure $\Psi(\bm r) =\sqrt n\, e^{i\bm q\cdot\bm r}$, and the solution with positive normalization has the   excitation energy~\cite{Fetter:1972,Pollock:1967,Rokhsar:1997b}  
\begin{equation}
\label{eq:move}
E_{\bm k}(\bm v) = \hbar\bm k\cdot\bm v + E_k,
\end{equation}
where $E_k$ is the Bogoliubov energy (\ref{eq:EBOG}) for a stationary condensate.  For small $v$, this excitation energy is  positive for any $k$.  When  $v$ exceeds the speed of sound $s$, however, then this eigenvalue will be negative in certain directions, which represents the onset of the Landau instability associated with spontaneous creation of phonons~\cite{Landau:1941}.   Although the energy instability implies that the system is formally unstable, the  absence of dissipation means that no mechanism exists to support the growth of the unstable process.  

For some special condensed systems (typically when BEC occurs in an excited single-particle state), the linearized Bogoliubov equations also have zero-norm solutions with  $\int dV\,(|u|^2-|v|^2)  = 0$.  In these  cases, the frequencies can be (and usually are) complex.  Such systems are said to exhibit a ``dynamical instability,'' and the unstable mode(s) will grow exponentially until various nonlinearities limit the evolution.    Such dynamically unstable behavior occurs in a condensate with a doubly quantized vortex~\cite{Pu:1999,Shin:2004},  in certain one-dimensional solitons~\cite{Muryshev:1999,Fedichev:1999,Feder:2000}, and  in one-dimensional optical lattices~\cite{Wu:2001}.

\section{Physics of One Vortex (a Few Vortices) in a Trap}

When the GP equation was  developed in 1961 to provide a tractable model for a vortex line in a  BEC,    rotating superfluid $^4$He was the only system of experimental interest (even though $^4$He is a dense highly correlated system).  Since 1995, however,  dilute ultracold trapped alkalai-metal gases have provided remarkable physical systems that indeed fit the GP picture in considerable detail.  

\subsection{One vortex in unbounded condensate}

\textcite{Gross:1961} and \textcite{Pitaevskii:1961}  introduced the GP equation specifically to describe a single straight vortex in an otherwise uniform dilute Bose-Einstein condensate with bulk particle density $n$. 
 The condensate wave function has the form $\Psi(\bm r) = \sqrt n \,\chi(\bm r)$, with 
 \begin{equation}
\label{eq:GPvortex}
\chi(\bm r) = e^{i\phi} f(r/\xi),
\end{equation}
where $r,\phi$ are two-dimensional plane-polar coordinates and $f \to 1$ for large $r$.  Equation (\ref{eq:v}) gives  the circulating velocity of a singly quantized long straight vortex line 
\begin{equation}
\label{eq:vel}
\bm v(\bm r) =\frac{\hbar}{Mr}\hat{\bm \phi};
\end{equation}
this hydrodynamic flow  has circular streamlines and a magnitude that diverges as $r\to 0$. 
The circulation $\kappa = \oint d\bm l\cdot \bm v = 2\pi\hbar/M$ can be transformed with Stokes's theorem as $\int d\bm S\cdot \bm \nabla\bm\times \bm v = 2\pi\hbar/M$, implying a singular  localized vorticity at the center of the vortex core
\begin{equation}
\label{eq:vorticity}
\bm \nabla\bm\times \bm v =\frac{2\pi \hbar}{M} \,\delta^{(2)}(\bm r) \,\hat{\bm z}.
\end{equation}

 The kinetic energy per unit length of the vortex is
 \begin{equation}
\label{eq:kevortex}
\frac{\hbar^2}{2M} \int d^2 r \,|\bm\nabla \Psi|^2 = \frac{\hbar^2\, n}{2M} \int d^2 r \left[\left(\frac{df}{dr}\right)^2 + \frac{f^2}{r^2}\right],
\end{equation}
where the first term arises from the density variation near the vortex core and the second term [$=~\frac{1}{2} M\int d^2 r\, n |f(r)|^2 v^2$] is the kinetic energy of the circulating flow.  With this kinetic-energy functional, the  Euler-Lagrange equation yields the nonlinear GP equation for the radial amplitude $f(r)$. The  resulting centrifugal barrier  forces the amplitude $f(r)$ to vanish linearly for $r\lesssim \xi$, which characterizes the vortex core, and $f(r) \approx 1$ for $r\gg \xi$.    In the present dilute limit, Eq.~(\ref{eq:dilute}) shows that the vortex core is significantly larger than the interparticle spacing. In addition,  the particle-current density $|\bm j(\bm r)| = n |f(r)|^2 \hbar/(Mr)$ vanishes both far from the vortex and at the center of the vortex, with a maximum near $r\approx \xi$.
 Finally, the speed of sound in a dilute Bose gas can be rewritten $s = \sqrt{gn/M} = \hbar/(\sqrt{2}M\xi)$, so that the circulating flow in Eq.~(\ref{eq:vel}) becomes supersonic for $r\lesssim  \xi$;  in this sense, local acoustic cavitation can be considered the  source  of the vortex core.  
 
In terms of the dimensionless scaled variable $x = r/\xi$,  the variational approximation $f_{\rm var}(x)  = x/\sqrt{x^2+2}$~\cite{Fetter:1969} provides a good fit to the numerical solution of the radial GP equation [see, for example, Fig.~9.1 in Sec.~9.2 of \textcite{Pethick:2002}].  The additional GP energy per unit length of vortex is $E_v\approx (\pi\hbar^2n/M)\ln(1.46 R/\xi)$, where $R$ is a large-distance cutoff.  Apart from the additive numerical constant, this result is essentially the integral of $\frac{1}{2}Mv^2n $ over a large circle of radius  $R$ with a short-distance cutoff of order $\xi$~\cite{Ginzburg:1958}.

\subsection{One vortex in a trapped condensate}

 A classical viscous fluid in a container that rotates  with angular velocity $\bm \Omega$   rapidly acquires the same angular velocity because of the microscopically rough walls.  From a theoretical view, the laboratory  is no longer  the  appropriate reference frame because the rotating walls are a time-dependent external potential that can do work on the system.  Thus it is essential to transform to the frame rotating with the walls, in which case the external potential becomes stationary.   Consider an originally static Hamiltonian $H(\bm r,\bm p)$ with an external trap potential $V_{\rm tr}(x,y,z)$ that in general is anisotropic about $\hat{\bm \Omega}$. When the potential rotates with angular velocity $\bm \Omega$, a direct transformation of the Lagrangian from the laboratory frame to the rotating frame eventually yields the Hamiltonian $H'(\bm r',\bm p') $ in the rotating frame 
\begin{equation}
\label{eq:rotH}
H'(\bm r',\bm p') = H(\bm r',\bm p') -\bm\Omega\cdot\bm L(\bm r',\bm p')
\end{equation}
where   $\bm r'$ and $\bm p'$ are the coordinates and canonical momenta in the rotating frame.  Here, the right-hand side involves the  nonrotating Hamiltonian $H(\bm r',\bm p')$  and angular momentum $\bm L(\bm r',\bm p')= \bm r'\bm\times \bm p'$, with  the original variables replaced by those in the rotating frame [see Sec.~39 of ~\textcite{Landau:1960} and Sec.~34 of~\textcite{Lifshitz:1980a}].\footnote{Note that $H(\bm r',\bm p')$ is not simply the  original Hamiltonian in the laboratory frame because of the primed variables.}

\subsubsection{Gross-Pitaevskii energy in a trap}

In the context of the GP equation, this transformation to a rotating frame yields a  modified energy functional 
\begin{eqnarray}
\label{eq:E'GP}
E'[\Psi] &= &\int dV\,\left[\frac{\hbar^2}{2M}\,|\bm \nabla\Psi(\bm r)|^2+ V_{\rm tr}(\bm r) |\Psi(\bm r)|^2  \right. \\ \nonumber 
&+ &\left.\frac{1}{2}\,g |\Psi(\bm r)|^4\right] -\bm\Omega\cdot \int dV\,\Psi^*(\bm r)\,\bm r\times \bm p\,\Psi(\bm r),
\end{eqnarray}
where all variables here are  in the rotating frame (written without primes for notational simplicity) [compare Eq.~(\ref{eq:rotH})].  The minimization of $E'[\Psi]$ now explicitly involves the angular velocity, and the last term  ($-\bm \Omega\cdot \bm L$) clearly favors states with nonzero (positive)  angular momentum.

 Most experiments on vortices in rotating dilute BECs occur in the TF regime.  Thus  I shall concentrate on this situation, which has several simplifying features.  Specifically, the dominant repulsive interactions imply that the mean condensate radius $R_0$ is large relative to the original mean oscillator length $d_0$.  Equation (\ref{eq:xiTF}) then implies that the vortex core size ($\sim \xi$) is smaller than $d_0$ by the same ratio $d_0/R_0$.  Thus the presence of  a few vortices  does not significantly affect the number density, which can be taken as the unperturbed TF shape in Eq.~(\ref{eq:parabolic}).  In most cases, the condensates are axisymmetric with radial and axial trap frequencies $\omega_\perp$ and $\omega_z$, so that the nonrotating TF density depends only on $r = \sqrt{x^2 + y^2}$ and $z$.
 
If an axisymmetric  condensate contains  a vortex at the center, the general wave function has the form~\cite{Dalfovo:1996,Lundh:1997,Sinha:1997} 
\begin{equation}
\label{eq:1vortex}
\Psi(\bm r) = e^{i\phi} \,|\Psi(r,z)|
\end{equation}
with the  phase $\phi$ appropriate for  a singly quantized vortex.  The hydrodynamic velocity is again $\bm v = \hbar\, \hat{\bm \phi}/(Mr) $, as in Eq.~(\ref{eq:vel}) for a singly quantized vortex in an asymptotically  uniform fluid.  The resulting centrifugal barrier near the axis of symmetry creates an axial node in the condensate wave function;  thus the presence of one vortex fundamentally alters the topology of the condensate density from ellipsoidal to toroidal because of the hole along the vortex core.  

It is important to generalize this wave function to an off-center vortex.  For simplicity, consider a three-dimensional disk-shaped condensate ($\omega_z \gg \omega_\perp$) with radial condensate dimension $R_\perp$.  In this case, a  vortex  at $\bm r_0 = (x_0, y_0)$ will remain 
essentially straight, and the principal effect  of the lateral translation is the altered phase
\begin{equation}
\label{eq:S}
S(x,y)  = \arctan\left(\frac{y-y_0}{x-x_0}\right),
\end{equation}
which is the azimuthal angle around the shifted origin $\bm r_0$.   In the TF regime, the density is taken to be unchanged (the logarithmically divergent circulating kinetic energy is cut off at the vortex core radius $\xi$).

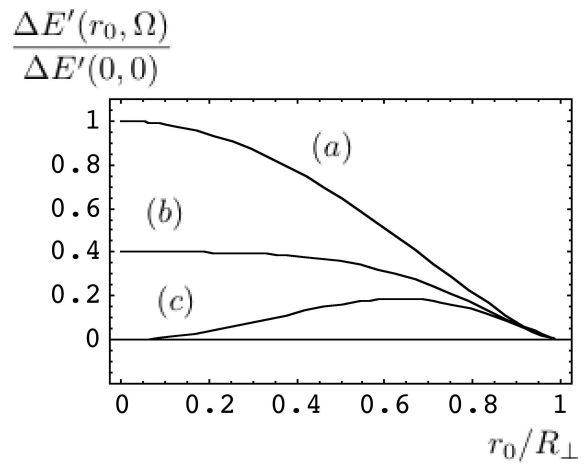
  \begin{figure}[ht] 
  \includegraphics[width=3.0in]{Fetter_fig02.epsf}
   \caption{Dimensionless scaled energy $\Delta E'(r_0,\Omega)/\Delta E'(0,0)$  [Eq.~(\ref{eq:GP1vortex})] of a singly quantized straight vortex in a rotating disk-shaped TF trap  as a function of the displacement $r_0$ from the symmetry axis.  Different curves represent different fixed values of the angular velocity $\Omega$:  (a) $\Omega = 0$;  (b) $\Omega = \Omega_m$, defined in Eq.~(\ref{eq:Omegam});   (c) $\Omega = \Omega_c$, defined in Eq.~(\ref{eq:Omegac}).}
 \label{fig:1v}
 \end{figure}

Use the energy functional $E'[\Psi]$ in Eq.~(\ref{eq:E'GP}) and let  $E'_0$ denote energy of a vortex-free state (it is independent of $\Omega$).  If $E'_1(r_0,\Omega)$ is the total GP energy   of a single off-center vortex in the rotating frame,  then  the difference  of the two energies $\Delta E'(r_0,\Omega) = E'_1(r_0,\Omega) - E'_0$ represents the formation  energy of the vortex.  A straightforward analysis with this model TF wave function~\cite{Svidzinsky:2000a} yields
\begin{equation}
\label{eq:GP1vortex}
\frac{\Delta E'(r_0,\Omega)}{\Delta E'(0,0)}= \left(1-\frac{r_0^2}{R_\perp^2}\right)^{3/2} - \frac{\Omega}{\Omega_c}\, \left(1-\frac{r_0^2}{R_\perp^2}\right)^{5/2},
\end{equation}
where  
\begin{equation}
\label{eq:Omegac}
\Omega_c = \frac{5}{2}\,\frac{\hbar}{MR_\perp^2}\,\ln\left(\frac{R_\perp}{\xi}\right),
\end{equation}
and the detailed form of $ \Delta E'(0,0)$ is unimportant here.
For fixed $\Omega$, this expression gives  the energy  in the rotating frame $\Delta E'(r_0,\Omega)$ as the vortex moves from the center of the condensate ($r_0= 0$) to the edge  ($r_0 = R_\perp$), shown 
 in Fig.~\ref{fig:1v} for various fixed  values of $\Omega$ [\textcite{Packard:1972} make a similar analysis for a vortex in an incompressible uniform superfluid; \textcite{Castin:1999} obtain comparable results numerically for a vortex in a trapped condensate]. 
 \begin{itemize}
\item[(a)]   Curve (a) shows the energy $\Delta E'(r_0,0)$ of a vortex in a nonrotating condensate as it moves from the center to the outer edge.  This energy decreases monotonically as $r_0$  increases, with negative curvature for $r_0\le R_\perp/\sqrt 2$.  If there is no dissipation, the distance $r_0$ determines  the  energy, and the vortex follows a circular trajectory at fixed $r_0$.  At low but finite temperature, however, weak dissipation slowly reduces the energy of the vortex, which thus  moves outward along curve (a), executing a  spiral trajectory toward the edge of  the condensate~\cite{Rokhsar:1997a,Rokhsar:1997b}.
\item[(b)]  With increasing   $\Omega$, the function $\Delta E'(r_0,\Omega)$ for fixed $\Omega$ flattens and the (negative) central curvature decreases.  Curve (b) has zero central curvature, which occurs for 
\begin{equation}
\label{eq:Omegam}
\Omega_m = \frac{3}{2}\,\frac{\hbar}{MR_\perp^2}\,\ln\left(\frac{R_\perp}{\xi}\right) = \frac{3}{5}\, \Omega_c.
\end{equation}
  This angular velocity $\Omega_m$ represents the onset of metastability for a vortex near the center of a  rotating condensate.  For $\Omega < \Omega_m$, the negative central curvature means that weak dissipation moves the vortex away from the center, and it eventually spirals out to the edge.  For $\Omega > \Omega_m$, however, the central curvature is now positive, and weak dissipation will move the vortex back to the center.  In this latter regime, the surrounding barrier makes the vortex  locally stable near the trap center, even though it is not globally stable  because the energy at the edge is definitely lower than the energy at the center.
\item[(c)]  Eventually, at $\Omega_c$ defined in Eq.~(\ref{eq:Omegac}), the central energy $\Delta E'(0,\Omega_c)$ vanishes [and thus equals $E'(R_\perp,\Omega_c)$].    For $\Omega>\Omega_c$, the central vortex is both locally and globally stable.  This value $\Omega_c$ is interpreted as the thermodynamic critical angular velocity for the creation of a singly quantized vortex in the TF limit.\footnote{For completeness, the corresponding thermodynamic critical angular velocity for a weakly interacting Bose-Einstein gas is $\Omega_c/\omega_\perp \approx 1-Na/(\sqrt{8\pi}\,d_z )$~\cite{Butts:1999,Linn:1999}. Note that this  result also holds for attractive interactions ($a<0$), but the resulting inequality $\Omega_c > \omega_\perp$ implies that the radial confinement disappears before the vortex would be  stable, in agreement with  earlier conclusions \cite{Dalfovo:1996,Wilkin:1998}.  Indeed, attractive interactions quite generally tend to  suppress BEC in favor of ``fragmented'' states with macroscopic occupation in more than one single-particle state~\cite{Pollock:1967,Nozieres:1982}.} As discussed below, experiments on rotating trapped gases generally find a different (larger) critical angular velocity for the creation of a vortex,  but similar analyses for incompressible fluids in cylindrical containers do provide a reasonable description of vortex formation in rotating superfluid $^4$He~\cite{Packard:1972,Yarmchuk:1982} [see also Secs.~2.5, 5.1 and 5.2  of~\textcite{Donnelly:1991}].
\end{itemize}

 \subsubsection{Dynamical motion of a trapped vortex}
 
 Although the GP energy functional Eq.~(\ref{eq:E'GP}) describes the equilibrium of a GP condensate with one  vortex (or more vortices), it also provides a very useful dynamical picture  of the vortex motion in a trapped condensate.  

 \paragraph{Lagrangian functional}
   The most direct approach relies on the basic observation that the time-dependent GP equation (\ref{eq:TDGP}) is the Euler-Lagrange equation for the time-dependent Lagrangian functional
 \begin{equation}
\label{eq:Lagrangian}
{\cal L}[\Psi] = \int dV\,\frac{i\hbar}{2}\left(\Psi^* \frac{\partial \Psi}{\partial t} - \frac{\partial \Psi^*}{\partial t} \Psi\right) - E'[\Psi]
\end{equation}
under variation of $\Psi^*$.  If the condensate wave function $\Psi$ depends on one or more parameters, the resulting Lagrangian functional  provides approximate Lagrangian equations of motion  for these parameters.  In the context of BECs, this approach was first applied with great success to the low-lying normal modes of a trapped stationary (nonrotating) condensate~\cite{PerezGarcia:1996}.  Subsequently, its application to a single straight vortex in a disk-shaped condensate yields an explicit prediction for the angular precession of such a trapped vortex~\cite{Lundh:2000,McGee:2001,Svidzinsky:2000a}. In this case, the position $\bm r_0$ of the off-center vortex serves as an appropriate parameter, and the Lagrangian formalism shows that the precession rate is proportional to (the negative of) the slope of the appropriate curve  in Fig.~\ref{fig:1v}.  For a nonrotating disk-shaped TF condensate, this analysis gives
\begin{equation}
\label{eq:precess}
\dot\phi = \frac{\Omega_m}{1-r_0^2/R_\perp^2},
\end{equation}
where $\Omega_m= \frac{3}{2}\,\hbar\,\ln(R_\perp/\xi) /(MR_\perp^2)$ is the critical rotation frequency for the onset of metastability given in Eq.~(\ref{eq:Omegam}).  
Note that a vortex in a nonrotating condensate precesses in the positive sense, which is the same sense as the circulating vortex velocity field.  

In this Lagrangian approach, the factor $1-r_0^2/R_\perp^2$ in the denominator arises from the parabolic radial  TF density profile.  It means that a vortex near the outer boundary precesses more rapidly than one near the center, although this simple picture omits bending of the vortex and thus  fails as $r_0\to R_\perp$.  

It is interesting to compare this TF result [Eq.~(\ref{eq:precess})] to the precession rate  for a long straight classical vortex in an incompressible fluid inside a circular cylinder of radius $R$.  In this latter case, the boundary   condition of zero radial flow velocity at the radius $R$ requires an external image vortex, yielding the result [see, for example, Sec.~9.4 of \textcite{Pethick:2002}]
\begin{equation}
\dot\phi_{\rm cl} = \frac{\hbar}{MR^2}\frac{1}{1-r_0^2/R^2}.
\end{equation} 
Although the  two expressions for the precession rate have the same denominators,  the precession here arises from the motion induced by the image (and there  is no large logarithmic factor).  Since  the TF density necessarily vanishes at the condensate radius, image vortices are generally omitted for trapped condensates, although this question remains partially unresolved~\cite{Anglin:2002,AlKhawaja:2005,Mason:2006}.

\paragraph{Matched asymptotic expansion}
An alternative approach to the dynamics of a vortex in a trapped condensate is to study the time-dependent GP equation (\ref{eq:TDGP})  itself; this analysis  relies on the method of matched asymptotic expansions~\cite{Pismen:1991,Rubinstein:1994} [see also Secs.~2.2 and 5.2 of \textcite{Pismen:1999}].  In the TF limit, the vortex core is small compared to other length scales, and it is possible to use the separation of length scales as follows~\cite{Svidzinsky:2000b}.  Assume that the straight vortex (located at $\bm r_0$) moves with a locally uniform velocity $\bm V(\bm r_0)$ that lies in the $xy $ plane. Transform to a locally comoving reference frame in which the vortex is stationary.  The problem is solved approximately in two different regimes:
\begin{enumerate}
\item Near the vortex core, the short-distance solution shows that the trap potential exerts a force proportional to $\bm \nabla_\perp V_{\rm tr}$ evaluated at $\bm r_0$.  This solution includes both the  detailed nonuniform structure of the vortex core and the asymptotic region $\xi \ll |\bm r - \bm r_0|$.
\item Far from the vortex, the core can be approximated as a line singularity, and the resulting solution also includes the region $\xi \ll |\bm r - \bm r_0|$. 
\end{enumerate}

The two solutions must agree in the common region of overlap,  and a lengthy analysis determines the translational velocity $\bm V(\bm r_0)$  of the vortex line near the center of the three-dimensional  disk-shaped trap
\begin{equation}
\label{eq:V}
\bm V(\bm r_0) =\frac{3\hbar}{4M\mu} \ln\left(\frac{R_\perp}{\xi}\right) \hat{\bm z} \bm\times \bm \nabla_\perp V_{\rm tr}(\bm r_0).
\end{equation}
To understand the structure of this result, note that the trap exerts a force $-\bm\nabla_\perp V_{\rm tr}$ on the vortex.  Since the vortex itself has intrinsic angular momentum, it  acts gyroscopically and hence moves at right angles to this applied force.  As a result, the motion follows an equipotential line along the direction $ \hat{\bm z} \bm\times \bm \nabla_\perp V_{\rm tr} $ and thus conserves energy, as expected for the GP equation at zero temperature.

In fact, this result is essentially the same as that in Eq.~(\ref{eq:precess}) for $r_0$ near the trap center.  The gradient of the axisymmetric potential $V_{\rm tr} = \frac{1}{2}M(\omega_\perp^2 r^2 + \omega_z^2 z^2)$ and the squared TF condensate radius $R_\perp^2 = 2\mu/(M\omega_\perp^2)$ readily yield the equivalent form for Eq.~(\ref{eq:V}) 
\begin{equation}
\label{eq:V'}
\bm V(\bm r_0) = \Omega_m r\, \hat{\bm \phi},
\end{equation}
involving the critical angular velocity $\Omega_m$ for the onset of metastability given in Eq.~(\ref{eq:Omegam}).    These expressions for the translational velocity $\bm V$  can be generalized to  treat a rotating  anisotropic harmonic trap with $\omega_x \neq \omega_y$, and to include the three-dimensional  effects of vortex curvature~\cite{Svidzinsky:2000a,Svidzinsky:2000b}.

Although the final result in Eq.~(\ref{eq:V}) is simple and physical, the details of the method of matched asymptotic expansions are not very transparent.  A more physical picture starts from the conservation of particles $\bm\nabla\cdot (n\bm v) = 0$ in the locally comoving coordinate system.  In the TF limit, the particle density is explicitly known in terms of the trap potential [compare Eq.~(\ref{eq:parabolic})].  Since $\bm v \propto \bm \nabla S$, this equation can be rewritten as $\bm\nabla n \cdot \bm \nabla S + n \nabla^2 S = 0$, which determines the phase $S$ of the condensate wave function in terms of the known density.   A perturbation expansion with the appropriate azimuthal angle [see Eq.~(\ref{eq:S})] as the leading term yields an additional phase proportional to $\ln|\bm r - \bm r_0|$~\cite{Svidzinsky:2000b,Sheehy:2004a,Sheehy:2004b}.    The gradient of this additional phase evaluated at $\bm r_0$ then fixes the precessional velocity of the vortex (\ref{eq:V}).    In this latter perspective, the precessional motion can be thought to arise from the gradient of the condensate density $n(\bm r)$ evaluated at the position of the vortex~\cite{Sheehy:2004a,Sheehy:2004b}
\begin{equation}
\label{eq:vprec}
\bm V(\bm r_0) =\frac{\hbar}{M} \,\frac{\hat{\bm z}\bm\times \left(-\bm \nabla n\right)}{2n}\,\ln\left(\frac{R_\perp}{\xi}\right),
\end{equation}
where the minus sign appears because the density decreases with increasing $r$ (this detailed result describes a two-dimensional condensate).  Apart from a numerical factor that reflects the difference between two and three dimensions, Eqs.~(\ref{eq:V}) and (\ref{eq:vprec}) describe the same physics  because the gradient of the  TF density $\bm \nabla n$ is proportional to $-\bm \nabla V_{\rm tr}$ [see Eq.~(\ref{eq:nTF})].

\paragraph{Bogoliubov equations and stability of a vortex}  Soon after the creation of BECs in dilute $^{87}$Rb~\cite{Anderson:1995}, theorists studied the  collective-mode spectrum of a nonrotating  stationary vortex-free trapped condensate using the GP equation (\ref{eq:GP}) for the condensate and the Bogoliubov equations (\ref{eq:BOGu}) and (\ref{eq:BOGv})  for the small-amplitude  normal modes.   As~\textcite{Edwards:1996} reported, the agreement between the theoretical predictions and the experimental observations was impressive [see   also Sec.~IV.A of \textcite{Dalfovo:1999}].

This success rapidly led to a similar numerical study~\cite{Dodd:1997} of the collective modes of a  singly quantized vortex in an otherwise nonrotating trapped condensate.  If the vortex is  on the axis of symmetry, the condensate  wave function is like that in Eq.~(\ref{eq:1vortex}), and the conservation of the  $z$ component of angular momentum greatly simplifies the analysis.  Remarkably, the spectrum of quasiparticle excitations for the vortex turns out to contain a very  unusual mode (now called ``anomalous'') with {\it positive} normalization integral [Eq.~(\ref{eq:BOGnorm})], {\it negative}  angular momentum $m_a=-1$ relative to that of the vortex, and {\it negative} frequency $\omega_a$ [\textcite{Rokhsar:1997a} has emphasized  this particular identification; see also \textcite{Fetter:1998}].  As noted in Sec.~II.C [see Eq.~(\ref{eq:BOGham})], such a negative eigenfrequency implies that a vortex in a nonrotating condensate is formally (energetically) unstable, but  the vortex cannot spiral out of the condensate in the absence of dissipation  [compare  the discussion concerning curve (a) in Fig.~\ref{fig:1v}].  Detailed studies~\cite{Linn:1999,Svidzinsky:2000a} show that $\omega_a \approx -\omega_\perp$ in the near-ideal limit and $\omega_a\approx -\Omega_m$ in the TF limit, so that $\omega_a $ remains negative for all coupling strengths.

The physics underlying the anomalous mode is especially  transparent in the weak-coupling (near-ideal)  limit, when the condensate with a singly quantized vortex at the origin has the radial wave function $\psi_1(r,\phi) \propto (x+i y)\exp(-\frac{1}{2} r^2) = re^{i\phi}\exp(-\frac{1}{2} r^2)$, with one unit of angular momentum and an excitation energy $\hbar\omega_\perp$.   Note that $\psi_1$ differs from  the true radial ground-state wave function $\psi_0 \propto \exp(-\frac{1}{2} r^2)$, which instead  has zero angular momentum and zero excitation energy.  In principle, the particles in the macroscopic  vortex condensate could make a transition to the  nonrotating ground state, with a change in angular momentum $m = -1$ and a change in energy $-\hbar \omega_\perp$;  this transition is just the anomalous mode in an ideal Bose-Einstein gas.  Such anomalous modes typically occur whenever the condensate occupies a single-particle state that is not the true ground state [compare the discussion of a moving condensate below Eq.~(\ref{eq:move})].

In the present case of a condensate with a single vortex, the anomalous mode has two very important consequences
\begin{enumerate}
\item The linearized solution for the $j$th mode of the time-dependent GP equation in Eq.~(\ref{eq:linGP})  immediately yields the corresponding perturbation in the density $\delta n_j = (\Psi^*u_j - \Psi v_j)e^{-i\omega_j t}$.  In particular,  a singly quantized vortex with $\Psi$ given in  Eq.~(\ref{eq:1vortex}) has an anomalous-mode density perturbation  $\delta n_a(\bm r,t) \propto  \exp[i(m_a\phi - \omega_a t)]$, with $m_a ~= ~-1$ and negative excitation frequency $\omega_a= -|\omega_a|$.  This density perturbation  $\delta n_a \propto \exp[i(|\omega_a|t-\phi)]$ precesses in the {\it positive} sense around the symmetry axis of the condensate at a rate $-\omega_a = |\omega_a|$.  In the TF limit when  $|\omega_a| = \Omega_m$~\cite{Svidzinsky:2000a}, this rate  agrees with that in  Eq.~(\ref{eq:precess}) and with experiments~\cite{Anderson:2000} discussed below in Sec.~III.D.1.

\item  When the condensate rotates, Eq.~(\ref{eq:rotH}) shows that the frequency $\omega_j $  of the $j$th mode in the laboratory frame  shifts to 
\begin{equation}
\label{eq:shifted}
\omega_j'(\Omega) = \omega_j - m_j\Omega
\end{equation}
 in the rotating frame, where $m_j$ is the corresponding angular quantum number.  For the anomalous mode, this result implies that 
$\omega_a(\Omega) = -|\omega_a| + \Omega$, which becomes positive and thus energetically stable when $\Omega$ exceeds $|\omega_a|$.  This conclusion confirms the critical rotation frequency $\Omega_m$ for the onset of metastability in the TF regime, as seen  in curve (b) of Fig.~\ref{fig:1v}.
\end{enumerate}

\subsection{Small vortex arrays in a trap}

In the TF limit, Eq.~(\ref{eq:S}) shows that the condensate wave function has a $2\pi$-phase singularity associated with a vortex at $\bm r_0$.  More generally, a vortex is simply a node in the condensate wave function~\cite{Feynman:1955}.  In the context of dilute ultracold BECs,  \textcite{Butts:1999}  exploit  this property in their studies  of small  vortex arrays [\textcite{Castin:1999} obtain similar vortex configurations with a somewhat different variational approach].

  \begin{figure}[ht] 
  \includegraphics[width=2.5in]{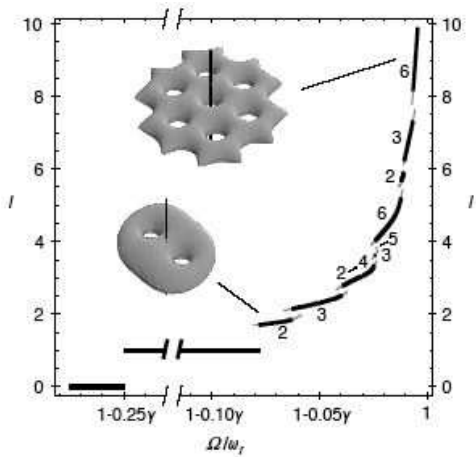}
   \caption{Angular momentum $l$ vs.~dimensionless angular velocity $\Omega/\omega_\perp$ for stable (black lines) and metastable (gray lines) states.  Here, $\gamma = (2/\pi)^{1/2} N a/d_z$ is assumed small. There are no stable states with $0< l<1$, and $l= 1$ represents a single vortex on the axis of symmetry.  For  larger $l$, the angular momentum changes smoothly because the positions of the  two or more vortices   vary continuously with $\Omega$ (for each stable branch, the integer denotes the rotational symmetry). Three-dimensional plots of the density for two  and seven vortices are remarkably similar to experimental images of \textcite{Yarmchuk:1979} for superfluid $^4$He (see Fig.~\ref{fig:Yarmchuk}) and of \textcite{Madison:2000a} for ultracold dilute $^{87}$Rb gas (see  Fig.~\ref{fig:ENS2}).   From~\textcite{Butts:1999}.}
 \label{fig:Butts}
 \end{figure}

To understand their procedure, assume that the condensate is tightly confined in the $z$ direction with a small axial  oscillator length $d_z \ll d_\perp$.  The remaining two-dimensional isotropic harmonic trap has the one-body Hamiltonian $H_0 = \frac{1}{2} \left(p^2/M + M\omega_\perp^2 r^2\right)$.  For positive angular momentum,   the low-lying normalized one-body eigenfunctions are 
\begin{eqnarray}
\label{eq:psim}
\psi_m(\bm r) &=&\frac{u^m e^{im\phi}}{ (\pi m!)^{1/2}\,d_\perp}\,\exp\left(-\frac{u^2}{2}\right) \nonumber \\
& =& \frac{\zeta^m}{ (\pi m!)^{1/2}\,d_\perp}\,\exp\left(-\frac{u^2}{2}\right),
\end{eqnarray}
where $u = r/d_\perp$,  $\zeta = (x+  iy)/d_\perp = ue^{i\phi}$, and $m\ge 0$.  Note that $\psi_m$ is also an eigenfunction of the dimensionless angular-momentum operator $l = -i  \partial /\partial \phi$ with eigenvalue $m$.  As a simple example  of a possible trial state, consider the linear combination $\Psi = c_0\psi_0 + c_1 \psi_1$, which vanishes at $\zeta_0   \equiv (x_0 + i y_0)/d_\perp  = -c_0/c_1$ (the only node).  The point $\zeta_0$ is  the position of the vortex, since the phase of $\Psi$ increases by $2\pi$ on going around $\zeta_0$  in the positive sense.

For a rotating condensate with  positive angular momentum, \textcite{Butts:1999} use a general linear combination of such states $\Psi = \sum_{m\ge 0} c_m \psi_m$ with normalization $\sum_{m\ge 0} |c_m|^2 = 1$.\footnote{Similar  solutions play a crucial role in the study of rapidly rotating condensates in Sec.~V.} The expectation value of the angular momentum per particle is $\hbar l = \hbar\sum_{m\ge0} m|c_m|^2$.  Furthermore, the expectation value of the one-body Hamiltonian  $\langle H_0\rangle =  \hbar\omega_\perp \sum_{m\ge0} m|c_m|^2 +\hbar\omega_\perp =\hbar\omega_\perp (l + 1) $ is essentially   that of the angular momentum.   Thus all states with the same value of $l$ have the same one-body energy, independent of the specific coefficients $c_m$.  To lift this degeneracy, it is necessary to consider the interaction energy  $E_{\rm int} = \frac{1}{2} g  \int d^3 r\,|\Psi|^4$ (assumed to be small).  In the limit of tight $z$  confinement, this energy reduces to a two-dimensional integral, which they evaluate numerically with  the linear-combination trial state $\Psi$.  For a given angular momentum $l$, the  equilibrium  state  minimizes  the total energy,  which also means the interaction energy because of the degeneracy. In this way, \textcite{Butts:1999} determine  the coefficients $c_m$  and thus the position of the vortices  (the nodes of $\Psi$)   as a function of $l$

Given the equilibrium energy $E_{\rm tot}(l)$, the thermodynamic relation $\hbar \Omega = \partial E_{\rm tot}/\partial  l$ fixes the corresponding angular velocity $\Omega$.  Extensive numerical studies~\cite{Butts:1999} then determine the resulting physical sequence of vortex states with increasing $\Omega$, as shown in Fig.~\ref{fig:Butts}.  The state with  $l = 1$ arises from a single vortex on the symmetry axis of the condensate, but  configurations of two or more vortices vary with $l$, so that the corresponding angular momentum does not generally take integer values.

\subsection{Experimental and theoretical studies in a trap}

Two basic experimental  methods have been most effective in  creating vortices in a BEC. The first approach~\cite{Matthews:1999b}  manipulates two hyperfine components of $^{87}$Rb, spinning up one component with an external coherent electromagnetic coupling beam~\cite{Williams:1999}.  This approach relies heavily on techniques from atomic physics, especially the optical Bloch equations for a two-level system [see, for example, Chaps.~7-11 of~\textcite{Feynman:1965}, Chap.~IV of \textcite{Cohen:1977},  Chaps.~1-3 of~\textcite{Allen:1987}, and Chap.~V of \textcite{Cohen:1998}].  The second  nearly simultaneous approach~\cite{Madison:2000a} is similar to the ``rotating-bucket'' method of conventional low-temperature physics [see, for example, Chaps.~2 and 5 of \textcite{Donnelly:1991}].  In the present context of BECs, a driven rotating deformation of the confining trap serves to spin up the whole condensate.

\subsubsection{Vortex behavior in two-component BEC mixtures (JILA)}

The search for quantized vortices in dilute  BECs started soon after the creation of trapped condensates. Since a mixture of two hyperfine components of $^{87}$Rb play an essential role in the first reported vortex~\cite{Matthews:1999b}, I first review the  physics of such systems.  The GP energy functional in Eq.~(\ref{eq:EGP}) is readily generalized for a  mixture of two BECs with condensate wave functions $\Psi_1$ and $\Psi_2$.   The total energy contains  (1)  the two separate energies $E_j[\Psi_j]$ of each component $j = 1, 2$, with appropriate selfinteraction parameters $g_j$ (I assume repulsive interactions  with $g_j >0$) and (2)   the interaction  energy $E_{12}[\Psi_1,\Psi_2] = \int dV\,g_{12}|\Psi_1|^2\,|\Psi_2|^2$ between species 1 and 2, where $g_{12}$ is the interspecies interaction parameter [see Sec.~12.1 of \textcite{Pethick:2002} and Sec.~21.11 of \textcite{Pitaevskii:2003}].  For the simplest case of two uniform  BECs in a box, if $g_{12}^2 > g_1g_2$, the two components are immiscible and  separate into nonoverlapping phases.  Otherwise, if $g_{12}^2 < g_1g_2$, they  form two uniform  interpenetrating BECs [in the context of dilute BECs, this result apparently was first obtained by~\textcite{Nepomnyashchii:1974}; see also~\cite{Colson:1978}].  

For trapped dilute BECs, the situation is more complicated because the nonuniform trap potentials  $V_j$ induce both kinetic and potential energies that affect the overall energy functional.  
The first such experiment~\cite{Myatt:1997}  at JILA (Boulder, CO) uses two different hyperfine states of $^{87}$Rb.  This isotope has nuclear spin $I = \frac{3}{2}$, which combines with the single valence electron ($S =\frac{1}{2}$) to give two hyperfine manifolds, a lower one  with $F = 1$ and an upper one with $F = 2$.  The typical trap has a local minimum in the magnetic field, and only some of these hyperfine states are weak-field seeking and hence stable [see Chaps.~3 and 4  of ~\textcite{Pethick:2002} and Chap.~9 of~\textcite{Pitaevskii:2003} for a description of magnetic and optical  traps].  In the common notation $|F,m_F\rangle$, the stable weak-field seeking states are $|1,-1\rangle$, $|2,1\rangle$, and $|2,2\rangle$.  

The experiment of~\textcite{Myatt:1997} cools state $|1,-1\rangle$ evaporatively and the interspecies interaction cools state $|2,2\rangle$ through sympathetic cooling.  Although the interaction parameters approximately  satisfy $g_{12}^2 \lesssim g_1g_2$~\cite{Cornell:1998}, implying  that two such uniform BECs would overlap, the trapped condensates in fact separate because of the differences in the two trap potentials and interaction parameters~\cite{Myatt:1997}.  A careful theoretical analysis~\cite{Esry:1997} that includes all the relevant effects finds a reasonable fit to the experiments.

To avoid the problem of different trap potentials,  subsequent experiments on these mixtures  use the pair of states $|1\rangle \equiv |1,-1\rangle$ and $|2\rangle \equiv|2,1\rangle$ that differ by an angular momentum $\Delta m = 2$.  Each component can be imaged selectively with appropriately tuned lasers.  Furthermore the state can be quickly changed from $|1\rangle$ to  $|2\rangle$ and back by a two-photon transition (microwave at $\approx 6.8$ GHz and rf at $\approx $ 2 MHz).  To understand the physics of this fascinating system, it is helpful to examine the coupled time-dependent GP equations for the two components~\cite{Hall:1998}
\begin{eqnarray}
\label{eq:2GP}
&\displaystyle{ i\hbar\frac{\partial }{\partial t} }
\left(
\begin{matrix}
\Psi_1 \\
\Psi_2
\end{matrix}
\right) =   \\[.2cm] \nonumber 
&\left(
\begin{matrix}
{\cal T}  + V_1 + V_{H1} +\frac{1}{2} \hbar\delta  & \frac{1}{2} \hbar\Omega(t)\,e^{i\omega_{\rm em}t} \\[.1cm]
 \frac{1}{2} \hbar\Omega(t)\,e^{-i\omega_{\rm em}t} & {\cal T}  + V_2 + V_{H2} -\frac{1}{2} \hbar\delta
 \end{matrix}
 \right)
 \left(
 \begin{matrix}
 \Psi_1\\
 \Psi_2
 \end{matrix}
 \right),
\end{eqnarray}
where ${\cal T} = -\hbar^2 \nabla^2 /(2M)$ is the kinetic energy, $V_j$ is the trap potential, and $V_{Hj}$ is the Hartree energy including the interaction with both densities $|\Psi_1|^2$ and $|\Psi_2|^2$.  The crucial new feature here is  the applied electromagnetic (microwave and rf) fields with combined frequency $\omega_{\rm em}$;  they  produce an  off-diagonal coupling  that involves the phase of each condensate wave function.  The   effective coupling $\Omega(t)$ is known as the  ``Rabi'' frequency, which increases with the strength of the applied electromagnetic fields [see Ch.~3 of~\textcite{Allen:1987}].  In general, this coupling depends explicitly on time (it vanishes when the fields are turned off).  Finally, $\delta$ is the detuning between the combined  frequencies of the driving fields and that of the atomic hyperfine transition $|1\rangle \leftrightarrow |2\rangle$.

This electromagnetically coupled two-component system provides a remarkable example of topological control of a quantum-mechanical state through an external parameter.  Specifically, when the near-resonant electromagnetic coupling $\Omega(t)$ is turned {\it off\/}, the interaction between the condensates involves only the densities (through $V_{Hj}$) with no phase information.  In this case, each condensate has its own complex  order parameter $\Psi_j$.  The associated  $U(1)$ symmetry [see, for example, Sec.~27 of~\textcite{Schiff:1968}]  has the topology of a circle or a cylinder.
 For $T\ll T_c$, the magnitude $|\Psi_j|$ is fixed, and each complex order parameter has  its own phase angle, yielding a quantized circulation that remains a  topological invariant.
 
The topology  is very different when the electromagnetic coupling is turned {\it on\/} because the  coupled condensates in Eq.~(\ref{eq:2GP}) now have a {\it single\/} two-component $SU(2)$ 
order parameter with the topology of a sphere, like that of a particle with spin $\frac{1}{2}$ [see Sec.~27 of~\textcite{Schiff:1968} and~\cite{Hall:1998}].    Two spherical-polar angles  suffice to characterize the coupled system, whose dynamics obeys the optical Bloch equations~\cite{Allen:1987}, similar to those of nuclear magnetic resonance (NMR).  Specifically, the transformation of $|1\rangle$ into $|2\rangle$ is analogous to a $\pi$ pulse in NMR that rotates a spin from up to down.  If the pulse is twice as long, the resulting $2\pi$ rotation essentially reproduces the initial $|1\rangle$ state up to a sign.

An elegant experiment~\cite{Hall:1998} shows the ability to manipulate  the  coupled condensates as a single  $SU(2)$ order parameter.  Application of a  $\pi/2$ pulse to a  pure state $|1\rangle$  produces a coherent mixture with equal population in both states that are spatially separated with a small region of overlap.  After a variable delay, a second $\pi/2$ pulse measures  the relative quantum phase between them.  A subsequent experiment demonstrates the three-dimensional character of the $SU(2)$ topology explicitly~\cite{Matthews:1999a,Williams:2000} as follows:   An external coupling field with a spatial gradient produces a differential twist along the length of the condensate.  With increasing time, the increasing phase difference  between the two ends of the condensate  relaxes periodically by  a three-dimensional  evolution of the coupled $SU(2)$ order parameter, leading to a collapse and recurrence of the structure.  Note that   such behavior would not occur for two uncoupled $U(1)$ condensate wave functions.

  \begin{figure}[ht] 
  \includegraphics[width=3.0in]{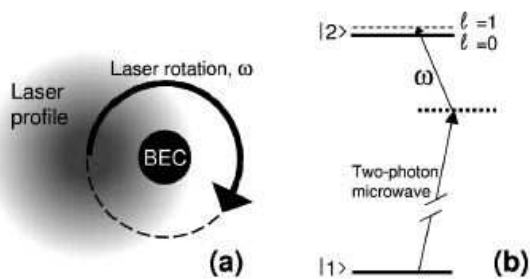}
 \caption{(a) Schematic illustration of the technique used to create a vortex:  An off-resonant laser provides a rotating force on the atoms across the condensate with a simultaneous  microwave drive of detuning $\delta$.  (b)  Level diagram showing the microwave transition to the sublevels of the state $|2\rangle$.  When  the rotation frequency of the laser beam $\omega$ is close to the detuning $\delta$, the perturbation selectively modulates the transition from the $l=0$ state  to the $l=1$ state.  From~\textcite{Matthews:1999b}.}
 \label{fig:JILA1}
 \end{figure}

 To create a quantized vortex, \textcite{Williams:1999}  propose applying an external laser beam in addition to  the external electromagnetic fields that   produce the two-photon transition.  This external laser beam in effect adds a rotating gradient to the coupling  fields with a frequency $\omega$ (see Fig.~\ref{fig:JILA1}).  Such a process starts from the ground state $|1\rangle$ and yields the state $|2\rangle$ with one unit of angular momentum.  Specifically, the theory predicts that the angular momentum of the  state $|2\rangle$  oscillates and increases with time.  Turning off the coupling at the appropriate time leaves the state $|2\rangle$  with a persistent  singly quantized vortex  because the associated order parameter for this state $|2\rangle$ now has $U(1)$ symmetry.  

\begin{figure}[ht] 
  \includegraphics[width=2.5in]{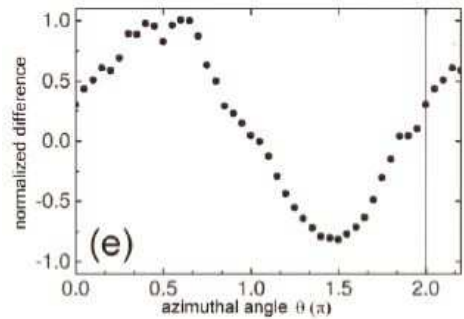}
 \caption{Cosine of the relative phase around the vortex, showing the sinusoidal variation of the azimuthal angle.   From~\textcite{Matthews:1999b}.}
 \label{fig:JILA2}
 \end{figure}

 In the experiment~\cite{Matthews:1999b}, the vortex in state $|2\rangle$ surrounds a nonrotating core consisting of state $|1\rangle$.     Furthermore, a $\pi$ pulse can interchange  $|1\rangle$ and  $|2\rangle$, so that the vortex can occur in either state.  There is one important difference between the states $|1\rangle$ and $|2\rangle$, because  the state  $|2\rangle$ with $m_F = 1$ is not maximally aligned.  Consequently  it can decay by spin relaxation with a measured lifetime $\sim 1$ s.  As  expected,  a vortex in  state $|1\rangle$ surrounding a core of state  $|2\rangle$ is more stable because  the state $|2\rangle$ core eventually decays,  leaving a simple one-component vortex in state $|1\rangle$.  The experiment  can image either the filled core or the surrounding vortex, but these images by themselves do not demonstrate the presence of quantized circulation.  Fortunately, the  interference procedure described above allows a clear  measurement of the relative phase between the two condensates at various positions,  demonstrating the sinusoidal variation predicted for a vortex (see Fig.~\ref{fig:JILA2}).

   \begin{figure}[ht] 
  \includegraphics[width=3.0in]{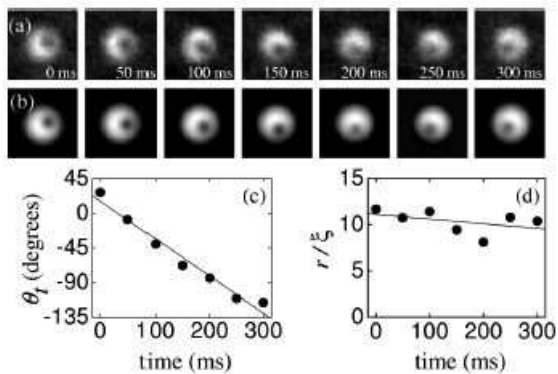}
 \caption{(a) Successive images of a two-component condensate with a vortex in state $|1\rangle$ surrounding a core in state $|2\rangle$.  (b) The image is then fit with a smooth Thomas-Fermi distribution.  (c) The azimuthal angle of the vortex core plotted against the time, giving a precession frequency of 1.3(1) Hz.  (d) The gradual decrease in the core radius $r$  (measured in units of the coherence length $\xi$) because of spin relaxation.    From~\textcite{Anderson:2000}.}
 \label{fig:JILA3}
 \end{figure}
 
One very important advantage of this  two-component experiment is the ability to follow the  motion of the vortex core with nondestructive time-lapse images~\cite{Anderson:2000}.  Figure \ref{fig:JILA3} shows the uniform precession of the filled vortex core in state $|2\rangle$ around the trap center,  determining the angular velocity of the circular motion. Note [Fig.~\ref{fig:JILA3}(d)] the gradual shrinkage of the core radius as state $|2\rangle$ decays through  spin-flip transitions.  The experimenters can also remove the core component with a brief resonant laser pulse, leaving a one-component vortex whose empty core (of order $\xi$) is too small to image in the trap.  To study the precession of such a one-component vortex, the experimenters first image the  two-component vortex, then remove the core, wait a variable time and finally  turn off the trap, imaging the expanded  vortex core.  In this way, they measure the precession rate of a one-component empty-core  vortex, yielding values in reasonable agreement with theoretical predictions~\cite{Fetter:2001}.  These data show no evidence that the vortex  spirals outward, suggesting that thermal damping is negligible on the time scale of $\sim 1$ s.

 \subsubsection{Vortex behavior in one-component BECs (ENS and subsequently others)}
 
 Most other experimental methods of creating vortices in dilute BECs rely on the rotation of an anisotropic potential, and I  briefly review the physics of such a system.  Consider a classical incompressible fluid in a long cylinder with elliptical cross section $|x|\le a$ and $|y|\le b$.  When the container rotates with angular velocity $\bm \Omega = \Omega \hat{\bm z}$, the moving walls (in the laboratory frame) exert forces on the fluid.  As a result, the fluid executes a purely irrotational motion with  velocity $\bm v_{\rm irr} = \bm \nabla \Phi_{\rm irr}$ [see Eq.~(\ref{eq:v})], where the associated velocity potential   $\Phi_{\rm irr} = \Omega \,xy\,(a^2-b^2)/(a^2+b^2)$ is proportional to $xy$ [see pp.~86-88 of \textcite{Lamb:1945}].  Remarkably, this irrotational  velocity field induces a nonzero angular momentum per unit length $L_{\rm irr} = L_{\rm sb} \,(a^2-b^2)^2/(a^2+b^2)^2$, where $L_{\rm sb}$ is the corresponding  angular momentum per unit length for a  fluid that executes solid-body rotation with $\bm v_{\rm sb}= \bm \Omega\bm\times \bm r$.  Note that  this irrotational angular momentum occurs without quantized vortices.  Such vortex-free angular momentum increases the critical angular velocity $\Omega_c$ for vortex formation in superfluid $^4$He~\cite{Fetter:1974},   as \textcite{DeConde:1975} verify by experiments on rotating elliptical and rectangular cylinders.

 The same irrotational flow occurs in rotating anisotropic trapped  BECs, where it appears in the phase of the condensate wave function $S=M\Phi_{\rm irr}/\hbar$ [see Eq.~(\ref{eq:v})] and leads to  unexpected  phenomena.    For example, a sudden  small rotation of the anisotropic trap potential  induces an  unusual oscillatory motion known as the ``scissors mode'' [see Secs.~14.3 and 14.4 of \textcite{Pitaevskii:2003} for a detailed description of this and related experiments].
 
   \begin{figure}[ht] 
  \includegraphics[width=3.0in]{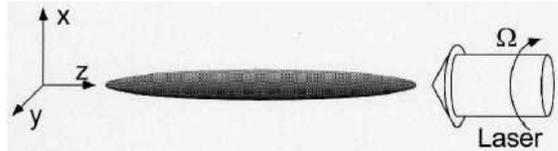}
 \caption{ A cigar-shaped condensate is confined by an axisymmetric magnetic trap   and stirred by an off-center far-detuned laser beam.  The laser beam propagates along the long axis and creates an anisotropic quadrupole  potential that rotates at an angular velocity $\Omega$.  From~\textcite{Madison:2000b}.}
 \label{fig:ENS1}
 \end{figure}
 
 The experimental group at the Ecole Normale Sup\'erieure (ENS) in Paris uses such a  direct approach (see Fig.~\ref{fig:ENS1}) to create one or more vortices in a one-component $^{87}$Rb cigar-shaped condensate~\cite{Madison:2000a}.  Starting with an axisymmetric condensate, they apply a toggled  off-center  stirring laser beam that slightly deforms the trap potential;  the plane of deformation then  rotates about the long axis with an applied angular velocity $\Omega/(2\pi)\lesssim 200$ Hz. Physically, the laser beam acts on each polarizable atom in the BEC with a ``dipole'' force\footnote{Here, ``dipole'' means the induced atomic electric dipole;  the actual deformation of the originally axisymmetric condensate usually  has quadrupole symmetry.} proportional to the gradient of the squared electric field of the laser [for the details of the stirring procedure, see~\cite{Madison:2000b}].  This dipole-force mechanism is the basis for optical trapping of atoms, as discussed briefly  in Sec.~III.E in connection with optical lattices [for a more detailed account, see Sec.~4.2 of \textcite{Pethick:2002} and Sec.~9.4 of \textcite{Pitaevskii:2003}].

\begin{figure}[ht] 
  \includegraphics[width=3.0in]{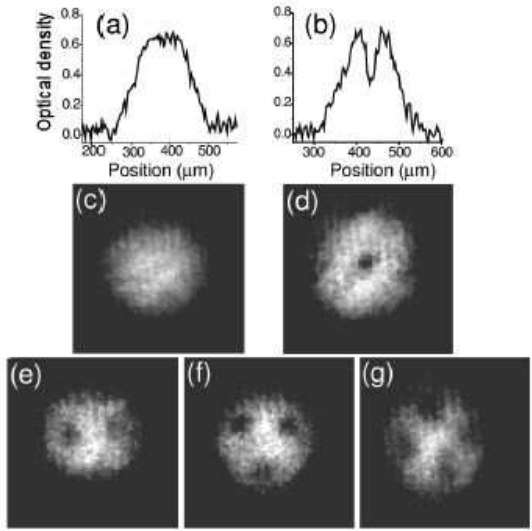}
 \caption{Absorption images of a BEC stirred with a laser beam; in all images the condensate has about $10^5$ atoms, the temperature is below 80 nK and the rotation rate $\Omega/(2\pi)$ increases from 145 Hz for (a) and (c) to 168 Hz for (g). Images (a) and (b) show the optical thickness for images (c) and (d), with the clear appearance of the vortex core. Images (e), (f), and (g) show states with two, three, and four vortices.   From~\textcite{Madison:2000a}.}
 \label{fig:ENS2}
 \end{figure}

\begin{figure} [htbp]
  \includegraphics[width=1.5in]{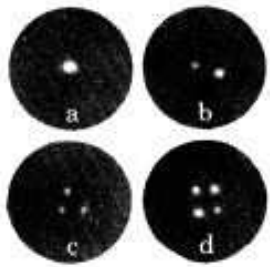}
  \caption{Photographs of stable vortex configurations in superfluid $^4$He, with up to four vortices.  Adapted from \textcite{Yarmchuk:1979}.}
  \label{fig:Yarmchuk}
 \end{figure}

 Since the vortex core radius $\sim \xi\approx 0.2-0.4 \ \mu$m is too small for direct visual detection, they turn off the trap and allow the condensate (including the vortex core) to expand.  The initial tight radial confinement means that this expansion is mostly in the radial direction, and the expanded condensate usually becomes disk-shaped.  The resulting images of the expanded condensate clearly show the presence of one or a few vortices (see Fig.~\ref{fig:ENS2}).  These latter images are remarkably similar to those in Fig.~\ref{fig:Yarmchuk} of vortices in rotating superfluid $^4$He~\cite{Yarmchuk:1979} obtained  by trapping charged particles on the vortex cores [see Sec.~4.5 and Chap.~5 of~\textcite{Donnelly:1991}].  \textcite{Hodby:2002} (Oxford) observe similar vortex patterns with a rotating magnetic deformation of the trap potential.  Their experimental arrangement  allows them to explore the physics of vortex nucleation for relatively large quadrupole deformations.
 
 Careful experimental studies~\cite{Madison:2000a,Madison:2000b,Chevy:2000,Madison:2001} show that the critical angular velocity $\Omega_c$ for the appearance of the first vortex is $\approx 0.7 \omega_\perp$, which is typically much larger than   that predicted in Eq.~(\ref{eq:Omegac}) for the relevant TF limit.  In fact, a rotating axisymmetric condensate has a quadrupole instability at an angular velocity $\Omega = \omega_\perp/\sqrt 2$, which agrees well with the measured value for the appearance of the first vortex.  Sections 14.5 and 14.6 of \textcite{Pitaevskii:2003}  discuss this intriguing behavior in detail.

  \subsubsection{Normal modes of an axisymmetric  condensate with one central  vortex}

\begin{figure}[ht]
  \includegraphics[width=3.0in]{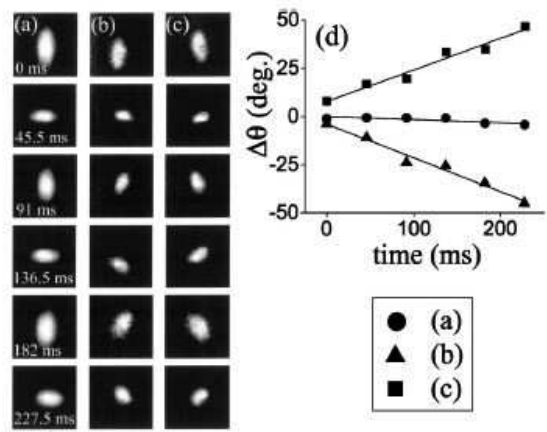}
 \caption{Nondestructive images of quadrupole mode of an  axisymmetric condensate.  Condensate (a) is vortex-free, whereas condensates in (b) and (c) have  straight vortices (with opposite sense of circulation)  aligned  perpendicular to the plane of the figure;  (d) shows a linear fit to the precession angle.  From~\textcite{Haljan:2001a}.}
 \label{fig:quad}
 \end{figure}

  A static condensate without vortices has numerous low-lying small-amplitude normal modes characterized by coupled density and velocity perturbations.  Many experimental studies have confirmed the various theoretical predictions in considerable detail [see pp.~178-195 of \textcite{Pethick:2002} and pp.~177-190 of \textcite{Pitaevskii:2003}].  A  direct treatment of the hydrodynamic equations~\cite{Stringari:1996}  for a nonrotating  axisymmetric condensate in the TF limit  yields the frequency of quadrupole modes with $l = 2$, $m = 0$ and $l = 2$, $m = \pm2$.  The  last  two modes with $m=\pm 2$ are degenerate and  together  produce an oscillating  quadrupole distortion of the condensate in the plane perpendicular to the original axis of symmetry.  In the absence of any perturbation that breaks time-reversal symmetry, the axes of the deformation remain fixed in space [see Figs.~\ref{fig:quad}(a)].  
    
  When the condensate has a vortex, however, its circulating velocity breaks time-reversal symmetry and splits the previously degenerate states with $\pm m$~\cite{Svidzinsky:1998,Zambelli:1998}.  This effect is analogous to the Zeeman splitting of magnetic sublevels.  In the usual TF limit, the fractional splitting is small (of order $|m| d_\perp^2/R_\perp^2$).  One dramatic consequence is that the quadrupole distortion now precesses slowly in the direction of the circulating vortex flow~\cite{Chevy:2000,Haljan:2001a}, as seen in Figs.~\ref{fig:quad}(b) and (c).

\textcite{Zambelli:1998} note that the magnitude of the splitting  provides a direct measure of the angular momentum of the condensate.  The ENS group~\cite{Chevy:2000} use this approach to detect the first appearance of a vortex in the condensate and to verify that its angular momentum  per particle is $\sim\hbar$.   It is remarkable that this macroscopic experiment  with $N \sim 10^6$ atoms measures a single-particle  fundamental  constant; it demonstrates the coherence of the macroscopic wave function [see Sec.~4.7 of \textcite{Leggett:2006} for a valuable perspective on this and related experiments].  Section 2.4 of \textcite{Donnelly:1991} discusses similar earlier macroscopic experiments on rotating  superfluid $^4$He.

The precession of the quadrupole mode provides a sensitive vortex detector (sometimes called ``surface-wave spectroscopy'') that complements the direct visibility of the vortex core.  The JILA group~\cite{Haljan:2001a} use this  method to track the large-amplitude  tilting motion of a  vortex in an axisymmetric condensate.  As a tilted vortex rotates  about the axis of symmetry, its empty core initially moves away from  the line of sight.  Periodically, however, the vortex returns to its original orientation, when the optical visibility  of the core and the quadrupole precession both attain  their initial values. The experiment indeed observes two recurrences, yielding good agreement with the predicted precession frequency~\cite{Svidzinsky:2000b}.

\subsubsection{Bent vortex configurations}

When a disk-shaped condensate rotates faster than the critical velocity $\Omega_m$ for creation of a metastable vortex [see Eq.~(\ref{eq:Omegam})], the initial equilibrium vortex is essentially straight.  For condensates with larger aspect ratios $R_z/R_\perp$, however, the equilibrium vortex shape can be curved, depending on the angular velocity and the specific aspect ratio.  Predictions of this phenomenon arise both from linearized stability analysis for a straight vortex~\cite{Svidzinsky:2000b,Feder:2001a} and from direct calculation of the state that minimizes the energy in the rotating frame~\cite{Garcia:2001a,Garcia:2001b,Aftalion:2001}. In a long cigar-shaped condensate, for example, the initial vortex can reduce its total energy by forming a small symmetrical loop near  the equator.  As the angular velocity $\Omega$ increases, the vortex loop  expands toward the axis of symmetry.   When these vortex loops become large,  ENS experiments~\cite{Rosenbusch:2002}  show that they can have both symmetric configurations (denoted ``U") and antisymmetric configurations [denoted ``(unfolded) N" although ``S'' or ``$\int$'' might be more appropriate].  \textcite{Modugno:2003} give a nice physical picture of how curvature can lower the total energy of the vortex.

\subsubsection{Kelvin (axial) waves on a vortex}

A long straight classical vortex line in an  incompressible fluid represents a dynamical equilibrium between the kinetic-energy density and the pressure, as follows from Bernoulli's equation.  In an influential article, \textcite{Kelvin:1880} (subsequently, he became  Lord Kelvin) studied the small oscillations of such a vortex;  he found a characteristic long-wavelength ($|k|\xi\ll 1$) dispersion relation
\begin{equation}
\label{eq:Kelvin}
\omega_k \approx \frac{ \hbar\,k^2}{2M}\,\ln\left(\frac{1}{|k|\xi}\right),
\end{equation}
where $k$ is the axial wave number, $\xi$ is the vortex core radius, and I  assume a quantized circulation $\kappa = 2\pi \hbar /M$.  For general $k$, these classical Kelvin modes are left-circularly polarized helical waves with velocity potential $\propto \exp\left[i(kz -\phi-\omega_k t)\right] K_1(kr)$, where $(r,\phi,z)$ are  cylindrical polar coordinates and $K_1(u)$ is a Bessel function of imaginary argument that vanishes for large $u$.  Each element of the vortex core executes a circular orbit with a sense opposite to the circulating velocity of the vortex line.  In a quantum-mechanical version, these modes  have quantized angular momentum with $m=-1$.  Note that this  negative helicity ($m=-1$)  is  independent of the direction of the wave propagation ($\pm k$).  The overall helical  behavior reflects the gyroscopic character of the vortex line, and the physics thus differs significantly from that of a stretched string.\footnote{For a detailed comparison, see \textcite{Fetter:1971}.}

Unfortunately, Kelvin's treatment is rather intricate,  for he allowed the unperturbed axisymmetric flow to have an arbitrary radial dependence (including both solid-body rotation and irrotational  vortex flow).  For  accessible discussions of these Kelvin modes, see Sec.~1.7  of \textcite{Donnelly:1991}, pp.~337-344 of \textcite{Fetter:1969}, pp.~138-139 of \textcite{Fetter:2006},  Sec.~29 of \textcite{Lifshitz:1980b} and Sec.~III.D of \textcite{Sonin:1987}.  

In a modern quantum-mechanical context,  \textcite{Pitaevskii:1961} used the Bogoliubov equations (\ref{eq:BOGu}) and (\ref{eq:BOGv}) to show that a singly quantized vortex in a dilute BEC  also has this Kelvin mode with the  same long-wavelength dispersion relation [Eq.~(\ref{eq:Kelvin})].
Although the existence of Kelvin waves are widely accepted, direct evidence for their presence in low-temperature superfluids is relatively meager.  Section 6.1 of~\textcite{Donnelly:1991} and Sec.~III.E of~\textcite{Sonin:1987} discuss the interpretation of relevant superfluid $^4$He experiments in some detail.  

The evidence is much more direct  in    dilute BECs.  \textcite{Bretin:2003} report  a clear signature of such quantized Kelvin modes.  They drive the condensate with a near-resonant quadrupole excitation ($m = \pm 2$) and observe preferential damping of the mode with $m = -2$.  Recall that  the  Kelvin modes have quantized angular momentum $m = -1$ relative to the circulation around the vortex, independent of the direction of $k$.  Thus the authors suggest that this particular excited state with $m = -2$ can decay by the creation of two Kelvin-wave excitations, each with $m = -1$ and opposite  momenta $\pm \hbar k$ along the axis  determined by the conservation of energy.  In contrast, the other quadrupole mode  $m = + 2$ has no such open channel for decay.  A theoretical analysis of the line shape~\cite{Mizushima:2003} confirms this interpretation.  Furthermore, transverse images of the expanded condensate excited  with  the quadrupole mode  $m = -2$ exhibit the expected  periodic spatial  structure associated with the Kelvin-mode deformation, whereas  those excited with $m =+ 2$ have no periodic structure~\cite{Bretin:2003}.

\subsubsection{Other ways  to create vortices}
As seen throughout  Sec.~III.D, experimentalists have devised many  procedures that  successfully create one or more vortices in a  dilute BEC.  Section III.D.1 discusses  dynamical phase-imprinting in a two-component condensate, and Sec.~III.D.2  considers the more common approach of rotating an anisotropic potential that  spins up the condensate.  A related technique (discussed below in Sec.~IV.B in connection with vortex arrays)  is to spin up the normal gas and then cool into the superfluid state~\cite{Haljan:2001b}.  One-dimensional solitons provide an additional unusual alternative, for they generally decay to form three-dimensional vortex rings~\cite{Dutton:2001,Anderson:2001}.  

\paragraph{transfer of orbital angular momentum from a laser} A Gaussian laser beam can be decomposed into  components with  orbital angular momenta $m$ relative to the propagation vector (the Laguerre-Gaussian modes).  Thus  a laser beam can, in principle,   create a vortex by coherently transferring orbital angular momentum to the condensate.  In a recent experiment,  \textcite{Andersen:2006} use a two-photon stimulated Raman process involving a counter-propagating linearly polarized Laguerre-Gaussian mode with $m = 1$ [see, for example \cite{Allen:1999,Allen:2003}] and a linearly polarized Gaussian mode with $m = 0$ (the linear polarization ensures that ``spin'' angular momentum does not cause the observed effect).  Some of the  atoms in an initial single $^{23}$Na condensate absorb a photon with  one unit of angular momentum from the Laguerre-Gaussian beam and emit a photon to the Gaussian beam, acquiring both angular momentum and linear momentum. Consequently, these atoms  form a new  coherent moving  condensate with a singly quantized vortex. After release from the trap, the resulting spatial separation between the new and the original condensates permits clear imaging  of the vortex core.  

\textcite{Andersen:2006} also  use a sequence of such pulses to  make vortices with higher quantized circulation;  as noted below, however, such vortices are essentially unstable with respect to decay into singly quantized vortices.  Recently,  similar techniques  create  persistent quantized flow  with one  unit of circulation in a toroidal trap~\cite{Ryu:2007}; the  $\sim 10$~s lifetime of this persistent flow arises from experimental factors and is not intrinsic.

\paragraph{merging of multiple BECs} Another interesting experiment creates a vortex by merging three $^{87}$Rb condensates in a single disk-shaped trap.  ~\textcite{Scherer:2007} apply a repulsive (blue-detuned) laser beam shaped as a three-fold optical barrier, dividing the noncondensed gas into three non-overlapping regions (see Sec.~III.E for  the effect of external laser beams).  As the gas cools below $T_c$, three independent and uncorrelated BECs form.   The optical barrier is then slowly removed, and  the three BECs merge.  After a variable delay time, they remove the  trap and image   the expanded condensate.   Depending on the initial relative  phases of the three condensates, a net circulating current can form, producing a clearly visible  vortex core in the expanded image.  For random phase differences, a vortex should appear with probability $0.25$, in reasonable agreement with the experimental observations for repeated trials.

\paragraph{imposition of topological phase} As a final topic in this discussion of  forming a single vortex, I consider the creation of a {\it doubly quantized} vortex through imposition of topological phase~\cite{Leanhardt:2002}.  The experiment follows suggestions by \textcite{Isoshima:2000} and \textcite{Ogawa:2002}.  It relies on a special form of magnetic confinement known as the Ioffe-Pritchard trap [for a detailed description, see Sec.~4.1.3 of \textcite{Pethick:2002}].  Here, the important feature is that the $x$ and $y$ components of the magnetic field have a quadrupole form, whereas the $z$ component can be taken as uniform.  

This MIT  experiment studies $^{23}$Na in the lower hyperfine state $|F = 1, m_F = -1\rangle$.  Initially, the $z$ component of magnetic field is large and positive, which aligns the hyperfine spin $\bm F$ vector with the direction of $\bm B \approx B_z \hat{\bm z}$. Thus this spin texture is uniform and  oriented, with variable number density  induced by the nonuniform trap potential.  The experiment then slowly  reverses the $z$ component of magnetic field, and the hyperfine spin vector $\bm F$ adiabatically follows the local magnetic field.   In particular, when $B_z$ vanishes, the hyperfine spin vector $\bm F$ takes the (nonuniform) quadrupole form of the remaining $x$ and $y$ components of $\bm B$.  As $B_z$ then becomes negative, the   topological  winding  of the spin texture continues adiabatically, leading to a nontrivial final texture.  Remarkably, at the end  of the  $B_z$ inversion, a detailed analysis  shows that the local spin texture has an additional factor  $e^{-i2m_F\phi}$, where $m_F = -1$.  Consequently, this final state represents a {\it doubly quantized} vortex with quantum number two.  \textcite{Leanhardt:2002} use surface-wave spectroscopy (precession of the quadrupole oscillation mode, see Sec.~III.D.3) to verify that the vortex indeed has angular momentum $\approx 2\hbar $ per particle.

The energy of a vortex arises principally from the total kinetic energy $\int dV\, \frac{1}{2}Mv^2|\Psi|^2$, which is proportional to the square of the circulation. In a multiply connected geometry like an annulus, such flow represents a macroscopic quantized circulation, and the presence of the inner wall ensures that the hydrodynamic flow generally remains stable (a persistent current) as long as the temperature is well below $T_c$.   The situation is very different for a free-standing  multiply quantized vortex, which should be unstable with respect to splitting into the appropriate number of singly quantized vortices.  As mentioned at the end of Sec.~II.C, \textcite{Pu:1999} study the Bogoliubov equations for a doubly quantized vortex and find imaginary eigenvalues that imply a dynamical instability [see also~\cite{Castin:1999}].  \textcite{Shin:2004}    investigate the stability of these doubly quantized vortices, with convincing images of the  relatively large initial vortex core decaying into two smaller cores in $\approx 75$~ms.  \textcite{Mottonen:2003} provide a theoretical analysis of the MIT experiments, emphasizing the unstable eigenvalues of the Bogoliubov equations.  

An alternative picture of the same topological transformation~\cite{Ogawa:2002,Leanhardt:2002}  relies on the idea of~\textcite{Berry:1984} phase that arises from an adiabatic alteration of the physical system (in this case, the overall direction of the magnetic field).  In the context of quantum mechanics, many recent textbooks provide a careful description of the Berry phase, for example, Sec.~10.3 of \textcite{Abers:2004}, Sec.~10.2 of \textcite{Griffiths:2005}, and pp.~592-607 of \textcite{Shankar:1994}.

\subsection{Laser trapping and optical lattices}

The response of an atom to   electromagnetic radiation is an old subject [see, for example,~\textcite{Allen:1987} and \textcite{Cohen:1998}], yet it has many important applications to the interaction between  ultracold dilute atoms and laser beams.  To understand the basic physics,  place an electric dipole with moment  $\bm p$ in an external electric field $\bm E$.  Standard electrostatics yields the  interaction energy 
$U_{\rm int} = - \bm p \cdot \bm E$,
where the minus sign means that the lowest-energy configuration has the dipole moment oriented along $\bm E$. 

 This expression holds for a fixed dipole moment $\bm p$,  but an atom typically acquires its electric dipole moment as a result of induced polarization.  In such a case,  the dipole  moment is proportional to the applied field, with $\bm p = \alpha \epsilon_0 \bm E$  in SI units, where $\epsilon_0\approx 8.85\times 10^{-12}$ farad/m is the permittivity of empty space and  $\alpha$ is the atomic polarizability. As a very simple classical model, a grounded conducting sphere of radius $b$ has a polarizability $\alpha = 4\pi b^3$ (three times its volume), and this picture provides a useful order-of-magnitude estimate for atomic polarizabilities.  The total interaction  energy  of the  atom with  the external field then builds up from $\bm E_i = \bm 0$, and integration yields the final expression 
\begin{equation}
\label{eq:Uint}
U_{\rm int} = - {\textstyle\frac{1}{2}} \alpha \,\epsilon_0\,E^2.
\end{equation}

If the electric field $\bm E(\bm r)$ is static and  nonuniform, then $U_{\rm int}(\bm r)  = - {\textstyle\frac{1}{2}} \alpha \,\epsilon_0\,E^2(\bm r) $ is also nonuniform.  Thus an atom experiences a force that seeks to minimize the interaction energy, placing the atom at the {\it maximum} of $E^2$ because the static polarizability is positive.  This result also holds for low frequencies, but the situation eventually changes because of  induced transitions among atomic energy levels.  The Drude model of a bound electron with resonant frequency $\omega_0$ and damping time $\tau$ subject to an oscillatory driving field  yields the frequency-dependent polarizability  
\begin{equation}
\label{eq:alpha}
\alpha(\omega) = \frac{\alpha(0)\,\omega_0^2}{\omega_0^2 -\omega^2-i\omega/\tau},
\end{equation}
where $\alpha(0)$ is the (positive) static polarizability.  Note that the real part of $\alpha(\omega)$  changes from positive to negative when the applied frequency $\omega$ exceeds the resonant frequency $\omega_0$. The combined relation $U_{\rm int}(\omega)$ from Eqs.~(\ref{eq:Uint}) and (\ref{eq:alpha}) is sometimes called the {\it ac Stark effect}.

These ideas apply directly to the optical trapping of atoms by a focused laser.
If the frequency $\omega$ of the laser is less than that of the atomic resonance (``red detuning'') with ${\rm Re}\, \alpha(\omega) > 0$, then Eqs.~(\ref{eq:Uint}) and (\ref{eq:alpha}) indicate
that the atoms  preferentially move toward the squared-field {\it maxima}, whereas if the laser frequency is higher than the resonant frequency (``blue detuning'') with ${\rm Re}\, \alpha(\omega)<0$, then the atoms  move toward the squared-field {\it minima}.  In practice, it is necessary to use far-off-resonance lasers to avoid dissipation.  For the alkali-metal atoms of interest here (Li, Na, K, Rb, Cs),   the lowest  $s$ to $p$ transition of the valence electron predominates, as in  the familiar yellow emission spectrum of Na. \textcite{Stamper:1998}  successfully trap a pre-existing $^{23}$Na BEC with  a focused infrared laser beam.  Unlike magnetic traps, which confine only certain hyperfine states, these optical traps confine all such states.  This experimental technique allows the study of  ``spinor'' condensates (see Sec.~VII.B.3) that involve coupled BECs, one for each hyperfine component of a given $\bm F$ multiplet~\cite{Ho:1998,Ohmi:1998}.

A  more subtle example is the formation of an optical standing wave to trap atoms in a periodic array. Consider a laser beam propagating along the $ z$ axis with electric field $\bm E(z,t) =  \bm E_0 \cos(kz-\omega t)$. If the laser beam  reflects normally off a mirror at $z = 0$, the resulting  standing wave $\bm E_0\left[\cos(kz-\omega t) -\cos(kz+\omega t)\right]= 2\bm E_0 \sin kz\sin \omega t$ yields the corresponding  squared electric field  $|E(z,t)|^2 =4 |E_0|^2 \sin^2 kz\,\sin^2\omega t$.  This periodic lattice has a spatial period $\frac{1}{2}\lambda$, where $\lambda = 2\pi/k$ is the wavelength of the original laser beam. 
For either sign of detuning, the atoms assume a spatially periodic configuration in this optical lattice, as emphasized by \textcite{Jaksch:1998}.   This seminal idea has  wide applications in the study of ultracold dilute gases, as discussed, for example, in Sec.~9.4 and Chap.~16  of \textcite{Pitaevskii:2003} and Sec.~IV of \textcite{Bloch:2007}. The strength of the laser field can be controlled externally,  so that  the height of the barrier for tunneling between adjacent  layers can vary from small to large.  This possibility leads to remarkable experiments that demonstrate the transition from superfluid to insulating behavior in three-dimensional optical lattices~\cite{Greiner:2002}.  In the present context of one-dimensional lattices, \textcite{Anderson:1998} are the  first to demonstrate such  atom trapping.  Their vertical lattice  experiences a gravitational potential difference $Mg\lambda/2$ between adjacent layers, with associated interlayer tunneling at a measured ac  Josephson frequency $\omega_J \approx Mg\lambda/(2\hbar)$~\cite{Josephson:1962}.

A particularly interesting situation arises when a  rotating condensate is subsequently confined in a one-dimensional optical lattice oriented along the rotation axis.  Consider an axisymmetric condensate with a single vortex on the axis of symmetry. Then adiabatically turn on an applied standing-wave laser field.  If the laser beams propagate   along the symmetry axis, the optical lattice  slices the rotating condensate into many circular disks.   The resulting periodic atomic array forms a set of effectively two-dimensional condensates that are coupled by tunneling between adjacent layers.  Each two-dimensional condensate contains a segment of the previously continuous vortex line.  Thus it becomes   a single pancake vortex, and the position of each  vortex center can fluctuate more-or-less independently,  depending on the strength of the laser field and the height of  resulting potential barriers.  \textcite{Martikainen:2003} use a variational trial function to study the resulting  axial modes in this discrete system, which are the analog of the Kelvin modes studied in Sec.~III.D.5.  At present, no direct experimental evidence for such behavior exists.

\subsection{Berezinskii-Kosterlitz-Thouless transition in  dilute BECs}

Another remarkable application of an optical lattice is the study of the equilibrium state  of a two-dimensional  dilute Bose-Einstein gas.   Although such a uniform  two-dimensional  system  cannot have a true condensate with long-range order~\cite{Mermin:1966,Hohenberg:1967}, nevertheless  \textcite{Berezinskii:1971} and \textcite{Kosterlitz:1973} predict that it can indeed become superfluid below a certain critical temperature $T_{BKT}$.  Above this temperature,  free single vortices can form easily.  The motion of such vortices  disrupts any  organized long-range phase coherence \cite{Leggett:2001} and hence  destroys the associated superfluid flow.  Below $T_{BKT}$, in contrast,  free vortices do not occur, although bound pairs of a vortex and an antivortex do  exist [in the present context of dilute ultracold gases, Sec.~17.5 of \textcite{Pitaevskii:2003} discusses this behavior in more detail].  As the temperature increases, so does  the size of the bound pairs, and  they eventually  become unbound at $T_{BKT}$, producing free vortices.
Experiments with thin superfluid $^4$He films~\cite{Bishop:1978}  and with Josephson-coupled superconducting arrays~\cite{Resnick:1981}  have both confirmed these predictions in considerable detail.

In a recent experiment at ENS (Paris), \textcite{Hadzibabic:2006} apply a standing-wave laser field perpendicular to the symmetry axis  of a cigar-shaped condensate, splitting the atoms into two essentially independent thin pancake strips with lateral dimensions $\approx 120\ \mu{\rm m}\times 10\ \mu{\rm m}$.  See Fig.~\ref{fig:ENS1} for the geometry (although there is  no rotating laser beam in the present case): the laser beams propagate in the $x$ direction and  the pancakes are parallel to the $yz$ plane.  After these two isolated Bose-Einstein gases reach thermal equilibrium, the confining traps are turned off.  Both pancakes then expand primarily in the $x$ direction  perpendicular to their planes  and  overlap spatially.  

\begin{figure}[ht] 
  \includegraphics[width=2.5in]{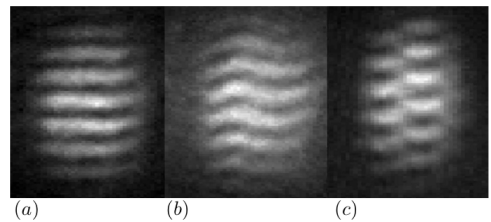}
 \caption{Interference fringes between  two pancake condensates (a) at low temperature, (b) at high temperature showing wavy interference fringes without vortices, (c) fringe dislocation  indicates the presence of a vortex in one or both condensates.  From~\textcite{Kruger:2007}.}
 \label{fig:BKT}
 \end{figure}

Soon after the first observations of BECs in dilute trapped gases, the MIT group~\cite{Andrews:1997} cut a cigar-shaped condensate in half with a blue-detuned laser, creating two separate three-dimensional condensates.  When the trap is turned off, these two condensates overlap and produce  clear straight  interference fringes.  Similarly, the ENS experiment here looks for and finds interference fringes between the two pancakes.  At low temperature [see Fig~\ref{fig:BKT}(a)], the fringes are indeed straight, indicating constant relative phase.  As the temperature increases, however, the fringes become wavy [Fig~\ref{fig:BKT}(b)];  furthermore, as seen in Fig~\ref{fig:BKT}(c), they sometimes contain one or more dislocations, which  indicates the presence of free vortices in one or both pancakes~\cite{Stock:2005,Simula:2006}.  They repeat the  experiment at various temperatures and use the presence of free vortices to determine $T_{BKT}$.  The resulting value agrees well with the theoretical predictions of \textcite{Berezinskii:1971} and \textcite{Kosterlitz:1973}, although the finite size of the present samples yields a  crossover instead of a sharp transition.

\textcite{Schweikhard:2007}  report an intriguing study of vortex formation in   a two-dimensional array of Josephson-coupled  BECs.  They start at  $T$ below $T_c$ with a disk-shaped condensate along with a residual normal cloud and then apply a hexagonal optical lattice, transforming the system into an array of coupled BECs that act like Josephson junctions.  Each condensate has a single effective phase, and the coupling energy between adjacent condensates has the form $J(1-\cos\Delta\phi)$ where $\Delta\phi$  is their relative phase.  Equivalently, the tunneling current between any pair of condensates is proportional to $J\sin\Delta\phi$  [for a careful discussion, see \textcite{Josephson:1962}, \textcite{Leggett:2001}, and Chap.~6 of \textcite{Tinkham:1996}].   Varying the amplitude  of the optical lattice  through the laser intensity tunes the coupling parameter $J$. 

The experiment starts by inferring the temperature $T$ from a nondestructive image and then ramps down the optical lattice, allowing the adjacent BECs to merge.  As in~\textcite{Scherer:2007}, discussed  in Sec.~III.D.6, vortices can form depending on the relative phases,  appearing as holes in the expanded gas cloud after  the trap is turned off.  For a  given temperature $T$, \textcite{Schweikhard:2007} measure the vortex density $\cal D$ near the center of the cloud as a function of the Josephson coupling parameter $J$.  For two different temperatures, a plot $\cal D$ as a function of $J/T$ shows nearly universal behavior:   $\cal D$ is negligibly small for $J/T \gg 1$ (low temperature), but it rises rapidly and saturates for $J/T\ll 1$ (high temperature), as expected for the BKT vortex unbinding crossover.

\section{Vortex Arrays in Mean-Field Thomas-Fermi (TF) Regime}

A rapidly rotating condensate has a dense array of  vortices, with a uniform  areal density~\cite{Feynman:1955}
\begin{equation}
\label{eq:nv}
 n_v = \frac{M \Omega}{\pi\hbar},
\end{equation}
 as follows from these  arguments:
   
     (1)  Consider a steady classical incompressible  flow with velocity $\bm v(\bm r)$ in a rigid container that rotates with angular velocity $\bm \Omega$.  If the fluid has number density $n(\bm r)$, then the relevant part of the energy in the rotating frame is 
     \begin{equation}
     \label{eq:E'noncl}
     E' = \int dV\,(\textstyle{\frac{1}{2}} M v^2 -M\bm\Omega\cdot \bm r\bm\times \bm v )\,n. 
     \end{equation}
     This result is equivalent to 
          \begin{equation}
     \label{eq:E'noncla}
     E' = \textstyle{\frac{1}{2}}M \int dV\, (\bm v -\bm \Omega\bm\times \bm r)^2n - E_{\rm sb}, 
     \end{equation}
     where $E_{\rm sb} = \frac{1}{2} M\int dV\,|\bm\Omega\bm \times \bm r|^2n = \frac{1}{2} I_{\rm sb}\Omega^2$ is the rotational energy for solid-body rotation with the corresponding moment of inertia $I_{\rm sb}$.  The first term of Eq.~(\ref{eq:E'noncla}) is non-negative, and hence the absolute minimum of $E'$ occurs when $\bm v  = \bm\Omega\bm\times \bm r$, namely solid-body rotation $\bm v_{\rm sb} = \bm\Omega\bm\times \bm r$;  in addition, the corresponding minimum value of $E'_{\rm min}$  has the classical  value $-E_{\rm sb}$.  Since $\bm \nabla\bm\times \bm  v_{\rm sb} = 2\bm\Omega$, the irrotational superflow cannot assume such a solid-body rotation, but it seeks to mimic this flow through the nucleation of quantized vortices. 
     
The angular momentum  $ L_z = \langle \bm r\bm \times\bm p \rangle $ follows from the thermodynamic relation
\begin{equation}
\label{eq:HF}
 L_z = -(\partial E'/\partial \Omega)_N, 
\end{equation}
[the Hellmann-Feynman theorem gives the same result.  See, for example, pp.~1192-1193 of \textcite{Cohen:1977}; for a brief history of this theorem, see p.~1 of \textcite{Feynman:2000}] .  Since the actual $E'$ exceeds the classical value $-\frac{1}{2} I_{\rm sb} \Omega^2$ by a positive correction that depends on $\Omega$, the equilibrium value of the angular momentum necessarily differs from the classical value $I_{\rm sb}\Omega$.   Evidently, this system exhibits a nonclassical moment of inertia [see Sec.~2.5 of \textcite{Leggett:2006} for a discussion of these quantum-mechanical effects].  In addition, it also has transverse shear modes (the Tkachenko modes, see Sec.~IV.B.4), and it ``melts'' for sufficiently large angular velocities to form a ``liquid'' state (see Sec.~VI).  Whether this system should be considered a ``supersolid''  in the conventional sense [see, for example, Sec.~8.3 of ~\textcite{Leggett:2006}] merits additional study.

   (2) For a superfluid, Eq.~(\ref{eq:vorticity}) shows that a  singly quantized vortex at a position $\bm r_0$  has a localized vorticity $\bm \nabla \bm\times \bm v= (2\pi \hbar/M)\,\delta^{(2)}(\bm r-\bm r_0) \,\hat{\bm z}$.  In a rotating superfluid, the vortices will be uniformly distributed to approximate the uniform classical vorticity.  Take a contour $\cal C$ containing ${\cal N}_v$ such vortices   in an area $\cal A$; the total circulation around $\cal C$ is $\Gamma_{\cal C} = (2\pi \hbar /M)\,{\cal N}_v$.  Stokes's theorem shows that the corresponding classical value would be $\Gamma_{\cal C}^{\rm cl} = 2\Omega {\cal A}$, and comparison  immediately yields the mean areal vortex density $n_v = {\cal N}_v/{\cal A} =  M\Omega/(\pi\hbar)$.  The  inverse $n_v^{-1}$  is the area per vortex $\pi \hbar/(M\Omega) \equiv \pi l^2$, which defines the radius 
   \begin{equation}
   \label{eq:r0}
   l = \sqrt{\frac{\hbar}{M\Omega}}
   \end{equation}
    of an equivalent circular cell.  Thus the intervortex separation is $\approx 2l$.  Note that $l$ decreases with increasing $\Omega$ (it is also the usual oscillator length for a frequency $\Omega$). 

\subsection{Physics of BEC in axisymmetric harmonic traps for rapid rotation}

The Feynman relation (\ref{eq:nv}) shows that the number of vortices increases linearly with $\Omega$ assuming that the geometry remains fixed.  Although this picture  indeed applies to superfluid $^4$He in a rotating bucket, the situation is quite different for a dilute gas in a harmonic trap, because the centrifugal forces expand the condensate radially (number conservation means that the  condensate  also shrinks along the axis of rotation).  As a result, the total number of vortices in a given condensate \begin{equation}
\label{eq:Nv}
{\cal N}_v(\Omega) \approx n_v \pi R_\perp^2(\Omega) = M\Omega R_\perp^2(\Omega)/\hbar = R_\perp^2(\Omega)/l^2
\end{equation}
 increases faster than linearly with $\Omega$.

To quantify the dependence of $R_\perp$ on $\Omega$,  it is convenient to focus on the TF limit of a large condensate, when the spatial variation of the condensate density is negligible compared to the other terms in the  energy functional.  Thus $\bm p \Psi = -i \hbar \bm \nabla \Psi = (\hbar\bm \nabla S) \Psi -i\hbar e^{iS}\bm\nabla |\Psi| \approx M \bm v\,\Psi $, and the GP energy functional in the rotating frame Eq.~(\ref{eq:E'GP}) simplifies to  
\begin{eqnarray}
\label{eq:E'TF}
E'[\Psi] \approx& \int dV\,\left[\left(\textstyle{\frac{1}{2} }M v^2-M\bm\Omega\cdot \bm r\bm\times \bm v +V_{\rm tr}\right) |\Psi|^2\right. \nonumber \\
& +\left.\textstyle{\frac{1}{2} } g|\Psi|^4\right].
\end{eqnarray}
Like the TF limit  $E_{\rm TF}[\Psi]$ in Eq.~(\ref{eq:ETF}) for a stationary condensate, this equation  depends only of $|\Psi|^2$ and $|\Psi|^4$, but it  now includes the hydrodynamic flow through $\bm v$.  

\begin{figure}[ht] 
  \includegraphics[width=3.5in]{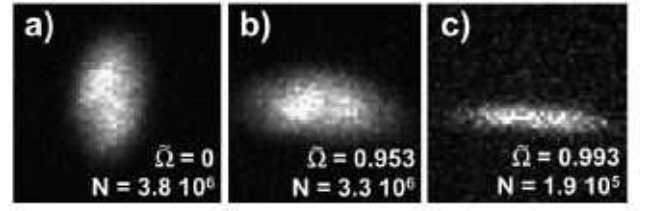}
 \caption{Side view of BECs in a trap. (a)  Static BEC, with aspect ratio $R_z/R_\perp = 1.57$.  Dramatically reduced aspect ratio is evident  for  (b)  $\Omega/\omega_\perp = 0.953$ and (c) $\Omega/\omega_\perp = 0.993$.  From~\textcite{Schweikhard:2004}.}
 \label{fig:aspect}
 \end{figure}

In the typical case of rotation about $\hat{\bm z}$, simple manipulations yield the equivalent expression
\begin{eqnarray}
\label{eq:E'TFa}
E'[\Psi] &\approx&\int dV \textstyle{\frac{1}{2} }M\left(\bm v-\bm v_{\rm sb}\right)^2 |\Psi|^2\nonumber \\
 &+&\int dV \left[ (V_{\rm tr} -\textstyle{\frac{1}{2} }\Omega^2 r^2) |\Psi|^2  + \textstyle{\frac{1}{2}} g|\Psi|^4\right]\nonumber \\
&\equiv& E_v' + E_{\rm TF}' ,
\end{eqnarray}
where $E_v'$ and $E_{\rm TF}'$ denote the  contributions in the first and second lines.   For a large condensate with many vortices,   the actual superfluid velocity $\bm v$ closely approximates the classical solid-body value $\bm v_{\rm sb}$, so that $E_v'$ [the term containing $(\bm v-\bm v_{\rm sb})^2$] vanishes to leading order.    The remaining terms $E_{\rm TF}'(\Omega)$  have exactly the same form as $E_{\rm TF}$ in Eq.~(\ref{eq:ETF}), but the radial part of the confining potential is altered to $\frac{1}{2}M(\omega_\perp^2 -\Omega^2)r^2$.  Variation  with respect to $|\Psi|^2$ gives an equation of the same form as Eq.~(\ref{eq:nonuniform}), but with the   reduced radial confining potential.  Thus the TF density again has the form of Eq.~(\ref{eq:parabolic}), but with modified TF condensate radii that depend on $\Omega$.  The normalization condition in three dimensions shows that 
\begin{equation}
\label{eq:mu(o)}
\mu_{\rm TF}(\Omega)=\mu_{\rm TF}(0)(1-\Omega^2/\omega_\perp^2)^{2/5}, 
\end{equation}
so that the chemical potential decreases continuously and formally vanishes for  $\Omega\to \omega_\perp$, as does the central density $\mu_{\rm TF}(\Omega)/g$ (because of the reduced radial confinement).  A combination of these results leads to the specific $\Omega$ dependence of the condensate dimensions~\cite{Butts:1999,Fetter:2001a,Raman:2001}
\begin{equation}
\label{eq:TFOmega}
\frac{R_\perp(\Omega)}{R_\perp(0)} = \left(1-\frac{\Omega^2}{\omega_\perp^2}\right)^{-3/10},\>\> \frac{R_z(\Omega)}{R_z(0)} =   \left(1-\frac{\Omega^2}{\omega_\perp^2}\right)^{1/5}.
\end{equation}
As anticipated, the external rotation expands the condensate  radially and shrinks it  axially, approaching a two-dimensional configuration.  The corresponding aspect ratio
\begin{equation}
\frac{R_z(\Omega)}{R_\perp(\Omega)} = \frac{\sqrt{\omega_\perp^2-\Omega^2}}{\omega_z}
\end{equation}
provides a convenient diagnostic tool to infer the actual angular velocity of the rotating condensate~\cite{Haljan:2001b,Raman:2001,Schweikhard:2004}, as seen in Fig.~\ref{fig:aspect}.

The two terms  in Eq.~(\ref{eq:E'TFa}) have quite different magnitudes in the mean-field TF regime. 
The dominant contribution $E_{\rm TF}'(\Omega)$ is  the TF energy of the rotating condensate with the modified squared radial trap frequency $\omega_\perp^2 \to \omega_\perp^2 -\Omega^2$.  This quantity follows directly from Eqs.~(\ref{eq:ETFN}) and  (\ref{eq:mu(o)}) 
\begin{equation}
\label{eq:E'TFb}
E_{\rm TF}' (\Omega)= \textstyle{\frac{5}{7}}\,\mu_{\rm TF}(\Omega) N\propto  (1-\Omega^2/\omega_\perp^2)^{2/5}.
\end{equation}
The thermodynamic relation in Eq.~(\ref{eq:HF}) yields the corresponding TF angular momentum of the rotating condensate, and some algebra gives the simple expression
\begin{equation}
\label{eq:LTF}
L_{\rm TF} = \textstyle{\frac{2}{7}}\,MNR_\perp^2(\Omega)\,\Omega = I_{\rm sb} (\Omega)\,\Omega,
\end{equation}
where $I_{\rm sb} (\Omega) = \textstyle{\frac{2}{7}}\,MNR_\perp^2(\Omega)$ is the solid-body moment of inertia for the deformed rotating condensate (this expected result follows from a separate calculation of $\langle r^2 \rangle_{\rm TF}$).

The remaining term $E_v'$ in Eq.~(\ref{eq:E'TFa}) gives a small  nonclassical contribution to the 
angular momentum of the rotating condensate.  If the volume integral is approximated as a sum of integrals over circular cylindrical cells of radius $l$ centered at each vortex, a straightforward analysis yields 
\begin{equation}
\label{eq:Lv}
E_v'(\Omega) \approx N \hbar \Omega\ln\left(l/\xi \right).
\end{equation}
 With logarithmic accuracy (ignoring the $\Omega$ dependence inside the logarithm), the corresponding nonclassical angular momentum then becomes 
 \begin{equation}
\label{eq:Lva}
L_v' = -\frac{\partial E_v'}{\partial \Omega}\approx -  N \hbar \ln\left(\frac{l}{\xi}\right).
\end{equation}
Thus the total angular momentum of the rotating condensate indeed displays  nonclassical behavior [see Secs.~2.5 and 8.3 of \textcite{Leggett:2006}]
\begin{equation}
\label{eq:nonclass}
L_z \approx  I_{\rm sb} (\Omega) \,\Omega -  N \hbar \ln \left(l/\xi\right).
\end{equation}
Note that the correction  reflects  the inability of the superfluid to rotate like a  solid body.  It arises directly from the quantization of circulation and would vanish in the classical limit  ($\hbar \to 0$) when $\kappa \to 0$ and $n_v = 2\Omega/\kappa\to \infty$.  The correction $L_v'$ also becomes small in the limit of a dense vortex array ($l\to \xi$), as discussed in  Sec.~V.B.1.

\subsection{Experimental and theoretical studies of vortex lattices in the TF regime}

The first images of one or a few vortices (Sec.~III.D) rapidly led to the creation of large vortex arrays.  Initially, the MIT group~\cite{Abo:2001} use a rotating deformation  (like the ENS technique---see Fig.~\ref{fig:ENS1}) to obtain  remarkable images of  triangular vortex lattices that are  strikingly  uniform, even near the outer edge of the condensate.  These arrays closely resemble  the classic triangular flux-line lattice  in type-II superconductors predicted by \textcite{Abrikosov:1957} and the triangular vortex lattice in rotating neutral incompressible superfluids predicted by \textcite{Tkachenko:1966}.

\begin{figure}[ht] 
  \includegraphics[width=1.5in]{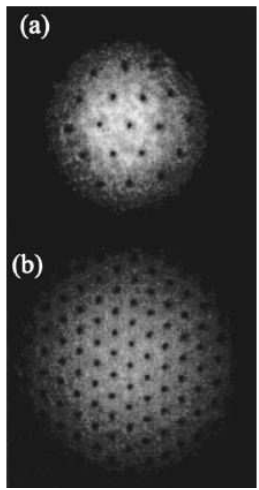}
  \caption{Images of expanded $^{87}$Rb condensates showing (a) small vortex array for slow rotation and (b)  large vortex array for rapid rotation.  Note the highly regular triangular form.  Adapted from~\textcite{Coddington:2004}.}
 \label{fig:JILA}
 \end{figure}
 
The JILA group~\cite{Haljan:2001b,Engels:2003} take a quite different approach to imparting angular  momentum to the condensate.  They  start by spinning-up  a normal cloud above $T_c$ with an elliptically  deformed rotating   disk-shaped trap.  To cool into the BEC regime, they adiabatically  alter their trap to cigar-shaped and then evaporate normal atoms near the  ends of the axis of symmetry (these atoms have small angular momentum).  This process  yields a rotating condensate with relatively large angular momentum per particle and good triangular arrays, as seen in Fig.~\ref{fig:JILA}.  These  rapidly rotating BECs  facilitate a series of detailed experiments on the properties of the vortex lattice.

\subsubsection{Collective modes of rotating condensates}

The study of collective modes of a static  condensate provides one of the  crucial tests of the  time-dependent GP equation, as discussed in Sec.~IV of \textcite{Dalfovo:1999} and Chap.~12 of \textcite{Pitaevskii:2003}. One particularly effective theoretical approach~\cite{Stringari:1996} linearizes the hydrodynamic equations (\ref{eq:cons}) and (\ref{eq:Bern}) in the small perturbations $\delta n$ and $\delta\bm v = \bm \nabla \delta \Phi$.  Since this formalism explicitly assumes irrotational flow, it fails for a rapidly rotating TF condensate with a large vortex lattice, where the average superfluid velocity has uniform vorticity $\bm \nabla\bm\times\bm v \approx 2\bm\Omega$.  

\textcite{Cozzini:2003} propose a simple generalization, based on the Euler equation of nonviscous hydrodynamics, retaining the full convective term $\partial\bm v/\partial t + (\bm v\cdot\bm\nabla) \bm v$.  A familiar vector identity yields the equivalent form of the Euler equation in the laboratory frame
\begin{equation}
\label{eq:rotn}
M\frac{\partial\bm v}{\partial t} +\bm \nabla\left(\frac{1}{2}M v^2 + V_{\rm tr} + gn\right) = M \bm v\bm\times (\bm \nabla\bm\times\bm v),
\end{equation}
which includes both the trap potential $V_{\rm tr}$ and the Hartree potential $gn$, but omits the quantum kinetic energy from Eq.~(\ref{eq:Bern}), as appropriate in the TF limit.    For  a nonrotating  irrotational condensate,  $\bm \nabla \bm\times \bm v $  vanishes, and Eq.~(\ref {eq:rotn}) then omits the right-hand side, reproducing the gradient of  Eq.~(\ref{eq:Bern}) in the  TF limit.  In contrast, a rotating TF condensate has  many vortices, with $\bm \nabla\bm\times \bm v \approx 2\bm\Omega$.  Thus \textcite{Cozzini:2003} replace the right-hand side of Eq.~(\ref{eq:rotn}) with $2M\bm v\bm\times \bm \Omega$, which they call the limit of  ``diffused  vorticity," although ``mean-field'' or coarse-grained'' vorticity may be more descriptive.  A detailed analysis gives the frequencies of the lowest $m = \pm 2$  quadrupole modes in the laboratory frame
\begin{equation}
\label{eq:quad}
\omega\,(m=\pm2) = \sqrt{2\omega_\perp^2-\Omega^2} \pm \Omega.
\end{equation}
In particular, the difference of the two frequencies is $2\Omega$, whereas the sum of the squared frequencies is $4\omega_\perp^2$, independent of the  rotation rate.  \textcite{Haljan:2001b} fully confirm both these detailed predictions.

As an extension of these experiments, \textcite{Engels:2002} study the dynamics of the (originally triangular) vortex lattice in a rapidly rotating condensate.  They can image in all three perpendicular directions, which allows them to confirm theoretical predictions~\cite{Feder:2001b,Garcia:2001b} that the vortices are nearly straight in such dense arrays.  Furthermore,  Eq.~(\ref{eq:quad}) shows that the $ m = \pm 2$ quadrupole modes   have several intriguing  features in the limit $\Omega\to \omega_\perp$.   In particular, \textcite{Engels:2002} note that the resulting $m= -2$ mode is small and hence nearly stationary in the laboratory frame, but the corresponding $m = 2$ mode is nearly stationary in the rotating frame because  $\omega(m = 2)' = \omega(m = 2) -2\Omega$ then becomes small [see Eq.~(\ref{eq:shifted})].

In a particularly interesting experiment, \textcite{Engels:2002} form a rotating vortex lattice with an angular velocity $\Omega/\omega_\perp\sim 0.95$.  They then apply a relatively small  trap distortion   with $(\omega_x^2-\omega_y^2)/(\omega_x^2 +\omega_y^2)\sim 0.036$ that is stationary in the  laboratory. Such a fixed  perturbation is nearly resonant with the slow $m = -2$ quadrupole mode.  As a result, the rotating condensate acquires a  stationary quadrupole distortion that grows to $\sim 0.4$  over a time of $\sim 300$ ms.  Hence the rapidly rotating triangular vortex lattice experiences an oscillating shear distortion with a period that is $\frac{1}{6}$ times the overall rotation period $2\pi/\Omega$ because of the six-fold symmetry.  Every time a basis vector of the rotating lattice lines up with the minor axis of the  fixed elliptical confining potential, the vortex lattice deforms significantly,   forming what are effectively  parallel stripes  of closely  spaced vortices. 

To understand this behavior in detail, recall  that the stationary elliptical  distortion  of the rotating condensate induces a quadrupolar irrotational flow (see the  discussion at the beginning of Sec.~III.D.2) that combines with the   overall rotation $\bm v_{\rm sb} = \bm \Omega\bm\times \bm r$ to produce the total flow velocity $\bm v$.  In the laboratory frame, the vortices  follow streamlines of this total local velocity.   \textcite{Cozzini:2003}  integrate the relevant dynamical equations of motion for the vortices.  They indeed find stripe-like patterns whenever  one of the lattice vectors lies along the minor axis of the distorted condensate. 

 In a related approach, \textcite{Mueller:2003} use a variational two-dimensional wave function.  Although their formalism strictly applies only in the extreme limit of $\Omega\to \omega_\perp$   when the interaction energy is small compared to the single-particle energy (the ``lowest Landau-level'' regime, discussed below in Sec.~V), their  trial wave function can describe a two-dimensional lattice of vortices with general symmetry, using Jacobian elliptic functions that are  closely related  to those in the  earlier work of \textcite{Tkachenko:1966}.  In their simulation, a small elliptical distortion of the trap potential at $t = 0$ induces an increasing distortion of the  condensate's aspect ratio.   The initial triangular vortex array remains a well-defined lattice  but  deforms following the combined velocity field of the rotational and  irrotational flow.  As time evolves, stripes appear periodically, producing  images that closely resemble those found experimentally~\cite{Engels:2002}.

\subsubsection{Uniformity of the vortex array}

Nearly 30 years ago, \textcite{Campbell:1979}  studied  numerically the various equilibrium arrangements of two-dimensional vortex  arrays in uniform incompressible ideal fluids inside  a fixed circular boundary (as a model for  superfluid $^4$He in a rotating bucket). They found predominantly  distorted triangular arrangements with concentric circles of vortices and  a depleted region near the outer edge. \textcite{Yarmchuk:1979} see similar patterns in rotating superfluid $^4$He (see Fig.~\ref{fig:Yarmchuk}), although  the small number of vortices precludes a definitive comparison.   In contrast, images of vortex lattices in rotating BECs are strikingly    triangular and regular, even out to the edge of the visible condensate, as seen in  Fig.~\ref{fig:JILA}.  

The treatment of the density profile in the mean-field Thomas-Fermi regime omits the term $(\bm v -\bm v_{\rm sb})^2$ entirely [see Eq.~(\ref{eq:E'TFa})],  which assumes a uniform vortex distribution and ignores the fine-grain structure of individual vortices.  \textcite{Sheehy:2004a,Sheehy:2004b} investigate these effects by considering a two-dimensional TF condensate with number density $n(\bm r) = n(0)\,(1-r^2/R_\perp^2)$, where $R_\perp$ grows with increasing  $\Omega$.  This rotating condensate contains an array of vortices at regular two-dimensional  lattice sites $\bm r_j$ with  mean density $\overline{ n}_v = M\Omega/(\pi\hbar)$ [the Feynman value in Eq.~(\ref{eq:nv})].  They then subject each vortex to  a small displacement field $\bm u(\bm r)$, so that the original site moves to $\bm r_j + \bm u(\bm r_j)$, and the vortex density becomes 
\begin{equation}
\label{eq:nvr}
n_v(\bm r) = \overline{n}_v\,\left[1-\bm \nabla\cdot \bm u(\bm r)\right]
\end{equation}
This distorted vortex density in turn induces a change in the mean background fluid flow based on the following integral
\begin{equation}
\label{eq:newv}
\overline{\bm v}(\bm r) \approx \frac{\hbar}{M}\int d^2 r' n_v(\bm r') \frac{\hat{\bm z}\bm\times(\bm r - \bm r')}{|\bm r-\bm r'|^2}=\bm \Omega\bm\times\bm r-2\bm \Omega\bm\times \bm u(\bm r),
\end{equation}
where $\bm \Omega = \Omega \hat{\bm z}$ and the second term is a correction to the solid-body value.

A detailed analysis with Eqs.~(\ref{eq:E'TFa}), (\ref{eq:nvr}) and (\ref{eq:newv}) expresses the deformation part of the energy $E'$ in terms of $\bm u$, and the corresponding Euler-Lagrange equation leads to the general expression 
\begin{equation}
\label{eq:u(r)}
\bm u(\bm r) \approx-\frac{l^2}{8}  \ln\left(\frac{l^2}{\xi^2}\right) \bm\nabla \ln n(\bm r),
\end{equation}
where $n(\bm r)$ is the condensate number density,  $l$ is the mean circular cell radius from Eq.~(\ref{eq:r0}), and $\xi$ is evaluated with the central density.\footnote{As a very simple application, assume a uniform particle density, so that  $\bm u$ vanishes. Equation (\ref{eq:nvr}) then indicates that the vortex density is also uniform.  Note that this formalism  relies on the TF energy functional and  has little connection to the numerical studies of vortex arrays in uniform incompressible fluids~\cite{Campbell:1979}, where a vortex-free region occurs near the outer wall.}  In the particular case of the two-dimensional TF density profile with $n(\bm r) \propto 1-r^2/R_\perp^2$, this equation yields 
\begin{equation}
\label{eq:u(r)TF}
\bm u(\bm r) \approx \frac{l^2}{4R_\perp^2} \ln\left(\frac{l^2}{\xi^2}\right) \frac{\bm r}{1-r^2/R_\perp^2}.
\end{equation}
 Note that the deformation of the regular vortex lattice is purely radial, as might be expected from the circular symmetry, and it increases with increasing distance from the center.    Furthermore, the quantity $R_\perp^2/l^2$ is the number ${\cal N}_v$ of vortices in the rotating TF condensate [see Eq.~(\ref{eq:Nv})], so that the nonuniform distortion is of order ${\cal N}_v^{-1}$  (at most a few \%), even though the TF density decreases significantly near the edge of the condensate.

 \begin{figure}[ht] 
  \includegraphics[width=2.5in]{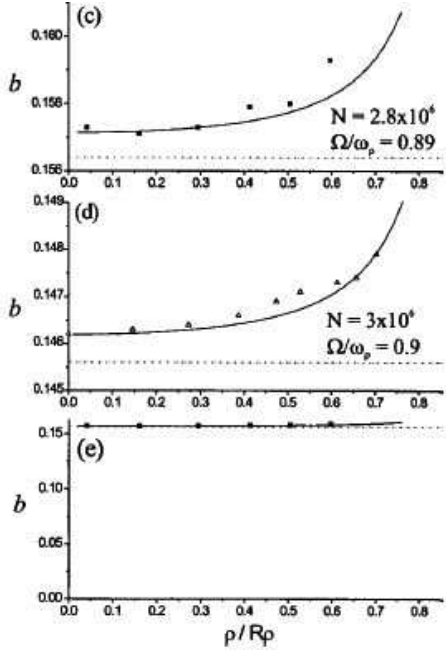}
 \caption{Lattice spacing as a function of the scaled radial position~$r/R_\perp$. The solid curve is the prediction (\ref{eq:nvrTF}) from \textcite{Sheehy:2004a,Sheehy:2004b}, with the rigid-body value as a dashed line.   Plots (c) and (d) show that the fractional amplitude decreases with increasing $\Omega$, and (e) restores the suppressed zero in (c), indicating that the overall effect is small but measurable.  The intervortex spacing increases by less than $2 \%$ over a region where the atomic density changes by $\sim 35\%$.  From~\textcite{Coddington:2004}.}
 \label{fig:distort}
 \end{figure}

 Finally, Eq.~(\ref{eq:nvr}) gives the corresponding  nonuniform axisymmetric vortex density
  \begin{equation}
\label{eq:nvru}
n_v(r) = \overline{n}_v +\frac{1}{8\pi}\ln\left(\frac{l^2}{\xi^2}\right) \nabla^2\ln n(\bm r),
\end{equation}
which becomes
\begin{equation}
\label{eq:nvrTF}
n_v(r)  \approx \overline{n}_v - \frac{1}{2\pi R_\perp^2} \ln\left(\frac{l^2}{\xi^2}\right)\frac{1}{(1-r^2/R_\perp^2)^2},
\end{equation}
for the two-dimensional TF condensate.  
The correction is again small, of order ${\cal N}_v^{-1}$.  \textcite{Coddington:2004} study this behavior in detail, as shown in Fig.~\ref{fig:distort}.  Comparison of images in Fig.~\ref{fig:distort}(c) and \ref{fig:distort}(d) show that the distortion decreases with increasing number of vortices (namely with increasing  $\Omega$), and Fig.~\ref{fig:distort}(e) restores  the suppressed zero in (c), indicating that the overall effect is indeed small.

 \subsubsection{Vortex core size for large rotation speeds}

The principal effect of rapid rotation is the altered aspect ratio seen in Fig.~\ref{fig:aspect}, but the rotation also affects the vortex core radius $\xi$.  In the TF limit of a large condensate, the basic definition in Eq.~(\ref{eq:xia}) shows that $\xi^2$ increases with increasing  $\Omega$ because the radial expansion  reduces the central  density $n(0)$ and thus the chemical potential $\mu = gn(0)$.  To quantify this behavior, it is helpful to consider the dimensionless ratio $\xi^2/l^2 = \hbar\Omega/(2\mu)$, which is roughly the fraction of each vortex cell that the core occupies.  A combination of Eqs.~(\ref{eq:Nad}), (\ref{eq:muTF}), (\ref{eq:xia}), and (\ref{eq:mu(o)}) yields the TF expression~\cite{Fetter:2001a} 
\begin{equation}
\label{eq:frac}
\frac{\xi^2}{l^2} = \frac{\hbar\Omega}{2\mu} = \frac{\bar{\Omega}}{(1 - \bar{\Omega}^2)^{2/5}}\left(\frac{\omega_\perp}{\omega_z}\frac{ d_\perp}{15 N a }\right)^{2/5},
\end{equation}
where $\bar{\Omega} = \Omega/\omega_\perp$ is a scaled dimensionless rotation speed.

Since $d_\perp/(Na)$ is small for a typical TF condensate, the ratio $\xi^2/l^2$ also remains small until $\bar{\Omega}$ approaches one, and it then grows rapidly because of the small denominator.  As  \textcite{Fischer:2003}  emphasize, this limit of rapid rotation requires a more careful treatment.  Specifically, they do not assume that $\xi^2$ is proportional to $ \mu^{-1}$.  Instead, they treat $\xi^2$ as a variational parameter and find that the ratio $\xi^2/l^2$ grows only until it reaches  $\approx \frac{1}{2}$, when the vortex core occupies a significant fraction of each unit cell in the vortex lattice.  Strictly, the TF limit no longer applies for this large $\bar{\Omega}\lesssim 1$ because the density variation now makes a significant contribution to the total kinetic energy.  \textcite{Baym:2004}  construct a more general variational theory that includes both the TF limit and the  lowest-Landau-level  regime described below in Sec.~V.

 \begin{figure}[ht] 
  \includegraphics[width=2.5in]{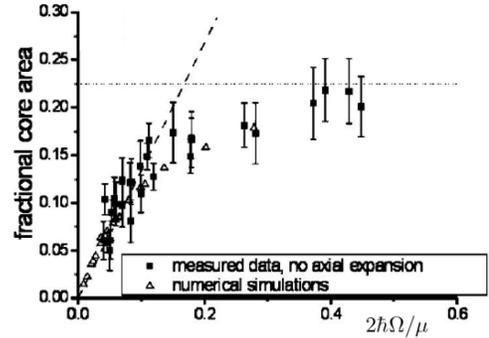}
 \caption{Fractional vortex core area as function of $2 \hbar\Omega/\mu$.  The dashed line is the TF theory and the dotted line is the lowest-Landau-level theory discussed below in Sec.~V.  From~\textcite{Coddington:2004}.}
 \label{fig:frac}
 \end{figure}

\textcite{Coddington:2004} study the size of the vortex core from   both experimental and numerical perspectives.  Because of  different fits to the vortex-core density profiles and different definitions of the fractional core area, their numerical values differ somewhat from those in Eq.~(\ref{eq:frac}) and in \textcite{Fischer:2003} and \textcite{Baym:2004}.  Figure \ref{fig:frac} shows typical experimental and numerical values of the fractional core area compared with the   ratio $ 2\hbar\Omega/\mu$.  For small  values of $ \Omega$, Eq.~(\ref{eq:frac}) indicates that  $\xi^2/l^2$ should vary linearly with $\Omega/\mu $ in the TF regime.  As $\Omega$ increases toward the critical value $\omega_\perp$,  however, the fractional core area  saturates at a constant value.  Detailed numerical studies~\cite{Cozzini:2006} of a rapidly rotating condensate yield a very similar picture of this behavior.
 
\subsubsection{Tkachenko oscillations of the vortex lattice}

In 1966, \textcite{Tkachenko:1966,Tkachenko:1966a} published two remarkable papers on the behavior of  arrays of straight vortices in an unbounded  rotating incompressible irrotational fluid (as a model for superfluid $^4$He).   The first paper shows that a triangular lattice has the lowest energy of all simple lattices (those with one vortex per unit cell).  The second paper studies small perturbations of a general  lattice, showing that the square lattice is unstable for waves in certain directions, but that the triangular lattice is stable for all normal modes.  For the latter structure, he also  determines the dispersion relation of the small-amplitude normal modes of a vortex lattice,  along with the corresponding eigenvectors for a given wave vector $\bm k$ lying in the $xy$  plane perpendicular to $\bm \Omega$.   The calculation is a {\it tour de force} of analytic function theory, involving several elliptic-type functions, especially the Weierstrass $\zeta$ function that  has a simple pole at each lattice site~\cite{Chandrasekharan:1985,Wolfram:2007}.  Ultimately, however,  the final result is simple.  At long wavelengths $kl\ll 1$, where $l = \sqrt{\hbar/(M\Omega)}$ is the vortex-cell radius, the wave is predominantly transverse with a linear dispersion relation
\begin{equation}
\label{eq:Tkachenko}
\omega(\bm k) \approx c_T k,
\end{equation}
where 
\begin{equation}
\label{eq:cT}
c_T =\frac{1}{2}\sqrt{\frac{\hbar \Omega}{M}} = \frac{1}{2}\, l\Omega
\end{equation}
is the speed of propagation.  This long-wavelength motion is effectively a transverse phonon in the vortex lattice.\footnote{Unlike Newtonian particles in a two-dimensional lattice, however, vortices obey first-order dynamical differential  equations.  Hence, there is only one normal mode for each wave vector $\bm k$.}

The significant  radial expansion of  rapidly rotating BECs means that the resulting vortices are essentially two-dimensional.   Thus   bending modes are irrelevant, but the nonzero compressibility  of these atomic gases  requires a significant modification to \textcite{Tkachenko:1966a}'s analysis.   \textcite{Sonin:1987} and \textcite{Baym:2003}  generalize Tkachenko's result in Eq.~(\ref{eq:Tkachenko}) to find the new long-wavelength expression 
\begin{equation}
\label{eq:Tka,comp}
\omega(k)^2 \approx c_T^2 k^2 \frac{s^2 k^2}{4\Omega^2 + s^2 k^2},
\end{equation}
where $s$ is the speed of sound.  This expressions assumes an infinite uniform system, but \textcite{Anglin:2002,Baym:2003,Baksmaty:2004,Sonin:2005} also include the nonuniformity of the trapped BEC.  If $sk \gg \Omega$ (the short-wavelength or incompressible limit), then this expression reduces to Tkachenko's $\omega(k) \approx c_T k$, but if $sk \ll \Omega$ (the long-wavelength or compressible limit), then the mode becomes soft with $\omega(k) \propto k^2$.  Similar softening of the collective-mode spectrum has important consequences for stability of the vortex lattice at large $\Omega$~\cite{Sinova:2002,Baym:2005}, as discussed in Sec.~VI.

\begin{figure}[ht] 
  \includegraphics[width=2.5in]{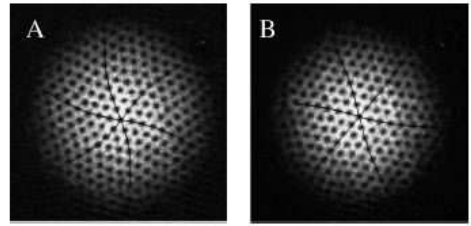}
 \caption{Lowest Tkachenko mode of the vortex lattice excited by atom removal taken (A)  500 ms after the end of the blasting pulse and (B) 1650 ms after the blasting pulse.  Lines are sine fits to the distortion of the vortex lattice.  From~\textcite{Coddington:2003}.}
 \label{fig:Tkachenko}
 \end{figure}

\textcite{Coddington:2003} observe these Tkachenko oscillations in considerable detail.  They start with an essentially uniform vortex lattice and then apply a weak perturbation (along with other approaches, they use a ``blasting pulse" that removes a small fraction of the atoms near the center of the trap).  At a scaled rotation frequency $\bar{\Omega} = 0.95$, they  wait a variable time and then turn off the trap, yielding an image of the resulting distorted  vortex lattice.  Figure \ref{fig:Tkachenko} shows the vortex lattice at one quarter and three quarters of the oscillation period, with the lines as a sine fit to the distortion of the vortex lattice.  In practice, the normal mode has the correct shape, but the measured frequency is significantly smaller than that predicted in Eq.~(\ref{eq:Tka,comp}).  \textcite{Anglin:2002a} suggest that this discrepancy arises from  reliance on a continuum theory that ignores the  increased vortex core radius for large $\Omega$.

\textcite{Tkachenko:1969} also evaluates the shear modulus of the triangular vortex lattice in an incompressible fluid, which is directly related to the long-wavelength oscillation spectrum.  Subsequently, \textcite{Cozzini:2006} use numerical methods to study the more general situation of a  vortex lattice in a rotating BEC, which  also includes both the incompressible (TF) limit and the compressible (LLL) limit \cite{Sonin:2005,Sinova:2002}.

\subsubsection{Rotating two-component BECs}

The  experimental creation of relatively large vortex arrays in single-components BECs in $^{23}$Na~\cite{Abo:2001} and $^{87}$Rb~\cite{Madison:2000a,Haljan:2001b,Hodby:2002} rapidly led to proposals for corresponding  two-component vortex arrays~\cite{Mueller:2002,Kasamatsu:2003}.  As discussed in  Sec.~III.D.1, the interspecies interaction parameter $g_{12}$ significantly affects the behavior, especially for trapped condensates (the intraspecies parameters $g_1$ and $g_2$ are assumed positive).   

\textcite{Mueller:2002} initiated the theory of these two-component condensates, assuming rapid rotation ($\Omega\lesssim\omega_\perp$).  This limit allows a considerable simplification (see Sec.~V) because the  condensate wave functions $\Psi_1$ and $\Psi_2$ can then be constructed from the lowest-Landau-level (LLL) single-particle states [the same as those used by \textcite{Butts:1999} in Sec.~III.C].  Apart from an overall Gaussian factor, the resulting trial solutions $\Psi_j$ are analytic functions of the complex variable $\zeta = x + i y$; they involve  quasiperiodic Jacobi theta functions that are closely related to other elliptic functions appearing in the treatment of  normal modes in a rotating vortex lattice~\cite{Tkachenko:1966,Tkachenko:1966a,Chandrasekharan:1985}, as mentioned in  Sec.~IV.B.4.  This analytical approach allows a relatively simple treatment of the change in the vortex lattice as the parameter $\alpha \equiv g_{12}/\sqrt{g_1g_2}$ varies from negative (when the interspecies interaction is attractive) to positive.  Specifically, for negative $\alpha$, the two components prefer to overlap and form a single triangular lattice.  As $\alpha$ becomes increasingly positive, however,  the overall repulsive interaction energy leads the two components  to separate spatially, with interlaced  triangular vortex arrays to minimize the overall density variation.  Eventually, the triangular lattices should distort, forming square or rectangular arrays when $\alpha \sim 1$.

Subsequently, \textcite{Kasamatsu:2003} study the behavior at lower angular velocity in the mean-field TF regime.  Their  extensive numerical analysis of the equilibrium vortex-lattice configuration as a function of the scaled dimensionless rotation rate $\bar \Omega= \Omega/\omega_\perp$ and the relative interaction strength $\alpha$ gives a qualitatively similar picture:  For any reasonable $\bar\Omega$,  the interpenetrating vortex arrays assume four-fold square symmetry as $\alpha$ approaches $1^-$ from below.  

\textcite{Schweikhard:2004} perform a series of experiments on these rotating two-component systems ($\bar\Omega \approx 0.75$).  They start with a single-component triangular vortex array in  the state $|1\rangle = |F = 1,m_F = -1\rangle$ (compare Sec.~III.D.1).   A short pulse with the two-photon  electromagnetic coupling fields transfers $\sim 80\%$ of the population to the state $|2\rangle = |F = 2,m_F = 1\rangle$. The initial triangular vortex lattice is now unstable, and after a turbulent stage ($\sim 2 $ s), it transforms to a square structure that persists for another $\sim 4$ s.  Ultimately, however, the state $|2\rangle$ decays through a spin-flip transition.  As a result, the remaining single-component condensate in state $|1\rangle$ again transforms back to a triangular array, similar to  the initial state.

\subsubsection{Rotating gas of paired fermions in tightly bound (BEC) limit}

One of the many remarkable new results from the study of ultracold gases is the superfluidity of bound pairs of fermionic atoms such as $^6$Li and $^{40}$K.  Although a  detailed discussion is inappropriate here, Secs.~I and VIII of \textcite{Bloch:2007}, Sec.~III of \textcite{Giorgini:2007},  and Sec.~8.4 of \textcite{Leggett:2006} provide excellent treatments of the behavior near a Feshbach resonance. Here, I  simply review the basic elements of scattering theory to explain how the $s$-wave scattering length can exhibit  a resonance associated with the appearance of a new bound state.

The simplest  description of a dilute atomic BEC assumes a weak repulsive effective interaction $g>0$ with a positive $s$-wave scattering length $a>0$ (this picture holds for the bosonic atoms $^{87}$Rb and $^{23}$Na).  Such a system is stable both in bulk and in a trap, with a variety of associated collective modes.  In contrast, the bosonic atom $^7$Li has an {\it attractive} effective interaction $g<0$ with a negative $a<0$.  In bulk, the negative interaction would yield an imaginary speed of sound, implying a long-wavelength instability.  In a trap, however, the  positive kinetic and potential energy can stabilize the condensate for sufficiently small $N$, but  the system will collapse~\cite{Ruprecht:1995,Bradley:1997} above a critical  $N_c$.  As an alternative picture of this stabilization, note that the quantized wavenumbers in a trap provide a lower bound $k_{\rm min}\sim \pi/R$, where $R$ depends on $N$. If the discrete analog of the Bogoliubov energy spectrum is real for all $k\ge k_{\rm min}$, then the trapped condensate is stable even for negative $a$.  

The  direct relation between the sign of $g$ and the sign of $a$ holds only for weak potentials that have no bound states.  The situation becomes more interesting when the potential can  support one (or more) $s$-wave bound states.  In this case, the $s$-wave scattering length exhibits divergent resonant behavior.
To be very specific, let us consider a spherical square-well potential $V(r) = V_0\,\theta(b-r)$ with height $V_0$ and range $b$, where $\theta$ denotes the unit positive step function. 
\begin{enumerate}
\item Bound states occur only for negative $V_0 = -|V_0|$.   A standard  analysis [see, for example, Sec.~33 of \textcite{Landau:1965} and pp.~164 and 349 of \textcite{Shankar:1994}] shows that there are no $s$-wave bound states if $|V_0|< \pi^2 \hbar^2/(8Mb^2)$;  as $|V_0|$ increases,  the $n$th $s$-wave bound state appears  when 
\begin{equation}
\label{eq:bound}
|V_0|= \frac{(2n-1)^2 \pi^2 \hbar^2}{8Mb^2}.
\end{equation}
\item Consider now the scattering of  a particle with positive energy $E=\hbar^2 k^2/(2M)$.  The  radial $l = 0$  wave function $\psi(r) =u(r)/r$ has the asymptotic  form $u(r) =  \sin(kr + \delta)$, where  $\delta(k)$ is  the $s$-wave phase shift.   A detailed analysis  for the same square-well potential $V(r)$ [see, for example Sec.~130 of \textcite{Landau:1965} and pp.~549-554 of \textcite{Shankar:1994}] yields the explicit expression 
\begin{equation}
\label{eq:delta}
\delta = -kb + \arctan\left(\frac{k}{k'}\tan k' b\right),
\end{equation}
where $k'^2 = 2M(E-V_0)/\hbar^2$ determines the wavenumber $k'$ inside the square well.\footnote{If $V_0>E$, then  $k'$ is written as $i\kappa$ and $\tan k'b /k'$ becomes $\tanh \kappa b/\kappa$.} It is evident physically that $\delta$ is negative for positive $V_0$ because the repulsive potential pushes out the wave function, whereas $\delta$ is positive for negative $V_0$ because the attractive potential pulls in the wave function.
This phase shift completely characterizes the $s$-wave scattering by this potential.

It is conventional and convenient to introduce the $s$-wave scattering length 
$a\equiv - \lim_{k\to 0} \delta(k)/k$; for the square-well potential, this result  gives 
\begin{equation}
\label{eq:a}
a = b - \frac{\tan k'b}{k'},
\end{equation}
where $k'^2 = -2M V_0/\hbar^2$ is now evaluated for $k = 0$ (and hence $E = 0$).\footnote{If $V_0$ is positive, then this relation is replaced by $\kappa^2 = 2MV_0/\hbar^2$; correspondingly,  Eq.~(\ref{eq:a}) becomes $a = b-\tanh \kappa b/\kappa$.}  If $V_0$ is positive and large, then $a \to b$, but the dependence can be very different in other situations.  For example, $a$ vanishes if $V_0 = 0$ (as expected).  More significant is the behavior for negative $V_0$, especially near the appearance of a bound state.  Equation (\ref{eq:bound}) shows that $k'b$ increases through $(n-\frac{1}{2})\pi$ when the $n$th  $s$-wave bound  state first forms.  Consequently, Eq.~(\ref{eq:a}) diverges to negative infinity  when the state is just unbound and then decreases from positive infinity when the trapped state is weakly bound.  Such resonant dependence  is a typical and generic feature of scattering by an attractive  potential that can support one or more bound states.  
\end{enumerate}

In the context of dilute ultracold gases, the attractive long-range van der Waals potential acts to bind the two fermionic atoms to form a molecule.  In general, there are several molecular states, each with  different magnetic moments.  Thus an applied magnetic field can shift a bound state in a closed channel across  the asymptotic continuum in an open channel, leading to a ``Feshbach resonance'' [see, for example, \textcite{Moerdijk:1995}, \textcite{Kohler:2006}, Sec.~2 of \textcite{Castin:2006}  and Secs.~I and VIII of \textcite{Bloch:2007} for more detailed discussions].  In essence, sweeping an external magnetic field through a Feshbach resonance transforms a tightly bound bosonic molecule of the two fermions into a weakly attracting fermionic pair.  The bosonic molecules can form a BEC that displays essentially all the properties of an atomic BEC, whereas the weakly attracting fermionic pairs are analogous to the Cooper pairs in the BCS theory of conventional low-temperature superconductors~\cite{deGennes:1966,Tinkham:1996}.  As a result, this transition through a Feshbach resonance is called the ``BEC-BCS crossover.''

\begin{figure}[ht] 
  \includegraphics[width=3.0in]{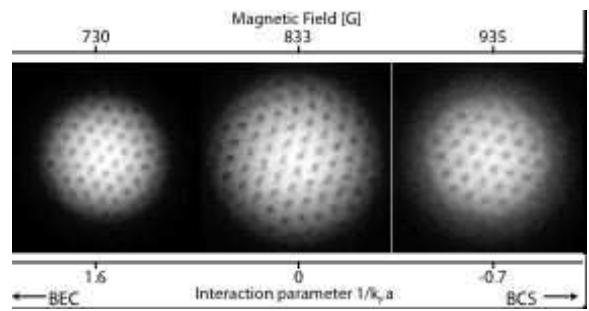}
 \caption{Vortex lattice in a rotating gas of $^6$Li fermions.  The left figure is on the BEC side, the central figure is on resonance, and the right figure is on the BCS side [note that the horizontal scale is $-(k_F a)^{-1}$, which is effectively $-(n^{1/3} a)^{-1}$ apart from a numerical factor].  From~\textcite{Zwierlein:2007}.}
 \label{fig:Zwierlein}
 \end{figure}

Particularly relevant in the present context is the experiment of \textcite{Zwierlein:2005} who  rotate a BEC of tightly bound bosonic molecules  with a deformed optical trap (the magnetically induced Feshbach resonance precludes the use of the usual magnetic trap).  They then sweep the magnetic field from the positive scattering length (the BEC side of  bound molecules) through the resonance to the negative scattering length (the BCS side of attracting fermionic pairs).  Figure \ref{fig:Zwierlein} shows the vortex lattice at three different values of the magnetic field.  Note that the scattering length $a$ diverges in the resonance region.   Consequently, it is conventional to use $(k_F a)^{-1}$ as the independent variable, which varies smoothly from positive in the BEC limit through zero near the crossover region to negative in the BCS limit. Here, $k_F$ is the Fermi wavenumber, but, apart from a numerical factor of order unity, it is the same as $n^{1/3}$, with $n$ the boson density.   The explicit persistence of quantized vortices through the BEC-BCS crossover  provides strong evidence for  superfluidity on the fermionic side, although the experiment did not reach  the weakly bound BCS limit [Sec.~VIII of \textcite{Giorgini:2007} gives a detailed discussion of  rotating ultracold Fermi gases]. 

\textcite{Sensarma:2006} use the  fermionic Bogoliubov-de Gennes equations [see, for example, Chap.~5 of \textcite{deGennes:1966} and  Sec.~10.1 of \textcite{Tinkham:1996}] to study the structure of a single quantized vortex in a uniform three-dimensional atomic Fermi  gas at zero temperature.  Here I consider only the  BEC side, since a full summary of   ultracold  Fermi gases is beyond the scope of this review. 

Let us focus on the vortex core radius (healing length) $\xi$ defined in Eq.~(\ref{eq:xia})
for a trapped condensate.  As noted in Sec.~IV.B.3, $\xi$ grows with increasing rotation speed $\Omega$ because of the reduced central density [and hence reduced chemical potential $\mu(\Omega)$], but Eq.~(\ref{eq:frac}) also exhibits the explicit dependence on the scattering length $a$.  Here, it is preferable to consider the slightly different dimensionless ratio 
\begin{equation}
\label{eq:BECxi}
\frac{\xi}{d_0} = \frac{1}{(1-\bar{\Omega}^2)^{1/5}}\left(\frac{d_0}{15 Na}\right)^{1/5},
\end{equation}
  which involves only the scaled rotation speed $\bar{\Omega} = \Omega/\omega_\perp$, the $s$-wave scattering length $a$, and the trap geometry through $d_0$ and the radial rap frequency $\omega_\perp$.  
  
   \textcite{Sensarma:2006} remark that the healing length in a uniform cold BEC Eq.~(\ref{eq:xi}) is proportional to  $(n a)^{-1/2}$ and decreases with increasing positive $a$ for fixed $n$.  Consequently, the vortex core size should shrink on approaching the resonance regime from the BEC side.  Note, however, that  inclusion of the trap  in Eq.~(\ref{eq:BECxi}) alters the fractional power   from $a^{-1/2}$ to $a^{-1/5}$.  Thus the dependence on $a$ may be difficult to measure in current experiments. It is also clear from Eq.~(\ref{eq:dilute}) that  the GP picture fails as $\xi$ approaches the interparticle spacing, which is expected to be the minimum value for $\xi$. On the BCS side, the qualitative relation $\xi \sim \hbar v_F/\Delta$ shows that the vortex core again grows on moving away from the resonance regime because the BCS energy gap $\Delta$ then becomes small~\cite{Sensarma:2006} [see also Sec.~8.4 of \textcite{Leggett:2006}].  
   
   On the BEC side, recall that  the condensate density $n(r)$ around a vortex decreases rapidly from its bulk value $n$ for $r\lesssim \xi$ inside the vortex core. Consequently, the circulating mass  current $j(r) = M n(r) v(r)  =  n(r) \hbar/r$ reaches a maximum ``critical'' value $j_c \sim n\hbar/\xi$ for  $r \approx \xi$ [compare the discussion below Eq.~(\ref {eq:kevortex})].  As $a$ increases on the BEC side toward the Feshbach resonance, the decreased $\xi$ implies that  this critical  mass current $j_c$ grows and saturates  when $\xi \sim n^{-1/3}$.  A more detailed analysis by \textcite{Sensarma:2006} shows that $j_c$ then again decreases on moving from the resonance regime  to the BCS side.  Thus both $\xi$ and $j_c$ display  non-monotonic dependence in traversing the BEC-BCS crossover region.
   
  \subsection{Effect of additional quartic confinement}
  
   The discussion of  rotating condensates in the mean-field Thomas-Fermi regime (Sec.~IV.A) makes it clear the a purely harmonic trap with radial frequency $\omega_\perp$ will cease to confine the system if the rotation speed $\Omega$ reaches or exceeds $\omega_\perp$. Indeed, the limit $\Omega\to \omega_\perp$  is singular, because the condensate radius and the total angular momentum both diverge.  In the  regime $\omega_\perp-\Omega \ll \omega_\perp$, the system is expected to undergo a  series of crossovers and complicated phase transitions (Secs.~V and VI) whose exact nature remains  under debate. 
   
    Thus it is interesting to consider the effect of adding an additional strongly confining  potential that rises more rapidly than $r^2$.  I here focus on the most common (quartic) model   $\propto r^4$~\cite{Fetter:2001a,Lundh:2002,Kasamatsu:2002,Kavoulakis:2003}.  Such an anharmonic trap will confine the condensate even for $\Omega>\omega_\perp$, thus allowing a more controlled study of possible new states (they are not generally the same as those  predicted in a harmonic trap in the limit $\Omega\to\omega_\perp$).  In terms of the usual dimensional quantities, the anharmonic radial trap potential has the form 
\begin{equation}
\label{eq:anharm}
V_{\rm tr}(r) = \frac{\hbar\omega_\perp}{2} \left(\frac{r^2}{d_\perp^2} + \lambda \frac{r^4}{d_\perp^4}\right),
\end{equation}
where $d_\perp= \sqrt{\hbar/(M\omega_\perp)}$ is the radial oscillator length and $\lambda$ is a dimensionless parameter that characterizes the admixture of quartic contribution.
    
    	For simplicity, I focus on the limit of rapid rotation  and
consider only a two-dimensional condensate that is uniform in the $z$ direction over a length $Z$~\cite{Kavoulakis:2003,Fetter:2005}, with $\Psi(\bm r,z) = \psi(\bm r)/\sqrt Z$.  It is also convenient to measure energies in units of $\hbar\omega_\perp$ and lengths in units of $d_\perp$, and to normalize the dimensionless condensate wave function $\psi$  with $\int d^2 r\, |\psi|^2 = N$.  In the mean-field TF regime, the dimensionless energy functional in the rotating frame [compare Eq.~(\ref{eq:E'TF})] becomes
\begin{eqnarray}
E'[\psi] = \int d^2 r\left[\textstyle{ \frac{1}{2}}\left(v^2 + r^2 +\lambda r^4\right) |\psi|^2 \right.\nonumber \\
\label{eq:E'2d}
\left. -\bm \Omega\cdot \bm r\bm\times \bm v\,|\psi|^2 + \textstyle{\frac{1}{2}}\bar g\,|\psi|^4\right],
\end{eqnarray}
where $\bar g = 4\pi  a/Z$ is a dimensionless two-dimensional interaction parameter.  An alternative model for the axial confinement uses a tight harmonic trap, in which case the interaction parameter is $\bar g = \sqrt{8\pi}a/d_z$, but the subsequent description is unaffected.

When the rotating condensate has many vortices, the total velocity $\bm v$ closely approximates the solid-body form $\bm \Omega\bm\times \bm r$, in which case Eq.~(\ref{eq:E'2d}) reduces to 
\begin{equation}
\label{eq:E'2da}
E'[\psi] = \int d^2 r\left\{\textstyle{ \frac{1}{2}}\left[\left(1-\Omega^2\right) r^2 +\lambda r^4\right] |\psi|^2
+ \textstyle{ \frac{1}{2}}\bar g\,|\psi|^4\right\}.
\end{equation}
As in Eq.~(\ref{eq:nonuniform}), variation with respect to $|\psi|^2$ at fixed normalization yields two-dimensional density
 \begin{equation}
 \label{eq:n2d}
\bar g\, |\psi(r)|^2 = \mu + \textstyle{\frac{1}{2}}\left[\left(\Omega^2-1\right) r^2 -\lambda r^4\right].
\end{equation}
 For $\Omega < 1$, the density has a local maximum at the center, but it changes to a local minimum for $\Omega > 1$.  In addition, the central density $|\psi(0)|^2$ is proportional to  the chemical potential $\mu$.  As in the previous Eq.~(\ref{eq:mu(o)}), $\mu$  depends  on $\Omega$ through the normalization and decreases continuously with increasing $\Omega$, reflecting the radial expansion of the condensate.

In contrast to the mean-field TF picture for a harmonic radial trap, however, the chemical potential here can vanish at a critical angular velocity $\Omega_h > 1$,  when the central density also vanishes.  This behavior indicates the formation of a central hole;  for $\Omega> \Omega_h$,  the chemical potential $\mu $ becomes negative and the condensate assumes an annular form [see \cite{Fetter:2005} for a detailed description of this unusual  structure].   

 As $\Omega$ continues to  increase, the system can, in principle, make a second transition to a different annular state with purely irrotational (vortex-free)  circulating flow, known as a ``giant vortex."  In this case, the circulating velocity $v(r)$ has the quantized form ${\cal N}_v/r$,  reflecting the presence of ${\cal N}_v$   phantom vortices in the hole
  (equivalently, the flow simply represents  ${\cal N}_v$-fold  quantized circulation in a multiply connected condensate).  \textcite{Kavoulakis:2003} suggest the possibility of such irrotational flow in a quadratic plus quartic potential [earlier, \textcite{Fischer:2003} propose a similar structure in a rotating rigid cylinder].  As noted at the start of Sec.~IV, solid-body rotation [the model used in obtaining Eq.~(\ref{eq:n2d})] gives the absolute minimum energy for a rotating system.  Consequently,  any irrotational giant vortex necessarily depends on the finite quantization of circulation that introduces  graininess in the superflow.  In practice, estimates of the critical angular velocity for formation of a giant vortex require considerable numerical work [see~\cite{Kavoulakis:2003,Fetter:2005,Kim:2005,Fu:2006} for more details].
 
In an elegant series of experiments, the ENS (Paris) group~\cite{Bretin:2004,Stock:2004}   study the effect of  a weak quartic confinement in addition to the usual quadratic trap potential [see also \cite{Stock:2005a} for a more general review of rotating BECs]. Specifically, they use a blue-detuned (repulsive) Gaussian laser beam propagating along the symmetry axis of their elongated condensate, with a resulting dipole potential 
\begin{equation}
\label{eq:Vd}
V_{\rm d}(r) = V_0 \,e^{-2r^2/w^2}\approx V_0 -\frac{2V_0}{w^2} r^2 + \frac{2V_0}{w^4}r^ 4 +\cdots.
\end{equation}
The positive constant term $V_0$ simply shifts to zero of energy,  the quadratic term reduces the original radial trap frequency,   and the quartic term  provides weak positive confinement.  The measured oscillation frequency of the center of mass of the condensate gives the effective radial trap frequency $\omega_{\rm eff} /2\pi = 64.8$ Hz~\cite{Stock:2004}, and the other trap parameters yield the quartic coupling constant $\lambda \approx 10^{-3}$, indicating that the added confinement in Eq.~(\ref{eq:anharm}) is indeed weak.

  \begin{figure}[ht] 
  \includegraphics[width=3.4in]{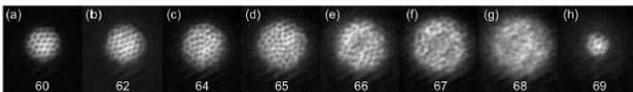}
 \caption{Pictures of the rotating gas taken along the rotation axis.  The number below each picture is the rotation frequency in Hz (note that $\omega_{\rm eff}/2\pi = 64.8$ Hz is the effective radial trap frequency).  Taken from~\cite{Bretin:2004}.}
 \label{fig:quartic}
 \end{figure}

For the dimensionless rotation speed $\Omega \approx 1$, the rotational deformation produces a nearly spherical condensate that remains stable up to $\Omega \lesssim 1.05$~\cite{Bretin:2004}.  Figure \ref{fig:quartic} shows  the expanded cloud for various rotation speeds both below and above the critical value  $\omega_{\rm eff}$.  Initially, a regular  vortex array appears, but it becomes  irregular in  Figs.~\ref{fig:quartic}(d)-(f).  In addition, for $\Omega> \omega_{\rm eff}$,  the condensate exhibits a clear local minimum in the density near the center, as expected from Eq.~(\ref{eq:n2d}) in the mean-field TF picture. Throughout the analysis, they fit the measured condensate shape to the TF prediction, which provides a direct determination of $\Omega$.  They also use surface-wave spectroscopy as discussed in Sec.~III.D.3~\cite{Zambelli:1998,Svidzinsky:1998,Cozzini:2003}   to obtain an independent measure of $\Omega$ (even though the number of visible vortices appears to be too small for $\Omega \ge \omega_{\rm eff}$).

The absence of visible vortices in Fig.~\ref{fig:quartic}(g) is puzzling.  One possible reason for the discrepancy is the three-dimensional character of the condensate, which allows the vortices to bend.  Also, the low temperature may make it difficult for the system to equilibrate.  Numerical studies~\cite{Aftalion:2004} suggest that much longer times are needed to attain a well-ordered vortex lattice in these regimes of large $\Omega$.  In any case, Fig.~\ref{fig:quartic}(h) implies that the condensate simply collapses, although  the reason remains unknown.

\section{Vortex Arrays in Mean-Field Lowest-Landau-Level (LLL) Regime}

In the mean-field Thomas-Fermi description of a rotating BEC, the basic approximation is the neglect of the ``kinetic energy'' associated with  the density variations.  Specifically, Eq.~(\ref{eq:E'TF}) omits the term $\hbar^2(\nabla|\Psi|)^2/(2M)$   whereas the usual kinetic energy  from the phase variation  $\frac{1}{2} M v^2|\Psi|^2$  remains.  As seen in Eq.~(\ref{eq:frac}), the vortex cores are  small for moderate values of the dimensionless rotation speed $\bar\Omega = \Omega/\omega_\perp$,  so that this approximation is valid.  As $\bar\Omega$ increases toward $1$, however, it fails when $1-\bar\Omega\sim d_\perp/(15 Na)\ll 1$ because the vortex cores then become comparable with the intervortex separation $\sim 2l$.   In this case, it becomes essential to return to the full GP energy functional $E'[\Psi]$  in Eq.~(\ref{eq:E'GP}).

In thinking about this new regime, \textcite{Ho:2001}  observes  that  the interaction energy per particle [of order $\mu(\Omega) = gn(0)$] is small  because the condensate expands radially [see Eq.~(\ref{eq:mu(o)})].  In essence, the condensate becomes  two-dimensional, and it is convenient to study a two-dimensional condensate that is uniform in the $z$ direction over a length $Z$.  Thus the original condensate wave function $\Psi(\bm r,z)$ can be written as $\psi(\bm r)/\sqrt{Z}$, where $\psi(\bm r)$ is a two-dimensional wave function with normalization $\int d^2 r\,|\psi|^2 = N$.  The GP energy functional then becomes 
\begin{eqnarray}
E'[\psi] = \int d^2 r\,\psi^*\left(\frac{p^2}{2M} + \frac{1}{2}M\omega_\perp^2 r^2 -\Omega  L_z \right.  \nonumber \\ 
\label{eq:E'2dosc}
  \left. +\frac{1}{2}g_{2D}|\psi|^2\right)\psi,
\end{eqnarray}
where $g_{2D} = g/Z$.  As in Sec.~IV.C, one can instead assume a tight Gaussian axial confinement, in which case $g_{2D} = g/( \sqrt{2\pi}\,d_z)$, with $d_z = \sqrt{\hbar/(M\omega_z)}$.

\subsection{Physics of lowest-Landau-level (LLL) one-body states}

The one-body Hamiltonian in Eq.~(\ref{eq:E'2dosc}) has the crucial advantage that it is exactly soluble with  very simple energy eigenvalues [see, for example, Sec.~12.14 of \textcite{Messiah:1961} and pp.~727-741 of \textcite{Cohen:1977}].  Consider the (nonrotating) two-dimensional isotropic harmonic oscillator with Hamiltonian 
\begin{equation}
\label{eq:H2d}
H_0 = \frac{p^2}{2M} + \frac{1}{2}M\omega_\perp^2 r^2.
\end{equation}
Classically, the motion is that of two independent oscillators (along $x$ and $y$) that can combine to give, for example, circular motion with frequency $\omega_\perp$ in the positive (counterclockwise) or negative (clockwise) sense, depending on the relative phases.  When viewed from a frame that rotates with angular velocity $\Omega$ in the positive sense,  the two circular  orbits with originally  degenerate frequencies  now have different  frequencies 
\begin{equation}
\label{eq:opm}
\omega_\pm = \omega_\perp\mp \Omega,
\end{equation}
 since the positive (negative) mode has a reduced (increased) frequency in the rotating frame.

The quantum-mechanical description relies on the familiar creation and annihilation operators 
\begin{equation} 
\label{eq:ax}
a_x = \frac{1}{\sqrt 2}\left(\frac{x}{d_\perp} + i\, \frac{p_x d_\perp}{\hbar}\right), \quad a_x^\dagger = \frac{1}{\sqrt 2}\left(\frac{x}{d_\perp} - i\, \frac{p_x d_\perp}{\hbar}\right), 
\end{equation}
along with similar operators $a_y$ and $a_y^\dagger$.  In the  present case, it is more appropriate to use circularly polarized states with  operators 
\begin{equation}
\label{apm}
a_\pm = \frac{a_x \mp i\,a_y}{\sqrt 2}, \qquad a_\pm^\dagger = \frac{a_x^\dagger \pm i\,a_y^\dagger}{\sqrt 2}
\end{equation} 
that obey the usual commutation relations  $[a_+,a_+^\dagger ] = [a_-,a_-^\dagger ] = 1$, with all other commutators vanishing.  A straightforward analysis [Sec.~12.14 of \textcite{Messiah:1961} and pp.~727-741 of \textcite{Cohen:1977}]  expresses $H_0$ in terms of these new operators
\begin{equation}
\label{eq:H0}
H_0 = \hbar\omega_\perp \left( a_+^\dagger a_+ + a_-^\dagger a_- + 1\right).
\end{equation}
Similarly the angular momentum $L_z = xp_y - y p_x$ becomes 
\begin{equation}
\label{eq:Lz}
L_z = \hbar\left( a_+^\dagger a_+ - a_-^\dagger a_-\right).
\end{equation}
The number operators $a_\pm^\dagger a_\pm$ have non-negative integer eigenvalues $n_\pm$; the corresponding operators $a_+^\dagger$ and $a_+ $ ($a_-^\dagger$ and $a_- $)  create and destroy one quantum with positive (negative) circular polarization and one unit of positive (negative) angular momentum.  For the nonrotating system with Hamiltonian $H_0$,  each quantum  simply has an energy $\hbar\omega_\perp$.

The principal advantage of this particular circularly polarized basis appears for the rotating system, with Hamiltonian $H_0' = H_0 - \Omega L_z$, because Eqs.~(\ref{eq:H0}) and (\ref{eq:Lz}) together imply 
 \begin{equation}
\label{eq:H0'}
H_0' = \hbar\omega_\perp + \hbar\omega_+  a_+^\dagger a_+ +\hbar\omega_- a_-^\dagger a_-,
\end{equation}
where $\omega_\pm = \omega_\perp\mp\Omega$ 
 are the frequencies in Eq.~(\ref{eq:opm})  for the positive and negative classical orbits as seen in the rotating frame.  Correspondingly, the energy eigenvalues are labeled with two non-negative integers $n_\pm$ 
\begin{equation}
\label{eq:enpm}
\epsilon(n_+,n_-) =   n_+\hbar(\omega_\perp -\Omega) + n_-\hbar (\omega_\perp+\Omega),
\end{equation}
where I omit the zero-point energy $\hbar\omega_\perp$ for simplicity.
In the limit of rapid rotations ($\Omega\to \omega_\perp$), these eigenvalues are essentially independent of $n_+$, which implies a large degeneracy.   The other integer $n_-$ then becomes the Landau-level index, with different Landau levels separated by an energy gap of $\sim 2\hbar \omega_\perp$.  

\begin{figure}[ht] 
    \includegraphics[width=3.5in]{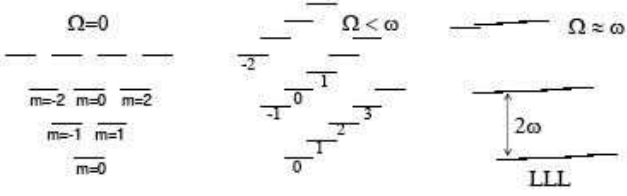}
 \caption{Excitation energy levels $\epsilon_{nm} = n \hbar\omega_\perp -m\Omega$  from Eq.~(\ref{eq:enm}) for various angular velocity $\Omega$ expressed in terms of quantum numbers   $n = n_++n_-$ for energy and  $m = n_+-n_-$ for angular momentum. Left side is for  $\Omega=0$, showing an inverted pyramid of states: for each non-negative integer $n$, there are $n+1$ degenerate angular-momentum  states ranging from $-n$ to $n$ in steps of $2$.  Center shows the situation for moderate $\Omega$.  States with $m = 0$ are unshifted, states with negative $m$  shift up whereas those with positive $m$ shift down.  Right side shows the extreme situation for $\Omega\to \omega_\perp$, when the states form essentially horizontal rows (the Landau levels).  The lowest Landau level has $n = m$, which means $n_-=0$, the first excited Landau level has $n - m = 1$ (namely $n_- = 1$),  {\it etc.} From~\textcite{Dalibard:2007}.}
 \label{fig:Dalibard}
 \end{figure}

For small angular velocity, it is often convenient to introduce a different set of quantum numbers  [pp.~727-741 of ~\textcite{Cohen:1977}]  $n = n_++n_-$ for the energy and $m = n_+-n_-$ for the angular momentum, so that the energy eigenvalues have the equivalent form 
\begin{equation}
\label{eq:enm}
\epsilon_{nm} \equiv \epsilon[\textstyle{\frac{1}{2}}(n+m),\textstyle{\frac{1}{2}}(n-m)]  = n \hbar\omega_\perp - m\hbar\Omega.
\end{equation}
Figure \ref{fig:Dalibard}  uses this new basis to  illustrate the energy spectrum   $\epsilon_{nm}$  for different values of $\Omega$.  
\begin{enumerate}
\item On the left side of the figure ($\Omega = 0$), the excitation energy is simply $n\hbar\omega_\perp$, independent of $m$, forming an inverted pyramid of states: for each non-negative integer $n$, there are $n+1$ degenerate angular-momentum  states ranging from $-n$ to $n$ in steps of $2$.  For small $\bar\Omega\ll 1$, the energy  quantum number $n$ determines the large splitting and the angular-momentum quantum number $m$ lifts the remaining degeneracy (physically, this Coriolis-induced splitting  is like the weak-field  Zeeman splitting of magnetic sublevels).  
\item The central figure  shows the situation for moderate $\Omega$, where the (now signficant) Coriolis effect eliminates the degeneracy of the nonrotating spectrum.  States with $m = 0$ are unshifted, states with negative $m$  shift up by $|m|\hbar\Omega$, whereas those with positive $m$ shift down by the same amount.  
\item The right side shows the extreme situation for $\Omega\to \omega_\perp$, when the states again become nearly degenerate,  forming essentially horizontal rows (the Landau levels).  The lowest Landau level has $n = m$, which means $n_-=0$ and $n_+ = m$; the first excited Landau level has $n - m = 1$ (namely $n_- = 1$ and $n_+ = m+1$),  {\it etc.}  In this limit ($\bar\Omega\lesssim 1$), the quantum numbers $n_+$ and $n_-$ of the  rotating basis are most convenient.
\end{enumerate}

For rapid rotation ($\bar\Omega \lesssim 1$), it is appropriate to focus on states in the lowest Landau level (those with $n_- = 0$, denoted LLL), when the single  quantum number $n_+ = m$ suffices to specify the one-body states.  The relevant energies are the gap $2\hbar \omega_\perp$ between adjacent  Landau levels and the mean interaction energy $gn(0) = \mu$, which is assumed small in the present limit.  Thus the basic expansion parameter of the theory is $\mu/(2\hbar\omega_\perp)$, although current experiments have only achieved values as small as 0.6~\cite{Schweikhard:2004}.  The corresponding two-dimensional eigenfunctions  $\psi_m(\bm r)$ of both $H_0'$ and $L_z$ have a very simple form in plane-polar coordinates [note that these are precisely the trial states  from Eq.~(\ref{eq:psim}) used   to study vortex arrays in a weakly interacting condensate~\cite{Butts:1999}]
\begin{equation}
\label{eq:psima}
\psi_m(\bm r) \propto r^m e^{im\phi} e^{-r^2/(2d_\perp^2)},
\end{equation}
where the normalization is unimportant for the present  purpose.  

In particular, the ground state $\psi_0$ is just a two-dimensional  isotropic Gaussian.  It represents the vacuum for both  circularly polarized modes 
\begin{equation}
\label{eq:vac}
a_\pm\,\psi_0 = 0. 
\end{equation}
 The higher states $\psi_m$ for positive $m$ are proportional  to $(a_+^\dagger)^{m}\psi_0$; they can be written in terms of  $\zeta = (x+ i y)/d_\perp = (r/d_\perp) e^{i\phi}$ as $\psi_m\propto \zeta^m \,e^{-r^2/(2d_\perp^2)}$, where the first factor  is just  a non-negative integral power of the dimensionless complex variable $\zeta$.
  For a given $m \gg 1$, the density $|\psi_m(\bm r)|^2$ is large only in a circular strip of radius $\sim \sqrt m\,d_\perp$ and width  $\sim d_\perp/\sqrt m$.  In addition, it is easy to see that the mean-square radius for any non-negative $m$ is given by 
 \begin{equation}
\label{eq:rsq}
\langle r^2\rangle_m = (m + 1)\,d_\perp^2.
\end{equation}

Since the same LLL  wave functions appear  in the fractional quantum Hall effect for  two-dimensional electrons in a strong magnetic field, this mean-field LLL regime is sometimes called the ``mean-field quantum Hall'' regime~\cite{Watanabe:2004}.  In fact,  related many-body states  from the fractional quantum Hall effect play an important role for a rotating dilute Bose gas at still higher angular velocity, where the physics will turn out to be  quite different (Sec.~VI).  Thus  it seems preferable  to reserve the name ``quantum Hall regime'' for this more unusual (nonsuperfluid) limit.

\subsection{Direct treatment of  LLL states}

Before using the LLL wave function to construct a GP condensate wave function, I briefly discuss an alternative approach to the LLL states.  It is natural to rewrite the single-particle Hamiltonian in the rotating frame $H_0' = p^2/(2M) + \frac{1}{2} M\omega_\perp^2 r^2 - \bm\Omega\cdot \bm r\bm\times \bm p$ by completing the square to obtain two equivalent forms 
\begin{eqnarray}
\label{eq:squarea}
H_0' &=& \frac{\left( \bm p - M\bm \Omega\bm\times \bm r\right)^2}{2M} + \frac{1}{2}M \left(\omega_\perp^2 - \Omega^2\right) r^2
\\[0.2cm]
\label{eq:squareb}
H_0' & = &  \frac{\left( \bm p - M\bm \omega_\perp\bm\times \bm r\right)^2}{2M} + \left(\bm \omega_\perp - \bm \Omega\right)\cdot \bm L,
\end{eqnarray}
where $\bm \omega_\perp = \omega_\perp\hat{\bm z}$.  Either form is reminiscent of the nonrelativistic Hamiltonian of a  particle with charge $q$ in a magnetic field  $\bm B$, where the momentum $\bm p$ and vector potential $\bm A$ appear in the gauge-invariant  combination  $\bm p - q\bm A$, with $\bm B = \bm \nabla\bm\times \bm A$. 

Landau solved the quantum problem of an electron in a uniform 
 magnetic field $\bm B$ [see, for example, Chap.~XV  of \textcite{Landau:1965}, but the vector potential has a different (asymmetric) form $\bm A = - B\,y\,\hat{\bm x}$].  In contrast,  pp.~742-764 of \textcite{Cohen:1977}  solve the same problem in the symmetric gauge $\bm A = \frac{1}{2} \bm B \bm\times \bm r$.  It is of course possible simply to quote those results to obtain the LLL  wave functions $\psi_m$, but  such an approach requires a  charge $q$, whose sign (positive or negative) determines some crucial conventions in the choice of quantum-mechanical states.  As written, Eq.~(\ref{eq:squareb}) suggests a unit positive charge with fictitious  vector potential $\bm A =  M\bm \omega_\perp\bm\times \bm r$; the resulting  uniform magnetic field is $\bm B = 2M\bm\omega_\perp$.  Unfortunately, the usual LLL picture  treats electrons with negative charge, and the corresponding one-body states involve reinterpretations of some of the quantum numbers, as discussed in  pp.~750-757 of \textcite{Cohen:1977}. For this reason, I choose to work directly from the original Hamiltonian $H_0'=H_0-\bm \Omega\cdot \bm L$, because the role of $n_+$ and $n_-$ is particularly transparent.
 
 Another interesting feature of the transformation to a quadratic form in $\bm p - q\bm A$ is the possibility of  topological gauge potentials.  In the context of   optical lattices, \textcite{Jaksch:2003} propose inducing various terms in the single-particle Hamiltonian that mimic the effect of an external magnetic field.  Specifically, when  a two-level atom  hops around a closed path  in the lattice, external lasers can generate a net phase change  that is essentially equivalent to the magnetic flux of an applied field [see,  \cite{Mueller:2004,Sorensen:2005,Satija:2006} for alternative approaches, including the possibility of ``non-Abelian" gauge potentials].  
 
 Similar  ideas  also apply to  a simple trapped condensate.  \textcite{Juzeliunas:2005} propose using a three-level system in the ``$\Lambda$'' configuration [two nearly degenerate lower levels, each coupled to a higher third state by external laser beams~\cite{Arimondo:1996};  for a ``dressed-atom approach'' to these three-level systems, see, for example,  pp.~451-454 of~\textcite{Cohen:1998}]. Such an interacting  three-level system has a ``dark'' state that effectively uncouples  from the remaining states.   If one of the external lasers has orbital angular momentum [see, for example, Sec.~52 of  \textcite{Gottfried:1966} and~\textcite{Allen:1999,Allen:2003}], then  the dark state experiences a nontrivial topological gauge potential that in principle can act like an essentially arbitrary magnetic field (and hence an  arbitrary applied vorticity for the trapped condensate).   Whether such methods can indeed generate angular momentum in the condensate  and the associated vortices remains uncertain.  More work (both theory and experiment) is clearly desirable.
 
 \subsection{LLL condensate wave functions}
 
 \textcite{Ho:2001} proposes a linear combination of the LLL single-particle states in Eq.~(\ref{eq:psima}) as the GP condensate wave function for the rapidly rotating two-dimensional BEC 
 \begin{equation}
\label{eq:LLL}
\psi_{LLL}(\bm r) = \sum_{m \ge 0} c_m\psi_m(\bm r) = f(\zeta) e^{-r^2/(2d_\perp^2)},
\end{equation}
where $f(\zeta) = \sum_{m \ge 0} c_m \zeta^m$ is an analytic  function of the complex variable $\zeta$ [compare Sec.~III.C; see also pp.~587-592 of \textcite{Shankar:1994}].  Specifically, for a truncated basis set, the analytic function $f(\zeta)$ is a complex polynomial and thus has a factorized form 
\begin{equation}
\label{eq:factor}
f(\zeta) \propto  \prod_j \left(\zeta - \zeta_j\right),
\end{equation}
apart from an overall constant factor.  Evidently, $f(\zeta)$ vanishes at each of the points $\zeta_j$, which are the positions of the nodes of the condensate wave function $\psi_{LLL}(\bm r)$. In addition, the phase of this wave function increases by $2\pi$ whenever $\zeta$ moves in the positive sense around any of these zeros.  Thus the  points $\zeta_j$ are precisely the positions of the vortices in the trial state $\psi_{LLL}$, and minimization with respect to the constants $c_m$ is effectively the same as minimization with respect to the position of the vortices~\cite{Butts:1999}.

\subsubsection{Unrestricted minimization}

This trial state has some very interesting and unusual properties.  Apart from the overall Gaussian envelope, the spacing of the vortices completely  determines  the spatial variation of the number density $n_{LLL}(\bm r) = |\psi_{LLL}(\bm r)|^2=|f(\zeta)|^2 e^{-r^2/d_\perp^2}$.  The  number density varies smoothly between adjacent vortices because $|f(\zeta)|^2$ consists of harmonic functions.  Hence the vortex  core size is comparable with  $l = \sqrt{\hbar/(M\Omega)}$, which is essentially $d_\perp$ in the rapidly rotating limit.   In contrast to the Thomas-Fermi limit at lower angular velocity, the present trial function $\psi_{LLL}$ automatically incorporates all the kinetic energy.

It is important to emphasize that the one-body state $\psi_{LLL}$ arises from the GP energy functional and that the corresponding many-body state 
\begin{equation}
\label{eq:PsiGP}
\Psi_{GP}(\bm r_1, \bm r_2, \cdots,\bm r_N) \propto \prod_{n = 1}^N \psi_{LLL}(\bm r_n)
\end{equation}
is just a Hartree product of these LLL states.  For any such ground state, the system exhibits  ``off-diagonal''  long-range order~\cite{Penrose:1956,Yang:1962}, and $\psi_{LLL}$ is the corresponding macroscopically occupied condensate wave function. Thus the mean-field LLL regime still has a BEC and, in general,  exhibits superfluidity.

 In the limit of rapid rotation, this LLL wave function can serve as a possible variational trial solution for the GP energy functional.    Let $\langle \cdots\rangle_{LLL}$ denote an expectation value evaluated with $\psi_{LLL}$.  Equations (\ref{eq:Lz}) and (\ref{eq:vac}) show that $\langle L_z\rangle_{LLL} / \hbar =\langle a_+^\dagger a_+\rangle_{LLL}$ because $\langle a_-^\dagger a_-\rangle_{LLL}$ vanishes for any LLL state.  Furthermore, the quantity $r^2 = x^2+y^2$ is easily expressed as $(x+iy)(x-iy) =d_\perp^2 (a_-+a_+^\dagger)(a_++a_-^\dagger)$, whose  expectation value gives $\langle r^2\rangle_{LLL} / d_\perp^2 =\langle a_+^\dagger a_+\rangle_{LLL} + 1$.  Comparison yields the remarkably simple relation between the angular momentum and the mean-square radius~\cite{Ho:2001,Watanabe:2004,Aftalion:2005}
 \begin{equation}
\label{eq:LzLLL}
\langle L_z \rangle_{LLL} = M\omega_\perp \langle r^2 \rangle_{LLL} - \hbar
\end{equation}
for any linear combination of single-particle states in the lowest Landau level [as discussed in Sec.~III.C, a slightly different form of this relation is important in the analysis of \textcite{Butts:1999}].  Note that this relation also follows directly from Eq.~(\ref{eq:rsq}).  
 
 A combination of Eqs.~(\ref{eq:H2d}) and (\ref{eq:LzLLL}) expresses the full GP energy functional  (\ref{eq:E'2dosc})  in the very intuitive and familiar  form
 \begin{eqnarray}
\label{eq:GP2d}
E'[\psi_{LLL}] =& \hbar\Omega + \int d^2 r \left[M\omega_\perp^2 \left(1-\bar \Omega\right)\right. r^2 |\psi_{LLL}|^2\nonumber \\[0.2cm]
&+\left. \textstyle{\frac{1}{2}} g_{2D}|\psi_{LLL}|^4\right],
\end{eqnarray}
where the integrand again involves only  $r^2|\psi_{LLL}|^2$ and $|\psi_{LLL}|^4$ [compare Eq.~(\ref{eq:ETF})].  Unrestricted minimization with respect to $|\psi_{LLL}|^2$ at  fixed normalization readily yields the number density 
\begin{equation}
\label{eq:nLLL}
|\psi_{\rm min}(r)|^2 = n_{\rm min}(r) = n_{\rm min}(0) \left(1-\frac{r^2}{R_{0}^2}\right),
\end{equation}
where the (two-dimensional) central density is $n_{\rm min}(0) = \mu_{\rm min}/g_{2D}$, and the condensate radius is given by 
\begin{equation}
\label{eq:RLLL}
R_{0}^2 = \frac{\mu_{\rm min}}{M\omega_\perp^2(1-\bar\Omega)}
\end{equation}
Remarkably, this expression has the same parabolic form as the usual TF result in Eq.~(\ref{eq:parabolic}), 
  even though the present functional explicitly includes all the kinetic energy.  It does, however, rely on  the restriction to the lowest Landau level. 

The normalization condition $\int d^2 r \,|\psi_{\rm min}|^2 = N $   yields $N = \frac{1}{2} \pi R_{0}^2 n_{\rm min}(0)$, which is equivalent to 
\begin{equation}
\label{eq:RLLL1}
R_{0}^2 =\frac{2 N g_{2D}}{\pi \mu_{\rm min}}.
\end{equation}
The product of Eqs.~(\ref{eq:RLLL}) and (\ref{eq:RLLL1}) gives an explicit expression for $R_{0}^4$
\begin{equation}
\label{eq:RL}
R_{0}^4 = \frac{2Ng_{2D}}{\pi M\omega_\perp^2(1-\bar\Omega)}= \frac{8Nad_\perp^4}{Z(1-\bar\Omega)}
\end{equation}
that explicitly shows the radial expansion for $\bar\Omega\to 1$.  Similarly, the ratio of the same two equations
yields the chemical potential 
\begin{equation}
\label{eq:muLLL}
\mu_{\rm min} = \sqrt{\frac{2Ng_{2D} M\omega_\perp^2 (1-\bar\Omega)} {\pi}}=\sqrt{\frac{8Na(1-\bar\Omega)}{Z}} \,\hbar\omega_\perp.
\end{equation}
The first form  is similar to Eq.~(\ref{eq:mu(o)}) in the mean-field TF regime (the different fractional powers arise from the different dimensions for the normalization integrals), whereas the second form gives  the condition $\mu_{\rm min} \lesssim 2\hbar\omega_\perp$ for the validity of the  mean-field LLL regime 
\begin{equation}
\label{eq:LLLOK}
1-\bar\Omega \lesssim \frac{Z}{2Na}.
\end{equation}
If $Z/2a\sim 100$ and $N\sim 10^4$, then a dimensionless rotation speed $\bar\Omega \sim 0.99$  just barely approaches the mean-field LLL regime. 

The energy of this approximate minimizing mean-field LLL density $|\psi_{\rm min}|^2$ follows immediately by evaluating Eq.~(\ref{eq:GP2d})
\begin{equation}
\label{eq:E'LLL}
\frac{E_{\rm min}' - \hbar\Omega}{N\hbar\omega_\perp} = \frac{2\mu_{\rm min}}{3\hbar\omega_\perp}= \frac{4}{3}\sqrt{\frac{2Na(1-\bar\Omega)}{Z}}. 
\end{equation}
As emphasized by \textcite{Aftalion:2005}, the approximate density in Eq.~(\ref{eq:nLLL}) arises from an unrestricted variation and is not within the LLL. Thus the actual energy will be higher, because  the  minimization then includes the  vortices through the zeros of the analytic function in Eq.~(\ref{eq:factor}).  As seen below,  the main effect of the vortices is to increase the numerical coefficient in Eq.~(\ref{eq:E'LLL}), keeping  the same   dependence on the various  parameters. 
 
The derivative of Eq.~(\ref{eq:E'LLL}) with respect to $\Omega$ yields the angular momentum of the minimizing mean-field LLL state
\begin{equation}
(L_z)_{\rm min} = \textstyle{\frac{1}{3}}MN R_0^2\,\omega_\perp - \hbar,
\end{equation}
which also follows from Eq.~(\ref{eq:LzLLL}).  Here, the first term is just the solid-body value (since direct integration with the parabolic $|\psi_{\rm min}|^2$ yields $\langle r^2\rangle_{\rm min}= \frac{1}{3}N R_0^2$).  Unlike the mean-field TF regime discussed in Sec.~IV.A,  the LLL reduction from the classical value is only  of order $N^{-1}$ and thus negligible in the thermodynamic limit.

\subsubsection{Inclusion of the vortices}

The minimizing parabolic density profile in Eq.~(\ref{eq:nLLL}) cannot provide a full description of the rotating LLL condensate, for it omits the fine-grained structure associated with the vortices.

\paragraph{renormalization of $g_{2D}$}
In the present mean-field LLL regime, the intervortex separation $\sim d_\perp$ is small compared to the size of the condensate.  Thus it is possible to write the number density obtained  from Eq.~(\ref{eq:LLL}) in the form $|\psi_{LLL}(\bm r)|^2 \approx |\bar\psi_{LLL}(\bm r)|^2\,|m(\bm r)|^2$ as the product of a slowly varying envelope function $ |\bar\psi_{LLL}(\bm r)|^2$ and a rapidly varying modulation function $|m(\bm r)|^2$ that vanishes at the center of each vortex~\cite{Baym:2004,Aftalion:2005,Watanabe:2004,Fetter:2001a,Fischer:2003}.  For a large vortex lattice, this modulating function is essentially periodic, and it is convenient to normalize it so that $\int_{\rm cell} d^2 r \,|m(\bm r)|^2 = 1$ over each unit cell.  Substitution of this approximate density into the LLL energy functional (\ref{eq:GP2d}) yields products of slowly varying functions involving powers of $|\bar\psi_{LLL}(\bm r)|^2$ and rapidly varying periodic modulating functions involving powers of $|m(\bm r)|^2$.  

The integral over the condensate separates into a sum of integrals over each unit cell, with the slowly varying quantities acting as locally constant factors. Thus the only effect of the vortex lattice is that the interaction term  acquires a numerical factor $\beta = \int_{\rm cell} d^2 r\,|m(\bm r)|^4$, which effectively renormalizes the interaction constant $g_{2D}\to \beta g_{2D}$ [\textcite{Aftalion:2005}  discuss these steps carefully]. For an unbounded triangular vortex lattice, the resulting $\beta \approx 1.1596$ is the numerical value for the corresponding Abrikosov vortex lattice~\cite{Abrikosov:1957,Kleiner:1964}. Consequently,  the radius $R_0$ in Eq.~(\ref{eq:RL}) expands by the factor $\beta^{1/4} \approx 1.0377$.

\paragraph{nonuniformity of the vortex array} To study the effect of  the vortices in  more detail, \textcite{Ho:2001} considers the logarithm of the particle density for a LLL condensate wave function of the form in Eq.~(\ref{eq:LLL}), obtaining 
\begin{equation}
\label{eq:nL}
\ln |\psi_{LLL}(\bm r)|^2 = \ln n_{LLL}(\bm r)  = -\frac{r^2}{d_\perp^2}+ 2\sum_j\ln|\bm r-\bm r_j|,
\end{equation}
where $\bm r_j$ is the position of the $j$th vortex.  The two-dimensional Laplacian of this equation yields
\begin{equation}
\nabla^2\ln n_{LLL}(\bm r) = -\frac{4}{d_\perp^2} + 4\pi \sum_j\delta^{(2)}\left(\bm r-\bm r_j\right)
\end{equation}
because $\nabla^2\ln |\bm r-\bm r_j|= 2\pi \delta^{(2)}\left(\bm r - \bm r_j\right)$.  The sum over the delta functions is precisely the two-dimensional vortex density $n_v(\bm r)$, which gives a striking relation between the {\it particle} density $n_{LLL}(\bm r)$ in the mean-field LLL regime and the {\it vortex} density~\cite{Ho:2001,Watanabe:2004,Aftalion:2005}
\begin{equation}
\label{eq:nvLLL}
n_v(\bm r) = \frac{1}{\pi d_\perp^2} + \frac{1}{4\pi} \nabla^2 \ln n_{LLL}(\bm r).
\end{equation}
The first term is an effective  vortex density $n_{\rm eff}  = M\omega_\perp/(\pi\hbar)$ for $\Omega =\omega_\perp$, and the second is a (small) correction.

To appreciate the implications of this elegant equation, first assume that the vortex density is uniform with the actual Feynman value $\pi n_v = m\Omega/\hbar$.  Equation (\ref{eq:nvLLL}) then has the Gaussian solution $n_{LLL}(r) = n_{LLL}(0) \exp(-r^2/\sigma^2)$, where $\sigma^{-2} = \pi(n_{\rm eff} -n_v)$; equivalently, the effective squared condensate radius is 
 \begin{equation}
\label{eq:sigma}
\sigma^2 = \frac{\hbar}{M(\omega_\perp-\Omega)} = \frac{d_\perp^2}{1-\bar\Omega}.
\end{equation}
As expected from the radial expansion, the length $\sigma$ diverges as $\Omega\to\omega_\perp$.

 In fact, \textcite{Cooper:2004} use numerical methods to show that a small distortion of the vortex lattice can both lower the energy and significantly change the mean density profile from Gaussian to one closely resembling the parabolic shape in Eq.~(\ref{eq:nLLL}).  Thus it is  better  to substitute this profile into  Eq.~(\ref{eq:nvLLL}) to find the approximate and slightly nonuniform vortex density~\cite{Watanabe:2004,Aftalion:2005} 
 \begin{equation}
\label{eq:nvdist}
n_v(r) \approx  \frac{1}{\pi d_\perp^2} - \frac{1}{\pi R_0^2} \frac{1}{\left(1-r^2/R_0^2\right)^2}.
\end{equation}
This expression  is similar to the result in the mean-field Thomas-Fermi limit in Eq.~(\ref{eq:nvrTF}), and the correction is again small, of order $1/{\cal N}_v$.

\subsubsection{Rapidly rotating anisotropic traps}

In practice, most traps are not exactly axisymmetric (namely, $\omega_x\neq \omega_y$, and I take $\omega_x < \omega_y$ for definiteness).  The rotational properties of such realistic  traps differ in certain significant ways from those of symmetric traps.  Similar problems arise in the study of deformed nuclei~\cite{Valatin:1956}.   If the anistropic trap is stationary, the noninteracting single-particle ground state is simply a product of two one-dimensional Gaussians in $x$ and $y$ with oscillator lengths $d_x$ and $d_y$.   With increasing  rotation speed $\Omega$, however, the coupling term $-\Omega L_z$ mixes the $x$ and $y$ components.  As a result, the ground-state density  elongates along the direction of  weak confinement  (here along $x$)  and diverges as $\Omega\to \omega_x$~\cite{Linn:2001}.  

In complete analogy with the operators $a_\pm$ and $a_\pm^\dagger$ for the symmetric trap [see Eq.~(\ref{apm})], a rotating anisotropic  trap has a similar set of creation and destruction operators, but their form is rather more intricate~\cite{Valatin:1956,Fetter:2007}.  Nevertheless, in the limit $\Omega\to \omega_x$, the lowest single-particle states constitute a nearly degenerate set that again acts like the lowest Landau level.  The corresponding LLL condensate wave function again involves a polynomial in a scaled complex variable (a linear combination of $x$ and $i \,y$).  Consequently, many  of the results derived for the mean-field LLL regime in an isotropic trap continue to apply, at least qualitatively, for the rapidly rotating  anisotropic trap.  

One important difference is the significant anisotropy of the condensate as $\Omega\to \omega_x$~\cite{Linn:2001,Fetter:2007}. Indeed, \textcite{Sanchez:2005} and \textcite{Sinha:2005} both model this situation with a one-dimensional effectively infinite strip of finite width, containing one or more rows of vortices.  As discussed in Sec.~VI  below, however, this behavior may well be unobservable.  Before reaching such a one-dimensional limit,  there may well be a quantum phase transition to a correlated many-body state (sometimes called a ``vortex liquid'') that is not superfluid and has no BEC.  Such questions currently remain unanswered and merit more study.

\section{Quantum Phase Transition to Highly Correlated States}

The mean-field LLL regime relies on the assumption that the interaction energy per particle $\sim \mu$ is small compared to the gap $2\hbar\omega_\perp$, leading to the lower bound for $\bar\Omega$ [compare Eq.~(\ref{eq:LLLOK})]
\begin{equation}
\label{eq:LLLOKa}
1-  \frac{Z}{2N\tilde a} \lesssim \bar {\Omega}
\end{equation}
where $\tilde a = \beta a$ is  the renormalized scattering length that includes the vortex-induced density variation  (note that $\beta\approx 1.1596$ is  simply a numerical factor, see Sec.~V.C.2 for more details).
 In practice, typical experiments require $\bar\Omega\gtrsim 0.99$ to reach this regime.  For the present purpose, it is important to  emphasize that  the mean-field LLL regime has a BEC with macroscopic occupation  and is indeed a superfluid.  Specifically, the associated many-body GP ground state is simply the Hartree product in Eq.~(\ref{eq:PsiGP}).

For still larger values of $\bar\Omega$, however, the mean-field LLL regime should eventually  disappear through a quantum phase transition, leading  to a wholly different highly correlated many-body ground state $\Psi_{\rm corr}$.  Unlike the mean-field LLL ground state, the new ground state $\Psi_{\rm corr}$ does {\it not} have a BEC  and is {\it not} a superfluid.  As evidence for such a transition, the exact two-dimensional ground state for bosons with small $N$ and relatively  large fixed angular momentum $L_z$ indeed has a correlated form~\cite{Wilkin:1998,Wilkin:2000}.  Furthermore, detailed numerical studies~\cite{Cooper:2001} explicitly exhibit such a phase transition.  

To understand the physics of this quantum phase transition, it is helpful to focus on the number of vortices ${\cal N}_v$.
In the mean-field LLL regime, the two-dimensional condensate has a radius $R_0$, and 
Eq.~(\ref{eq:Nv}) yields the number of vortices  ${\cal N}_v = R_0^2/l^2\approx R_0^2/d_\perp^2$.  Use of   Eq.~(\ref{eq:RL}) and the discussion in Sec.~V.C.2 give
\begin{equation}
\label{eq:NvLLL}
{\cal N}_v \approx  \frac{R_{0}^2}{d_\perp^2} = \sqrt{\frac{8N\tilde{a}}{Z(1-\bar\Omega)}},
\end{equation}
  As an alternative derivation of this result~\cite{Bloch:2007}, note that the $m$th LLL state $\psi_m$ has a mean-square radius $\sim m d_\perp^2$ for $m \gg 1$ [Eq.~(\ref{eq:rsq})].  If the sum in Eq.~(\ref{eq:LLL}) extends to $m_{\rm max}$, then the condensate has an effective squared radius $\langle r^2 \rangle_{m_{\rm max}} \approx m_{\rm max} d_\perp^2$.  The number of vortices (the degree of the polynomial) is ${\cal N}_v = m_{\rm max}$, which is the same as Eq.~(\ref{eq:NvLLL}) found directly from the Feynman relation for the vortex density.
The parameters used below Eq.~(\ref{eq:LLLOK}) yield ${\cal N}_v\approx 200$.  

It is valuable to consider the  ratio $\nu \equiv N/{\cal N}_v$~\cite{Wilkin:2000,Cooper:2001} known as the ``filling factor''  or ``filling fraction'' because of a similar quantity in the fractional quantum Hall effect [see, for example, Chap.~1 of \textcite{Prange:1987}]. 
  In the mean-field LLL regime,  it has the simple form
\begin{equation}
\label{eq:nuLLL}
\nu \equiv \frac{N}{{\cal N}_v} = \sqrt{\frac{Z(1-\bar\Omega)N}{8\tilde a}},
\end{equation}
but it also  remains well defined for  general many-body ground states of rotating bosons [and for charged fermions in a magnetic field, which is its original role~\cite{Prange:1987}].

In  the mean-field LLL approach for weakly interacting bosons in a rotating trap, typified by the work of \textcite{Butts:1999} and discussed in Secs.~III.C and V.C, the equilibrium state is a vortex array that breaks rotational symmetry and is not an eigenstate of $L_z$.
As an alternative, \textcite{Cooper:2001} use exact diagonalization to study the  ground states for  ${\cal N}_v = 4, 6$, and $8$ and for many   values of $N$  [see also \cite{Wilkin:2000,Cooper:1999,Viefers:2000} for less extensive studies].  This numerical analysis uses a toroidal geometry with periodic boundary conditions in both directions, allowing a study of bosonic systems with relatively  large integral  values of both $N$ and ${\cal N}_v$.   For a particular aspect ratio of the torus that can accommodate a triangular array with ${\cal N}_v = 8$, they investigate filling fractions $\nu$ ranging through several  rational values from $\frac{1}{2} $ up to $9$.  They also study the mean-field GP  ground state containing a vortex lattice with ${\cal N}_v = 8$ for the same values of $\nu$.  Comparison of these two classes of states shows that the GP vortex lattice is the ground state for $\nu_c\gtrsim 6$, where the specific numerical value depends on the aspect ratio of the torus (and hence the number of vortices that can fit into the periodic structure).  Other choices of this aspect ratio yield critical values in the range $\nu_c \sim 6$-$10$.

A combination of Eq.~(\ref{eq:LLLOKa}) for the validity of the mean-field LLL regime and the
 lower bound $\nu\gtrsim 1 $ for the mean-field GP ground state  yields the  inequalities that define the allowed range of the mean-field LLL state~\cite{Bloch:2007}
 \begin{equation}
\label{eq:bounds}
 1- \frac{Z}{2N\tilde a}\lesssim \bar\Omega \lesssim 1-  \frac{8 \tilde a}{ZN} 
\end{equation}
The right-hand inequality for $\bar\Omega$ illustrates   the experimental difficulty of studying the quantum phase transition.    Since $Z/a$ typically exceeds $100$,  Eq.~(\ref{eq:nuLLL}) makes it clear that reaching even $\nu \sim 10$ will require significant reduction of $N$ (note that the high rotation speed  $\bar\Omega \sim 0.999 $  presents a daunting experimental challenge).

In contrast to the vortex lattices for $\nu > \nu_c$,  the ground states for smaller $\nu < \nu_c$ are rotationally symmetric incompressible vortex liquids that are eigenstates of $L_z$.  They have close similarities to the bosonic analogs of the \textcite{Jain:1989} sequence of fractional quantum Hall states that describe a  two-dimensional electron gas in a strong magnetic field.  The simplest of these many-body ground states is the bosonic Laughlin state~\cite{Laughlin:1983}
\begin{eqnarray}
\Psi_{\rm Laughlin}(\bm r_1,\bm r_2,\cdots,\bm r_N)\qquad\qquad\qquad    \nonumber\\
\label{eq:Laughlin}
\qquad \propto \prod_{n<n'}^N \left(z_n-z_{n'}\right)^2 \exp\left(-{\textstyle\frac{1}{2}} \sum_{n = 1}^N |z_n|^2\right),
\end{eqnarray}
where $z_n = (x_n + i y_n)/d_\perp$ refers to the $n$th particle.  Here the power $2$ ensures that this state is symmetric under the interchange of any two particles (in contrast to the fermionic case with power $3$). 
 \textcite{Wilkin:1998,Cooper:1999,Wilkin:2000} note that this state occurs for $\nu = \frac{1}{2}$. 
 
 It is worth emphasizing the difference between this Laughlin ground state and the GP Hartree ground state in Eq.~(\ref {eq:PsiGP}).  For the Laughlin state in Eq.~(\ref{eq:Laughlin}), the first (product) factor means that there is no off-diagonal long-range order~\cite{Yang:1962,Penrose:1956} and hence no BEC.   In addition, the Laughlin state vanishes whenever two particles come together, enforcing the many-body correlations.  Indeed, the usual short-range two-body potential $V(\bm r-\bm r') = g\delta(\bm r-\bm r')$ has zero expectation value in this correlated state~\cite{Trugman:1985}.  In contrast, the GP ground state has every particle in the same single-particle Hartree state $\psi_{LLL}$;  thus particles in the GP condensate tend to overlap wherever $|\psi_{LLL}(\bm r)|$ is large.
 
 \textcite{Sinova:2002} provide an appealing physical picture of the onset of this quantum phase transition. Using various approximate descriptions, they find that the long-wavelength spectrum of collective modes becomes quadratic in the limit of rapid rotations, in contrast to the familiar linear Bogoliubov spectrum in Eq.~(\ref{eq:phonon}).  This softened spectrum implies that the vortex lattice melts for a critical filling fraction $\nu_c\approx 8$, in reasonable agreement with the estimates of \textcite{Cooper:2001}.  An alternative description of the vortex-lattice melting~\cite{Baym:2005} relies on the softening of the Tkachenko vortex-lattice modes, as seen in Eq.~(\ref{eq:Tka,comp}).  The connection between these apparently distinct pictures remains uncertain.
 
The periodic toroidal geometry of \textcite{Cooper:2001} has the strong advantage of avoiding boundary effects;   \textcite{Regnault:2003,Regnault:2004}  use a spherical geometry to perform a similar numerical analysis.  Both groups propose specific sequences of known quantum Hall states for  the rotating dilute Bose gas when $\frac{1}{2} < \nu< \nu_c$, although there is currently no experimental evidence for such sequences.  Nevertheless,  \textcite{Cooper:2001} find strong  overlap between  their results from exact diagonalization and the Laughlin state ($\nu = \frac{1}{2}$) and other similar but more complicated quantum Hall states for the particular filling fractions  $\nu = 1, \frac{3}{2}, 2,\frac{5}{2},6$.   For a more detailed discussion of these various quantum Hall states, see, for example, \cite{Cooper:2001,Regnault:2003,Regnault:2004} and  Sec.~VII.C of \textcite{Bloch:2007}.

What is the physics of this ground-state transition from a vortex lattice to a vortex liquid? Why is the critical value of the filling fraction  $\nu_c \sim 5-10$, as opposed to $\nu_c \sim 1$ or $\nu_c \sim 0.1$?  To think about this question, it is helpful to consider $N$ bosonic particles in a plane, with $2N$ degrees of freedom.  Vortices appear as the system rotates,  and the corresponding vortex coordinates provide ${\cal N}_v $  collective degrees of freedom.\footnote{Note that vortices obey first-order dynamical equations, so that the associated $x$ and $y$ coordinates are not independent;  in particular, they can serve as a pair of Hamiltonian canonical variables~\cite{Fetter:1967,Haldane:1985}.}  This relation  between particle and collective degrees of freedom is familiar in many areas of many-body theory:  For example, the Debye theory of specific heat uses the phonon normal modes instead of the particle positions in the crystal, and the original number of particle degrees of freedom ($3N$) also determines the cutoff frequency for the phonon spectrum [see, for example, Chap.~23 of~\textcite{Ashcroft:1976}].  Similarly,  the theory of a  charged electron gas often relies on collective density modes (plasma oscillations)  instead of the  particle coordinates~\cite{Pines:1961}.

 For a slowly rotating Bose-Einstein gas in a plane, the $2N$ particle coordinates provide a convenient description.    In principle, the ${\cal N}_v$ collective vortex degrees of freedom should reduce the  original total $2N$ degrees of freedom to $2N - {\cal N}_v$, but the effect is unimportant as long as ${\cal N}_v \ll N$.  When ${\cal N}_v$ becomes comparable with $N$, however, the depletion of the particle degrees of freedom becomes crucial, ultimately producing a phase transition to  a wholly different ground state.  With this picture, the critical value $\nu_c \sim 5-10$ is not surprising, since ${\cal N}_v/N= 1/\nu$ is still relatively small.

If  experiments can indeed create these correlated states, how might they be detected?  Section VII.C of \textcite{Bloch:2007} proposes several techniques, including (1) a reduced three-body recombination rate  [because of the anti-correlation between the bosons inherent in the first factor in Eq.~(\ref{eq:Laughlin})] and (2) the characteristic density profile of these various quantum Hall states.  Another possibility would be to study the ensemble averaged  density-density noise correlations~\cite{Altman:2004,Brown:1956}, which should exhibit a different structure in the correlated state from that in the mean-field LLL regime.  

\section{Outlook}

As this review demonstrates in  detail, experimental and theoretical studies of rotating Bose-Einstein condensates generally agree well for  scaled angular velocities $\bar\Omega \equiv \Omega/\omega_\perp$ up to $\sim 0.995$. 

\subsection{Successes}  These accomplishments include several different regimes.  
\begin{enumerate}
\item For slow rotation $\bar \Omega \equiv \Omega/\omega_\perp\ll 1$ (Sec.~III), the condensate  includes only a few vortices, and  centrifugal effects are negligible.  Thus the overall density profile is essentially that of the nonrotating condensate, apart from the small holes associated with the vortex cores.  Experimental observations of vortex precession in one- and two-component superfluids (see Sec.~III.D.1) provide the only direct measurements of vortex dynamics in  dilute BECs.
\item  As the rotation increases, the centrifugal effects become important, leading to significant changes in the condensate's aspect ratio (Sec.~IV).  For not too large $\bar\Omega$, however, the intervortex spacing $\sim l= \left[\hbar/(M \Omega)\right]^{1/2}$  remains large compared to the vortex core size $\xi$.  In this mean-field Thomas-Fermi regime, the gradient energy associated with  density variation remains negligible, and  the hydrodynamic  flow velocity $\bm v = \hbar \bm\nabla S/M$ predominates in the energy balance.  Typically this case holds for $0.75\lesssim \bar\Omega\lesssim 0.99$.
\item Finally, for very large rotation speed $0.99 \lesssim \bar\Omega \lesssim 0.999$, the vortex cores expand, filling much of the available space.  The energy associated with spatial variation of the density now becomes  comparable to the hydrodynamic flow energy.  In this mean-field lowest-Landau-level regime (Sec.~V),  the single-particle part of the energy in the rotating frame is exactly soluble, leading to the rotational analog of the Landau levels for an electron in a uniform magnetic field.  In particular, there is an energy  gap $\sim 2\hbar \omega_\perp$ between the nearly degenerate lowest Landau level and the first excited Landau level.  In addition,  the condensate expands radially, decreasing both the central particle density and the mean interaction energy $\mu$, which thus becomes smaller than the energy gap $2\hbar\omega_\perp$.  The set of lowest-Landau-level states then provides a convenient and flexible description of  the equilibrium of the rapidly rotating condensate.  Although current experiments  with $\mu/(2\hbar\omega_\perp)\approx 0.6$ only just reach this  mean-field LLL  regime (see Sec.~V.A), it is reasonably well understood. 
\end{enumerate}

\subsection{Challenges}

Despite the many remarkable achievements of the past decade, several theoretical predictions of considerable interest remain unverified experimentally.

\subsubsection{Rapid rotation of both Bose and Fermi gases in the quantum Hall regime at small filling fraction}
For dilute ultracold  trapped single-component Bose gases of the sort considered in this review, one obvious challenge for future experiments is  the quantum phase transition from a superfluid  with a Bose-Einstein condensate to a nonsuperfluid correlated state. As discussed in  Sec.~VI, formation of this new state requires large angular momentum and smaller particle number $N$ than is typical in most current experiments.  Whether such states can indeed be realized is uncertain.  In principle, one might start with an anharmonic trap of the sort considered in Sec.~IV.C.  Once the condensate rotates with $\bar\Omega \gtrsim 1$, it might be feasible  adiabatically to turn off  the external laser beam (and hence  the anharmonic component of the trap potential), raising the effective $\omega_\perp$ and leaving the system with high angular velocity $\bar \Omega \lesssim 1$.

Another possibility is be to rely on topological gauge potentials (Sec.~V.B), in which applied laser fields with orbital angular momentum can produce effective nonuniform  topological ``magnetic''  fields that are more general than the usual constant $\bm B$. To my knowledge, all  studies of rapidly rotating trapped bosons rely on the detailed form of the  operator $-\bm \Omega\cdot \bm L = -\bm A \cdot \bm p$, with $\bm A =  \bm \Omega \bm \times \bm r$ as the equivalent vector potential.   The existence of a nonsuperfluid correlated many-body state reflects the high degeneracy of the lowest Landau level, which in turn  depends on the specific form of the effective vector potential ($\bm A = \bm \Omega\bm \times \bm r$, apart from a possible gauge transformation).   It is not  obvious that  such nonsuperfluid correlated many-body states occur for more general  topological gauge potentials.  This question certainly merits additional  detailed study. 

A related area for future activity is the study of rapidly rotating Fermi gases.  At present, this theoretical topic is only partially explored [see, for example,~\cite{Zhai:2006,Moller:2007,Bhongale:2007}], and much work remains. The only relevant experiments  are those of \textcite{Zwierlein:2005} on the vortex lattice in rotating fermionic $^6$Li (see Sec.~IV.B.6).  The experiment starts with a BEC of tightly bound bosonic molecules and then sweeps across the Feshbach resonance toward the more weakly bound fermionic regime.  Unfortunately, any move  far to the BCS side of the transition at fixed low temperature necessarily  encounters the severely reduced critical temperature of the weak-coupling BCS limit. 

As a pure theoretical argument, let us consider the weak-coupling fermionic ground state as a function of $\bar\Omega = \Omega/\omega_\perp$.  For intermediate $\bar\Omega$, a vortex lattice forms, as in \textcite{Zwierlein:2005}.  What happens for larger $\bar\Omega\,$? Is there a  transition to a normal state similar to that at $H_{c2}$ in a type-II superconductor [see, for example, Secs.~4.7, 4.8, and 5.3.3 of \textcite{Tinkham:1996}]?  Or does a transition to a correlated quantum Hall state intervene?  If so, for reasonable experimental parameters, how  does the filling fraction $\nu$ relate to $\bar\Omega\,$?  These intriguing questions deserve careful study.

\subsubsection{Rotation of ultracold dilute gases with long-range dipole-dipole interactions}

Dilute quantum gases with long-range anisotropic dipole-dipole interactions constitute a fascinating generalization of the usual case with  short-range isotropic interactions characterized by a scattering length $a$.  Although the  general dipole-dipole potential is rather complicated, experiments  usually use a strong uniform  field to orient the dipoles (either magnetic or electric) along the $\hat{\bm z}$ direction.  Such fields lead to the simpler interaction potential [see, for example, the  review by \textcite{Menotti:2007} and the experimental studies of $^{52}$Cr~\cite{Griesmaier:2005,Lahaye:2007,Koch:2007}]
\begin{equation}
\label{eq:Vdd}
V_{\rm dd}(\bm r - \bm r') = \frac{d^2}{|\bm r-\bm r'|^3} \left[1-3\left(\hat{\bm z}\cdot\hat{\bm n}\right)^2 \right],
\end{equation}
where $d$ is the relevant dipole moment and $\hat{\bm n}$ is a unit vector along the direction $\bm r-\bm r'$ joining the two atoms.  

The dipole-dipole potential in Eq.~(\ref{eq:Vdd}) is repulsive if $\hat{\bm z}\cdot\hat{\bm n}$ vanishes (namely if the oriented dipoles lie in the $xy$ plane), whereas it is attractive if $(\hat{\bm z}\cdot\hat{\bm n})^2  > \frac{1}{3}$ (in particular,  for oriented dipoles located  along the $\hat{\bm z}$ axis).   Since a BEC with attractive interactions  tends to collapse, a long cigar-shaped condensate with dipole-dipole interactions should be unstable,  but a flat disk-shaped condensate should be stable.  \textcite{Koch:2007} use a Feshbach resonance to decrease the $s$-wave scattering length, thus enhancing the relative role of $V_{\rm dd}$.  These experiments  confirm the qualitative expectations in great detail.

Recent theoretical studies have predicted many properties of the ground states of rotating BECs with dipolar interactions.  In the weak-interaction limit, \textcite{Cooper:2005} [see also \cite{Zhang:2005}] propose that  an increased  dipolar coupling strength induces a series of transitions in the  mean-field (large filling fraction $\nu$) condensate   involving vortex lattices of varying symmetry (triangular, square, stripe, and ``bubble'' phases).  They also study the behavior for small $\nu$, when the system is no longer superfluid.  Subsequently, \textcite{Komineas:2007}  analyze the  more general situation as a function of both the dipolar interaction and the chemical potential relative to the $s$-wave contact interaction. 

Another system of great interest is the rapidly rotating dipolar Fermi gas, whose predicted ground state~\cite{Baranov:2005}  has  a  simple incompressible form for filling fraction $\nu = \frac{1}{3}$.  For  smaller $\nu$, however, theory predicts  that the  quantum Hall states ultimately  become unstable  to the formation of crystalline Wigner states~\cite{Osterloh:2007}. Experimental studies of such nonsuperfluid fermionic states involve many technical issues that remain for  future investigation.

\subsubsection{Rotating spinor condensates}

The development of optical laser traps that confine all the magnetic substates~\cite{Stamper:1998} allows a direct study of spin-one BECs (as well as higher-spin cases).  These experiments use a focused  (red-detuned) infrared laser  that attracts the BEC to the waist of the beam.  For a general theoretical  introduction, see Sec.~III.E, along with Sec.~12.2 of \textcite{Pethick:2002}, Sec.~3.4 of \textcite{Fetter:2002}, and the original papers  \textcite{Ho:1998} and \textcite{Ohmi:1998}.  In contrast to  the two-component mixtures of the  different hyperfine states $F=1$ and $F=2$ (see Secs.~III.D.1 and IV.B.5), the rotational invariance of the interaction between two  atoms with the same $F=1$ now imposes  special restrictions on the allowed interaction potential.  The macroscopic order parameter becomes a spin-one object with three components
\begin{equation}
\label{eq:spinor}
\bm\Psi = \left(
\begin{matrix}
\Psi_1\\
\Psi_0 \\
\Psi_{-1}
\end{matrix}
\right)
\end{equation}
labeled by the $m_F$ value.  In the low-energy limit where only $s$-wave scattering is relevant, the effective short-range interaction has the form 
\begin{equation}
\label{eq:Vspinor}
V_{\rm int} (\bm r_1 -\bm r_2) = \left(g_0 + g_2 \bm F_1\cdot \bm F_2\right) \, \delta^{(3)}(\bm r_1-\bm r_2),
\end{equation}
where $\bm F_j$ is the hyperfine spin of the $j$th atom.  The coupling constant $g_0$ is generally positive, but $g_2$ is proportional to $a_2-a_0$ (the difference of the scattering lengths in the two symmetric channels $\bm F = \bm F_1+\bm F_2 = 0$  or 2).  Thus $g_2$ can be either positive  or negative, depending on the details of the two-body scattering.

A mean-field description~\cite{Ho:1998}  uses an effective energy functional and writes the spinor order parameter in a factored form $\Psi_\alpha (\bm r) = \sqrt{n(\bm r)} \,\zeta_\alpha (\bm r)$.  Here, $\alpha$ runs over the three allowed values $\alpha = $ ($1, 0, -1$), $n(\bm r)$ is the assumed common density for all three spin components, and $\zeta_\alpha (\bm r)$ is a normalized spinor with $\bm \zeta^\dagger\cdot\bm \zeta = 1$.  Apart from the gradient term in the effective energy, the ground-state spinor follows by minimizing the spin-dependent part of the energy density $\frac{1}{2} n^2 g_2 \langle \bm F\rangle ^2$, where $\langle \bm F \rangle = \sum_{\alpha\beta} \zeta_\alpha^*\, \bm F_{\alpha \beta} \,\zeta_\beta$.  

If $g_2$ is positive, then the minimum energy occurs for $|\langle\bm F\rangle| = 0$.  Such states are known as  ``polar'' or ``antiferromagnetic,'' and they follow by spatial rotations of the hyperfine state $|m_F = 0\rangle$ [this situation applies for  $^{23}$Na~\cite{Stenger:1998}].  If, however, $g_2$ is negative, then the minimum energy   is for a ``ferromagnetic" state with $|\langle\bm F\rangle |= 1$, obtained by spatial rotations of the state $|m_F = 1\rangle$ [this situation applies  for $^{87}$Rb~\cite{Sadler:2006,Vengalattore:2007}].

The multicomponent structure of the spinor order parameter in Eq.~(\ref{eq:spinor}) is  reminiscent of the order parameter for superfluid $^3$He, where each atom has a nuclear spin $1/2$ and zero net electronic spin. Low-temperature experiments show that $^3$He forms a $p$-wave paired superfluid, with the two atoms in a triplet nuclear spin state and unit relative orbital angular momentum [see, for example, \textcite{Vollhardt:1990}].  In particular,  the $^3$He order parameter has several components, analogous to  those for a spinor BEC. Many groups have studied the structure of  vortices in rotating $^3$He, with various  proposed candidates, both for a single vortex and for vortex lattices  [for detailed reviews, see  \textcite{Fetter:1986,Salomaa:1987}].  

Unfortunately,  the semi-phenomenological Ginzburg-Landau theory for superfluid $^3$He is often inadequate to predict the detailed experimental phase diagram for various vortices as a function of pressure and temperature.  Specifically,   the energy difference between many of the vortex states  is small,  and  the relevant phenomenological Ginzburg-Landau  parameters have considerable uncertainties their experimental values [see, for example, Secs.~7.5 and 7.6 of \textcite{Vollhardt:1990}].   Consequently, experimental studies provide  essential guidance for theoretical analyses.  \textcite{Eltsov:2005} and \textcite{Finne:2006} review recent experimental work on vortices and other related structures in superfuid  $^3$He.

In an unbounded dilute trapped BEC with a single-component order parameter, the vortex core size and structure arise from the balance of the kinetic energy and the interaction energy (see Sec.~III.A).  For a trapped BEC with a single component, the new feature is the additional energy of the confining trap, and various regimes occur depending on the dimensionless interaction parameter $Na/d_0$ (see Secs.~II.B and III.B.1).  If $Na/d_0\lesssim 1$, then the vortex core is comparable with the condensate radius (both  of order $d_0$, the mean  oscillator length) or the  intervortex separation (see Sec.~III.C).  In  the more typical Thomas-Fermi regime with $Na/d_0\gg 1$, however, the  vortex  core radius $\xi$  (the healing length) is much smaller than both  $d_0$ and the still larger mean condensate radius $R_0$.

For a three-component spinor condensate,  the spatial variation of each component provides  additional  freedom in minimizing the overall free energy. Since the resulting coupled GP field equations are nonlinear, several solutions can exist, at least in principle, each corresponding to a local minimum of the GP energy [see Sec.~7.5.7 of \textcite{Vollhardt:1990} for the similar situation involving the Ginzburg-Landau theory of superfluid $^3$He].  In practice, it turns out that $|g_2|\ll g_0$ for both $^{23}$Na and $^{87}$Rb~\cite{Yip:1999}, which means that gradient energies and other interaction energies play a significant role in determining the  condensate wave function.  This near degeneracy is especially important in understanding the core of a quantized vortex in a  spinor condensate, because these various energies compete with the energy of the circulating flow.  \textcite{Yip:1999} proposes several different vortex structures, and many other authors have amplified these considerations, both for single vortices  [see, for example,  \cite{Mizushima:2002,Bulgakov:2003,Isoshima:2003} and references therein] and for vortex lattices [see, for example,  \cite{Mizushima:2004} and references therein].  One experimental challenge in  studying a rotating spinor BEC  is ensuring both the full axisymmetry  of the  optical trap and its alignment with the axis of rotation;  otherwise the lifetime decreases significantly.    These difficulties appear in  the experiments on  a  rotating paired Fermi gas, where  the magnetic field associated with a Feshbach resonance also requires  an optical trap [see Sec.~IV.B.6 and~\textcite{Zwierlein:2005}].

\begin{acknowledgments} I thank A. Aftalion, E. Demler,  M. Krusius,  D. Roberts,  N. Wilkin,  and E. Zaremba  for valuable  discussions and suggestions.  Part of this work was completed during a stay at the Institut Henri Poincar\'e - Centre Emile Borel, Paris,  and at the Kavli Institute for Theoretical Physics, Santa Barbara, where the research was supported in part by the National Science Foundation under Grant No. PHY05-51164.  I thank these institutions for their hospitality and support.
\end{acknowledgments}

\bibliographystyle{apsrmp}
\bibliography{rotating}
\end{document}